\documentclass[aps,prd,twocolumn,showpacs,superscriptaddress,groupedaddress,amsmath,amssymb]{revtex4}  
\usepackage{graphicx}  
\usepackage{dcolumn}   
\usepackage{bm}        
\usepackage{amssymb}   
\usepackage{multirow} 
\usepackage{graphics}
\usepackage{enumerate} 

\usepackage[usenames]{color}

\begin{document}

\newcommand {\Bd} {\ensuremath{B^0}}
\newcommand {\Bs} {\ensuremath{B^0_s}}
\newcommand {\Bq} {\ensuremath{B^0_q}}
\newcommand {\Ds} {\ensuremath{D_s}}
\newcommand {\barBd} {\ensuremath{\bar{B}^0}}
\newcommand {\barBs} {\ensuremath{\bar{B}^0_s}}
\newcommand {\barBq} {\ensuremath{\bar{B}^0_q}}
\newcommand {\asld} {\ensuremath{a^d_{\mathrm{sl}}}}
\newcommand {\asls} {\ensuremath{a^s_{\mathrm{sl}}}}
\newcommand {\aslq} {\ensuremath{a^q_{\mathrm{sl}}}}
\newcommand {\aslb} {\ensuremath{A^b_{\mathrm{sl}}}}
\newcommand {\aint} {\ensuremath{A^{\mathrm{int}}_{S}}}
\newcommand {\amix} {\ensuremath{A^{\mathrm{mix}}_{S}}}
\newcommand {\aslbip} {\ensuremath{A^b_{\mathrm{sl, IP}}}}
\newcommand {\ktomu} {\ensuremath{K \to \mu}}
\newcommand {\pitomu} {\ensuremath{\pi \to \mu}}
\newcommand {\ptomu} {\ensuremath{p \to \mu}}
\newcommand {\ks} {\ensuremath{K_S}}
\newcommand {\kstp} {\ensuremath{K^{*+}}}
\newcommand {\kstpm} {\ensuremath{K^{*\pm}}}
\newcommand {\kstneu} {\ensuremath{K^{*0}}}
\newcommand {\phin} {\ensuremath{\phi(1020)}}
\newcommand {\pteta} {\ensuremath{(p_T, |\eta|)}}
\newcommand {\dgg} {\ensuremath{\Delta \Gamma_d / \Gamma_d}}
\newcommand {\mixBd} {\ensuremath{B^0}\mbox{-}\ensuremath{\bar{B}^0}}
\newcommand {\mixBs} {\ensuremath{B^0_s}\mbox{-}\ensuremath{\bar{B}^0_s}}

\widetext

\hspace{5.2in} \mbox{FERMILAB-PUB-13-445-E}

\title{
Study of CP-violating charge asymmetries of single muons and like-sign dimuons \\
in $\bm{p \bar p}$  collisions
}
\affiliation{LAFEX, Centro Brasileiro de Pesquisas F\'{i}sicas, Rio de Janeiro, Brazil}
\affiliation{Universidade do Estado do Rio de Janeiro, Rio de Janeiro, Brazil}
\affiliation{Universidade Federal do ABC, Santo Andr\'e, Brazil}
\affiliation{University of Science and Technology of China, Hefei, People's Republic of China}
\affiliation{Universidad de los Andes, Bogot\'a, Colombia}
\affiliation{Charles University, Faculty of Mathematics and Physics, Center for Particle Physics, Prague, Czech Republic}
\affiliation{Czech Technical University in Prague, Prague, Czech Republic}
\affiliation{Institute of Physics, Academy of Sciences of the Czech Republic, Prague, Czech Republic}
\affiliation{Universidad San Francisco de Quito, Quito, Ecuador}
\affiliation{LPC, Universit\'e Blaise Pascal, CNRS/IN2P3, Clermont, France}
\affiliation{LPSC, Universit\'e Joseph Fourier Grenoble 1, CNRS/IN2P3, Institut National Polytechnique de Grenoble, Grenoble, France}
\affiliation{CPPM, Aix-Marseille Universit\'e, CNRS/IN2P3, Marseille, France}
\affiliation{LAL, Universit\'e Paris-Sud, CNRS/IN2P3, Orsay, France}
\affiliation{LPNHE, Universit\'es Paris VI and VII, CNRS/IN2P3, Paris, France}
\affiliation{CEA, Irfu, SPP, Saclay, France}
\affiliation{IPHC, Universit\'e de Strasbourg, CNRS/IN2P3, Strasbourg, France}
\affiliation{IPNL, Universit\'e Lyon 1, CNRS/IN2P3, Villeurbanne, France and Universit\'e de Lyon, Lyon, France}
\affiliation{III. Physikalisches Institut A, RWTH Aachen University, Aachen, Germany}
\affiliation{Physikalisches Institut, Universit\"at Freiburg, Freiburg, Germany}
\affiliation{II. Physikalisches Institut, Georg-August-Universit\"at G\"ottingen, G\"ottingen, Germany}
\affiliation{Institut f\"ur Physik, Universit\"at Mainz, Mainz, Germany}
\affiliation{Ludwig-Maximilians-Universit\"at M\"unchen, M\"unchen, Germany}
\affiliation{Panjab University, Chandigarh, India}
\affiliation{Delhi University, Delhi, India}
\affiliation{Tata Institute of Fundamental Research, Mumbai, India}
\affiliation{University College Dublin, Dublin, Ireland}
\affiliation{Korea Detector Laboratory, Korea University, Seoul, Korea}
\affiliation{CINVESTAV, Mexico City, Mexico}
\affiliation{Nikhef, Science Park, Amsterdam, the Netherlands}
\affiliation{Radboud University Nijmegen, Nijmegen, the Netherlands}
\affiliation{Joint Institute for Nuclear Research, Dubna, Russia}
\affiliation{Institute for Theoretical and Experimental Physics, Moscow, Russia}
\affiliation{Moscow State University, Moscow, Russia}
\affiliation{Institute for High Energy Physics, Protvino, Russia}
\affiliation{Petersburg Nuclear Physics Institute, St. Petersburg, Russia}
\affiliation{Instituci\'{o} Catalana de Recerca i Estudis Avan\c{c}ats (ICREA) and Institut de F\'{i}sica d'Altes Energies (IFAE), Barcelona, Spain}
\affiliation{Uppsala University, Uppsala, Sweden}
\affiliation{Lancaster University, Lancaster LA1 4YB, United Kingdom}
\affiliation{Imperial College London, London SW7 2AZ, United Kingdom}
\affiliation{The University of Manchester, Manchester M13 9PL, United Kingdom}
\affiliation{University of Arizona, Tucson, Arizona 85721, USA}
\affiliation{University of California Riverside, Riverside, California 92521, USA}
\affiliation{Florida State University, Tallahassee, Florida 32306, USA}
\affiliation{Fermi National Accelerator Laboratory, Batavia, Illinois 60510, USA}
\affiliation{University of Illinois at Chicago, Chicago, Illinois 60607, USA}
\affiliation{Northern Illinois University, DeKalb, Illinois 60115, USA}
\affiliation{Northwestern University, Evanston, Illinois 60208, USA}
\affiliation{Indiana University, Bloomington, Indiana 47405, USA}
\affiliation{Purdue University Calumet, Hammond, Indiana 46323, USA}
\affiliation{University of Notre Dame, Notre Dame, Indiana 46556, USA}
\affiliation{Iowa State University, Ames, Iowa 50011, USA}
\affiliation{University of Kansas, Lawrence, Kansas 66045, USA}
\affiliation{Louisiana Tech University, Ruston, Louisiana 71272, USA}
\affiliation{Northeastern University, Boston, Massachusetts 02115, USA}
\affiliation{University of Michigan, Ann Arbor, Michigan 48109, USA}
\affiliation{Michigan State University, East Lansing, Michigan 48824, USA}
\affiliation{University of Mississippi, University, Mississippi 38677, USA}
\affiliation{University of Nebraska, Lincoln, Nebraska 68588, USA}
\affiliation{Rutgers University, Piscataway, New Jersey 08855, USA}
\affiliation{Princeton University, Princeton, New Jersey 08544, USA}
\affiliation{State University of New York, Buffalo, New York 14260, USA}
\affiliation{University of Rochester, Rochester, New York 14627, USA}
\affiliation{State University of New York, Stony Brook, New York 11794, USA}
\affiliation{Brookhaven National Laboratory, Upton, New York 11973, USA}
\affiliation{Langston University, Langston, Oklahoma 73050, USA}
\affiliation{University of Oklahoma, Norman, Oklahoma 73019, USA}
\affiliation{Oklahoma State University, Stillwater, Oklahoma 74078, USA}
\affiliation{Brown University, Providence, Rhode Island 02912, USA}
\affiliation{University of Texas, Arlington, Texas 76019, USA}
\affiliation{Southern Methodist University, Dallas, Texas 75275, USA}
\affiliation{Rice University, Houston, Texas 77005, USA}
\affiliation{University of Virginia, Charlottesville, Virginia 22904, USA}
\affiliation{University of Washington, Seattle, Washington 98195, USA}
\author{V.M.~Abazov} \affiliation{Joint Institute for Nuclear Research, Dubna, Russia}
\author{B.~Abbott} \affiliation{University of Oklahoma, Norman, Oklahoma 73019, USA}
\author{B.S.~Acharya} \affiliation{Tata Institute of Fundamental Research, Mumbai, India}
\author{M.~Adams} \affiliation{University of Illinois at Chicago, Chicago, Illinois 60607, USA}
\author{T.~Adams} \affiliation{Florida State University, Tallahassee, Florida 32306, USA}
\author{J.P.~Agnew} \affiliation{The University of Manchester, Manchester M13 9PL, United Kingdom}
\author{G.D.~Alexeev} \affiliation{Joint Institute for Nuclear Research, Dubna, Russia}
\author{G.~Alkhazov} \affiliation{Petersburg Nuclear Physics Institute, St. Petersburg, Russia}
\author{A.~Alton$^{a}$} \affiliation{University of Michigan, Ann Arbor, Michigan 48109, USA}
\author{A.~Askew} \affiliation{Florida State University, Tallahassee, Florida 32306, USA}
% ...
%
%
\collaboration{The D0 Collaboration\footnote{with visitors from
$^{a}$Augustana College, Sioux Falls, SD, USA,
$^{b}$The University of Liverpool, Liverpool, UK,
$^{c}$DESY, Hamburg, Germany,
$^{d}$Universidad Michoacana de San Nicolas de Hidalgo, Morelia, Mexico
$^{e}$SLAC, Menlo Park, CA, USA,
$^{f}$University College London, London, UK,
$^{g}$Centro de Investigacion en Computacion - IPN, Mexico City, Mexico,
$^{h}$Universidade Estadual Paulista, S\~ao Paulo, Brazil,
$^{i}$Karlsruher Institut f\"ur Technologie (KIT) - Steinbuch Centre for Computing (SCC)
and
$^{j}$Office of Science, U.S. Department of Energy, Washington, D.C. 20585, USA.
}} \noaffiliation
\vskip 0.25cm
\date{October 1, 2013}

\begin{abstract}
We measure the inclusive single muon charge asymmetry and the like-sign
dimuon charge asymmetry in $p \bar{p}$ collisions
using the full data set of 10.4 fb$^{-1}$
collected with the D0 detector at the Fermilab Tevatron.
The standard model predictions of the
charge asymmetries induced by CP violation
are small in magnitude compared to the current experimental precision, so
non-zero measurements could indicate
new sources of CP violation.
The measurements differ from the standard model predictions
of CP violation in these asymmetries
with a significance of $3.6$ standard
deviations. These results are interpreted in a framework of $B$ meson mixing
within the CKM formalism
to measure
the relative width difference $\dgg$ between the mass eigenstates of the $\Bd$ meson system,
and the semileptonic charge asymmetries $\asld$ and $\asls$
of $\Bd$ and $\Bs$ mesons respectively.
\end{abstract}

\maketitle

\section{Introduction}
\label{Introduction}

The D0 collaboration has published three measurements of the
like-sign dimuon charge asymmetry in $p\bar{p}$ collisions at a center-of-mass
energy of $\sqrt{s}=1.96$~TeV at the Fermilab Tevatron \cite{D01,D02,D03}.
All these measurements have consistent results. The asymmetry obtained
with 9 fb$^{-1}$ of integrated luminosity \cite{D03} deviates from the standard model (SM) prediction
by 3.9 standard deviations, assuming that the only
source of charge asymmetry is CP violation in
meson-antimeson mixing of neutral $B$ mesons.

In this article we present the final measurement of
the like-sign dimuon charge asymmetry using the full data set with
an integrated luminosity of 10.4 fb$^{-1}$ collected
from 2002 until the end of Tevatron Run II in 2011.
We use $6 \times 10^6$ like-sign
dimuon events in our analysis.
We obtain the raw like-sign dimuon charge asymmetry
$A \equiv (N^{++} - N^{--})/(N^{++} + N^{--})$ by counting
the numbers $N^{++}$ and $N^{--}$ of events
with two positive or two negative muons, respectively.
We identify several background processes producing the detector-related charge asymmetry
$A_{\rm bkg}$.
We obtain the residual like-sign dimuon charge asymmetry
$A_{\rm CP}$, which is the asymmetry from CP-violating processes,
by subtracting the asymmetry $A_{\rm bkg}$ from the raw asymmetry $A$.

We also collect events with at least one muon. The number
of events in this sample is $2 \times 10^9$.
We obtain the raw inclusive single muon charge asymmetry
$a \equiv (n^+ - n^-)/(n^+ + n^-)$ by counting the numbers $n^+$
and $n^-$ of positive and negative muons, respectively.
We measure the detector-related charge asymmetry $a_{\rm bkg}$
contributing to the raw asymmetry $a$.
The residual inclusive single muon charge
asymmetry $a_{\rm CP}$
is obtained by subtracting the background asymmetry $a_{{\rm bkg}}$ from $a$.
The asymmetry $a_{\rm CP}$ is found to be consistent with zero, and provides an important
closure test for the method to measure the background asymmetries $a_{\rm bkg}$
and $A_{\rm bkg}$.

The dominant contribution to the inclusive single muon and like-sign
dimuon background asymmetries $a_{\rm bkg}$ and $A_{\rm bkg}$ comes from the charge asymmetry of the
muons produced in the decay in flight of charged kaons
$K^- \to \mu \bar \nu$
\cite{cpconj}
or kaons that punch-through the absorber material of the D0 detector into the outer muon system.
The interaction cross-sections of positive and negative kaons with the detector material
are different \cite{PDG}, resulting in positive kaons having a longer inelastic interaction
length than negative kaons. Positive kaons hence have a higher probability to decay,
or to punch-through and produce a muon signal before they are absorbed in the detector material.
Therefore, a critical measurement
in this analysis, the fraction of muons from
kaon decay or punch-through, is measured in data.

The detector-related systematic uncertainties of $a_{\rm bkg}$ and $A_{\rm bkg}$ are
significantly reduced in our measurement
by a special feature of the D0 experiment -- the reversal of magnets
polarities. The polarities of the toroidal and solenoidal magnetic fields were reversed
on average every two weeks so that the four solenoid-toroid polarity
combinations were exposed to approximately the same
integrated luminosity. This allows for a cancellation of first-order
effects related to the instrumental charge asymmetries~\cite{D01}.


The main expected source of like-sign dimuon events in $p \bar{p}$ collisions
are $b \bar{b}$ pairs. One $b$ quark
decays semileptonically to a ``right-sign"
muon, i.e., to a muon of the same charge sign as the parent $b$ quark at
production. The other $b$ quark can produce a ``wrong-sign" muon with its
charge opposite to the charge of the parent $b$ quark. The origin of this
``wrong-sign" muon is either due to $\mixBd$ or $\mixBs$ oscillation,
or the sequential decay $b \to c \to \mu^+$.
These processes produce CP violation in both mixing \cite{Grossman}
and in the interference of $\Bd$ and $\Bs$ decay amplitudes with and
without mixing \cite{cpv-source}.
CP violation in interference was not considered
in \cite{D01,D02,D03}, while it is taken into account in this paper.

An example of a process in which CP violation in mixing can occur is \cite{Branco}
\begin{eqnarray}
p \bar{p} & \rightarrow & b \bar{b} X, \nonumber \\
b & \rightarrow & b~\mbox{hadron} \rightarrow \mu^- (\mbox{``right-sign" }\mu), \nonumber \\
  \bar{b} & \rightarrow & B^0_{(s)} \rightarrow \bar{B}^0_{(s)}
\rightarrow \mu^-  (\mbox{``wrong-sign"}\mu);
\end{eqnarray}
and its CP-conjugate decay resulting in $\mu^+ \mu^+$, where the probability of
$B^0_{(s)} \to \bar B^0_{(s)}$ is not equal to the probability of $\bar B^0_{(s)} \to B^0_{(s)}$.

An example of a process in which CP violation in interference can occur is  \cite{cpv-source}
\begin{eqnarray}
p \bar{p} & \to & b \bar{b} X, \nonumber \\
b & \rightarrow & b~\mbox{hadron} \rightarrow \mu^- (\mbox{``right-sign"}\mu), \nonumber \\
\bar{b} & \rightarrow & B^0 (\to \bar{B}^0) \rightarrow D^+ D^-,  \nonumber \\
  D^-  & \rightarrow  & \mu^- (\mbox{``wrong-sign"}\mu);
\end{eqnarray}
and its CP-conjugate decay resulting in $\mu^+ \mu^+$, where the probability of
$\Bd ( \to \barBd) \to D^+ D^-$ is not equal to the probability of $\barBd (\to \Bd) \to D^+ D^-$.



The SM prediction of the like-sign dimuon charge asymmetry, and its uncertainty,
are small in magnitude compared
to the current experimental precision \cite{Nierste,cpv-source}.
This simplifies the search for new
sources of CP violation beyond the SM which could contribute to the
like-sign dimuon charge asymmetry. 
Currently, the only established source of CP violation is the complex
phase of the Cabibbo-Kobayashi-Maskawa (CKM) matrix \cite{ckm}.
Although the CKM mechanism is extremely successful in describing all known CP
violating processes studied in particle physics \cite{hfag},
it is insufficient to explain the dominance of matter in the universe \cite{newcp}.
The search for new sources of CP violation beyond the SM
is therefore important in current and future particle physics experiments.

Many features of the present measurement remain
the same as in our previous publications,
so that all details not described explicitly in this paper can be found in Refs.~\cite{D02,D03}.
The main differences of the present analysis with respect to \cite{D03} are:
\begin{itemize}

\item The muon quality selections are the
same as in \cite{D03} except for the requirement of the
number of track measurements in the silicon microvertex tracker (SMT).
This change is discussed in Section~\ref{selection}.

\item The main emphasis of the present measurement
is on the dependence of the charge asymmetry on
the momentum of the muons transverse to the
beam, $p_T$, on the muon pseudorapidity, $\eta$
\cite{eta}, and
on the muon impact parameter in the transverse plane, IP
\cite{ip}.
The reason is to identify the detector-related effects
that contribute to the observed asymmetry, and to help
understand the origin of the asymmetry.

\item In Refs. \cite{D02, D03} we measured the $K \to \mu$ fraction
\cite{ktomu}
by reconstructing the decays
$K^{*0}(892) \rightarrow K^+ \pi^-$ with
$K^+ \rightarrow \mu^+ \nu$,
$K^{*+}(892) \rightarrow K_S \pi^+$, and
$K_S \rightarrow \pi^+ \pi^-$.
This method requires a correction
for muons with large IP that is described
in Section~\ref{secfk}. We have now also developed
an independent method to
obtain the background fractions using local measurements
of the muon momentum by the muon identification system.
This method, described in Section~\ref{sec_local},
is inherently insensitive to the muon IP.
The comparison between these two methods provides
an important validation of our measurement technique and
estimate of the systematic uncertainties.

\item We present the results in terms of model independent residual asymmetries
$a_{\rm CP}$ and $A_{\rm CP}$
and the deviation of these asymmetries from the SM prediction.
Assuming that the only sources of the like-sign dimuon
charge asymmetry are CP violation in mixing and interference
of neutral $B$ mesons,
we measure the quantities determining
these two types of CP violation: the semileptonic charge asymmetries
$\asld$ and $\asls$ of $\Bd$ and $\Bs$
mesons, respectively, and the relative width difference $\Delta \Gamma_d/\Gamma_d$ of
the $\Bd$ system. These quantities are defined in Section~\ref{expectation}. 
Because our measurements are inclusive,
other as yet unknown sources of CP violation could contribute
to the asymmetries $a_{\rm CP}$
and $A_{\rm CP}$ as well. Therefore, the model-independent asymmetries $a_{\rm CP}$ and
$A_{\rm CP}$ constitute the main result of our analysis. They are presented in a form
which can be used as an input for alternative interpretations.
\end{itemize}

The outline of this article is as follows:
the method and notations are presented in Section~\ref{Method};
the details of data selection
are given in Section~\ref{selection};
the Monte Carlo (MC) simulation used in this analysis is
discussed in Section~\ref{simulation}. The parameters obtained
from data are presented in Sections~\ref{sec_bck} and \ref{sec_abkg}.
The measurement of residual charge asymmetries, after subtracting
all background contributions, is presented in
Section~\ref{sec_results},
the SM contributions to these asymmetries are discussed
in Section~\ref{expectation}, and the interpretation of this measurement
in terms of CP violation in mixing and interference
of neutral $B$ mesons is discussed in Section~\ref{sec_interpretation}.
Finally, the conclusions are collected in Section~\ref{sec_conclusions}.
Appendix~\ref{sec_fit} presents the details of the fitting
procedure used in this analysis.

\section{Method}
\label{Method}

\begin{table}
\caption{\label{IP0}
Definition of the IP samples for inclusive muons.}
\begin{ruledtabular}
\begin{tabular}{cc}
IP sample & IP \\
\hline
1 & 0 -- 50 $\mu$m \\
2 & 50 -- 120 $\mu$m \\
3 & 120 -- 3000 $\mu$m \\
\end{tabular}
\end{ruledtabular}
\end{table}

\begin{table}
\caption{\label{IP1IP2}
Definition of the (IP$_1$,IP$_2$) samples
for like-sign dimuons.}
\begin{ruledtabular}
\begin{tabular}{ccc}
(IP$_1$,IP$_2$) sample & IP$_1$ & IP$_2$ \\
\hline
11 & 0 -- 50 $\mu$m & 0 -- 50 $\mu$m \\
12 & 0 -- 50 $\mu$m & 50 -- 120 $\mu$m \\
13 & 0 -- 50 $\mu$m & 120 -- 3000 $\mu$m \\
22 & 50 -- 120 $\mu$m & 50 -- 120 $\mu$m \\
23 & 50 -- 120 $\mu$m & 120 -- 3000 $\mu$m \\
33 & 120 -- 3000 $\mu$m & 120 -- 3000 $\mu$m \\
\end{tabular}
\end{ruledtabular}
\end{table}

\begin{figure}[!t]
\begin{center}
\includegraphics[width=0.48\textwidth]{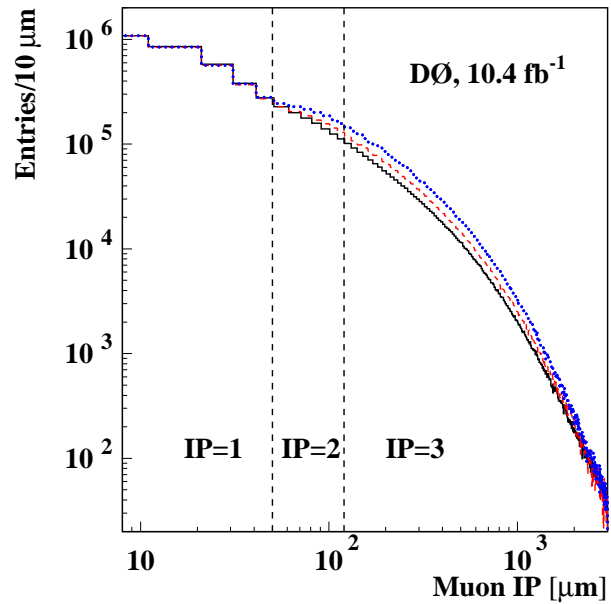}
\caption{IP distributions of one muon in the like-sign dimuon sample when
the other muon has IP in the IP=1 (full line), IP=2 (dashed line), and IP=3 (dotted line) range.
The distributions are normalized to have the same number of entries in the first bin
$[0,10]~\mu$m (only a fraction of this bin is shown in the figure).
The vertical dashed lines show the definition of boundaries of the IP samples.}
\label{fig-ip}
\end{center}
\end{figure}

\begin{table}[!t]
\caption{\label{pT_eta}
Bins of $(p_T, |\eta|)$. Global kinematic requirements are
$1.5 < p_T < 25$ GeV,
($p_T > 4.2$ GeV or $|p_z| > 5.4$ GeV), and $|\eta| < 2.2$.}
\begin{ruledtabular}
\begin{tabular}{ccc}
$(p_T, |\eta|)$ bin & $|\eta|$ & $p_T$ (GeV) \\
\hline
1 & $< 0.7$ & $< 5.6$ \\
2 & $< 0.7$  & 5.6 to 7.0 \\
3 & $< 0.7$  & $> 7.0$ \\
\hline
4 & 0.7 to 1.2 & $< 5.6$ \\
5 & 0.7 to 1.2 & $> 5.6$ \\
\hline
6 & $> 1.2$ & $< 3.5$ \\
7 & $> 1.2$ & 3.5 to 4.2 \\
8 & $> 1.2$ & 4.2 to 5.6 \\
9 & $> 1.2$ & $> 5.6$ \\
\end{tabular}
\end{ruledtabular}
\end{table}

\begin{figure}
\begin{center}
\includegraphics[width=0.48\textwidth]{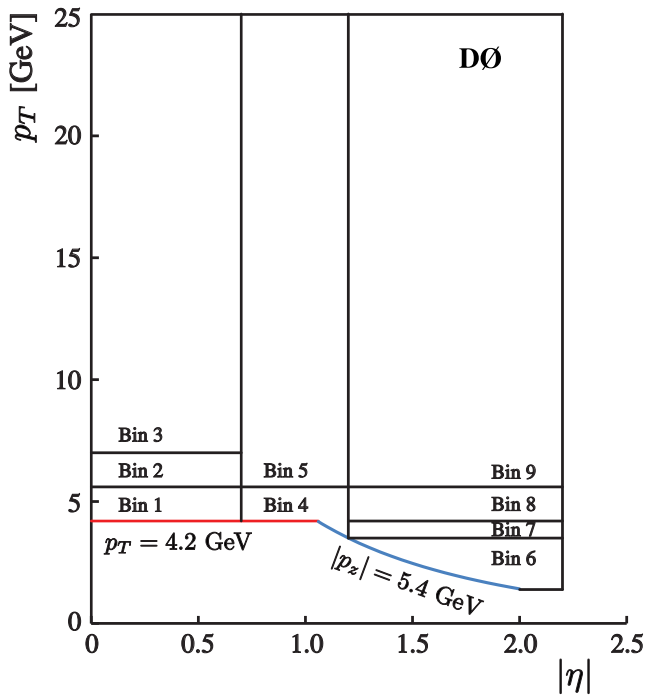}
\caption{Definition of the nine $\pteta$ bins. Global kinematic requirements are
$1.5 < p_T < 25$ GeV,
($p_T > 4.2$ GeV or $|p_z| > 5.4$ GeV), and $|\eta| < 2.2$.}
\label{fig-bins}
\end{center}
\end{figure}

The expressions used in this analysis are described in detail in Ref.\ \cite{D02}.
Here we emphasize the changes to our previous procedure.
We use two sets of data:
\begin{enumerate}[i]
\item the \textit{inclusive muon} data, collected
with inclusive muon triggers, which include all events with
at least one muon candidate passing quality and kinematic requirements described
below;
\item the \textit{like-sign dimuon} data,
collected with dimuon triggers, which include all events
with two muon candidates passing the same
quality and kinematic requirements and the additional
dimuon requirements described in Section~\ref{selection}.
\end{enumerate}

We select muons with $1.5 < p_T < 25$ GeV and $|\eta| < 2.2$.
In addition, we require either $p_T > 4.2$ GeV or $|p_z| > 5.4$ GeV,
where $p_z$ is the momentum of the muon in the proton beam direction.
This selection is applied to ensure that the muon
candidate is able to penetrate all three layers of the central or forward muon detector \cite{D03}.
The upper limit on $p_T$ is applied to suppress the
contribution of muons from $W$ and $Z$ boson decays.
Other muon requirements are discussed in Section~\ref{selection}.

To study the IP, $p_T$, and $|\eta|$ dependence of the
charge asymmetry, we define
three non-overlapping samples of inclusive muons according to the IP value,
or six non-overlapping samples of like-sign dimuons according to the (IP$_1$,IP$_2$) values of the two muons.
Here, IP$_1$ and IP$_2$ are the smaller and larger IP of the two muons, respectively.
The definitions of these samples are given in Tables~\ref{IP0} and \ref{IP1IP2}.
Figure~\ref{fig-ip} shows the IP distributions of one muon in the like-sign dimuon sample when
the other muon has IP in the IP=1, IP=2 or IP=3 range
\cite{ip1}. Note that the two IP's are correlated, and that the IP
distributions span more than four orders of magnitude.
Figure~\ref{fig-ip} also shows the
definition of the boundaries of the IP samples.

These IP samples are additionally divided into
nine exclusive bins of $\pteta$.
Table~\ref{pT_eta} and Figure~\ref{fig-bins} show the definition
of the nine $\pteta$ bins which may have non rectangular shapes due to the $p_T$ and $p_z$
kinematic requirements.


\subsection{Inclusive single muon charge asymmetry}
For a particular IP sample, the raw muon charge asymmetry in each $\pteta$
bin $i$ is given by
\begin{equation}
a^i \equiv \frac{n^+_{i} - n^-_{i}}{n^+_{i} + n^-_{i}}.
\label{ai}
\end{equation}
Here, $n^+_{i}$ ($n^-_{i}$) is the number of positively (negatively) charged muons
in bin $i$. This and all of the following equations are given for a particular
IP sample. However, to simplify the presentation, we drop the index IP from all of them.

The expected inclusive single muon
charge asymmetry, in a given IP sample, can be expressed as
\begin{equation}
\label{inclusive_mu_a}
a^i = a^i_{\rm CP} + a^i_{{\rm bkg}}.
\end{equation}
Here $a^i_{\rm CP}$
is the contribution from CP violation effects in
heavy-flavor decays to muons, and
$a^i_{\rm bkg}$ is the contribution from different background sources not related to
CP violation.

The background contributions come from muons
produced in kaon and pion decay, or from hadrons that punch through the
calorimeter and iron toroidal magnets to reach the outer muon detector.
Another contribution is related to muon detection and identification.
All these contributions are measured with data, with minimal input
from simulation. Accordingly, the background asymmetry $a_{\rm bkg}^i$ can be expressed \cite{D02} as
\begin{equation}
\label{abkg}
a^i_{\rm bkg} = a_\mu^i + f^i_{K} a^i_K + f^i_{\pi} a^i_\pi + f^i_{p} a^i_p.
\end{equation}
Here, the quantity $a_\mu^i$ is the muon detection and identification asymmetry described
later in this section.
The fractions of muons from kaons, pions and protons
reconstructed by the central tracker
\cite{mucand}
in a given
$\pteta$ bin $i$ and misidentified as muons are $f^i_{K}$, $f^i_{\pi}$ and $f^i_{p}$.
Their charge asymmetries are $a^i_K$, $a^i_\pi$ and $a^i_p$, respectively.
We refer to these muons as ``long" or ``$L$" muons, since they are produced by particles traveling long
distances before decaying within the detector.
The tracks of $L$ muons in the central tracker are generally produced by the parent hadron
that subsequently decays at a large radius.
The charge asymmetry of these muons results from
the difference in the interactions of positively and negatively charged particles with the detector material,
and is not related to CP violation.
For charged kaons this difference arises
from additional hyperon production channels in $K^-$-nucleon reactions,
which are absent for their $K^+$-nucleon analogs. Since the interaction
probability of $K^+$ mesons is smaller, they travel further than $K^-$ in the
detector material, and have a greater chance of decaying to muons, and a
larger probability to punch-through the absorber material thereby
mimicking a muon signal. As a result, the asymmetry $a_K$ is positive.


The muon detection and identification asymmetry $a_\mu^i$ can be expressed as
\begin{equation}
\label{deltai}
a_\mu^i \equiv (1-f^i_{\rm bkg}) \delta_i.
\end{equation}
The background fraction $f^i_{\rm bkg}$ is defined as $f_{\rm{bkg}}^i = f^i_{K} + f^i_{\pi} + f^i_{p}$.
The quantity $\delta_i$ is the charge asymmetry of
single muon detection and identification.
Due to the measurement method, this asymmetry does not include the possible
track reconstruction asymmetry. A separate study presented in Ref.~\cite{D02}
shows that track reconstruction asymmetry is consistent with zero within
the experimental uncertainties, due to the regular reversal of the magnet polarities
as discussed in Section~\ref{selection}.

The background charge asymmetries $a^i_K$, $a^i_\pi$ and $a^i_p$ are measured in the
inclusive muon data, and include the detection and identification asymmetry. The parameters
$\delta_i$ are therefore multiplied by the factor $1-f^i_{\rm bkg}$.

The residual asymmetry $a_{\rm CP}^i$ is obtained from Eq.~(\ref{inclusive_mu_a})
by subtracting the background asymmetry $a_{\rm bkg}^i$ from the raw asymmetry $a^i$.
To interpret it in terms of CP violation in mixing, the asymmetry
$a^i_{\rm CP}$ is expressed as
\begin{equation}
\label{aCP}
a^i_{\rm CP} = f^i_{S} a_S.
\end{equation}
Here the quantity $f^i_{S}$ is
the fraction of muons from weak decays of $b$ and $c$ quarks and $\tau$ leptons, and from
decays of short-lived mesons ($\phi, \omega, \eta, \rho^0, J/\psi, \psi'$, etc.)
and Drell Yan
in a given {\pteta} bin $i$. We refer to
these muons as ``short" or ``$S$" muons, since they arise from the decay of particles
within the beam pipe at small distances from the $p \bar p$ interaction point.
The quantity $a_S$ is the charge asymmetry associated with these $S$ muons.

Since $S$ muons originate from inside the beam pipe, their production
is not affected by interactions in the detector material, and
once residual tracking, muon detection, and identification charge imbalances are removed,
the muon charge asymmetry $a_{S}$ must therefore be produced only through
CP violation in the underlying physical processes.
Its dependence on the CP violation in mixing is discussed in Section~\ref{expectation}.

By definition the fractions $f_K^i$ and $f_\pi^i$ in Eq.~(\ref{abkg}) include only those
background muons with the reconstructed track parameters corresponding to the track
parameters of the kaon or pion, respectively.
Such muons are mainly produced by $K^\pm$ and $\pi^\pm$ mesons that decay after passing through the tracking detector or punch-through the absorber material.
The method used to measure the fractions $f_K^i$ and $f_\pi^i$ corresponds to this definition,
see Section~\ref{sec_bck} for details.
In addition, there are background muons with reconstructed track parameters corresponding
to the track parameters of the muon from the
$K^\pm \to \mu^\pm \nu$ and $\pi^\pm \to \mu^\pm \nu$ decay. Such muons
are mainly produced by the kaon and pion decays
in the beam pipe and in the volume of the
tracking detector.
Technically, the muons produced in such decays should be treated
as $S$ muons, since the parent hadron does not travel a long distance in the detector material
and, therefore, these muons do not contribute to the background asymmetries. However,
direct CP violation in semileptonic kaon or pion decay is significantly smaller than the experimental sensitivity
\cite{Gronau} and is assumed to be zero.
Therefore, such muons do not contribute to the asymmetry $a_S$.

To take into account the contribution of these $S$ muons from kaon and pion decay,
we introduce the coefficients $C_K$ and $C_\pi$ \cite{D02,D03}.
They are defined as
\begin{equation}
\label{ck}
C_K   \equiv \frac{\sum_{i=1}^9 f_K^i}  {\sum_{i=1}^9 (f_K^i +{f'}_K^i)},~~~
C_\pi \equiv \frac{\sum_{i=1}^9 f_\pi^i}{\sum_{i=1}^9 (f_\pi^i + {f'}_\pi^i)}.
\end{equation}
Here, ${f'}_K^i$ and ${f'}_\pi^i$
are the fractions of background muons with reconstructed
track parameters corresponding to the track parameters of the muon from the
$K^\pm \to \mu^\pm \nu$ and $\pi^\pm \to \mu^\pm \nu$ decay,
respectively.
The coefficients $C_K$ and $C_\pi$ reduce the fractions $f_S^i$
\cite{ck}
because, by definition
\begin{equation}
f_S^i + \frac{f_K^i}{C_K} + \frac{f_\pi^i}{C_\pi} + f_p^i \equiv 1.
\end{equation}
In this expression we assume that the coefficients $C_K$ and $C_\pi$ are
the same for each $\pteta$ bin $i$. The variation of $C_K$ and $C_\pi$ in
different $\pteta$ bins produces a negligible impact on our result.
The coefficients $C_K$ and $C_\pi$ are determined in simulation, which is discussed
in Section~\ref{simulation}.
They are typically in the range 85\% -- 99\%, except
at large IP, see Table~\ref{tab5}.

The total inclusive single muon charge asymmetry $a$, in a given IP sample, is
given by the average of the nine individual measurements $a_i$ in $(p_T, |\eta|)$
bins $i$, weighted by the fraction $f_\mu^i$ of muons in each bin $i$:
\begin{equation}
\label{atot}
a = \sum_{i=1}^9 f^i_\mu a^i = a_{\rm CP} + a_{\rm bkg},
\end{equation}
where
\begin{eqnarray}
\label{atot1}
a_{\rm CP} & \equiv & \sum_{i=1}^9 f^i_\mu a_{\rm CP}^i = \sum_{i=1}^9 f^i_\mu f_S^i a_S = f_S a_S, \\
a_{\rm bkg} & \equiv & \sum_{i=1}^9 f^i_\mu a_{\rm bkg}^i.
\end{eqnarray}
The quantities $f_S$ and $f^i_\mu$ are defined as
\begin{eqnarray}
f_S & \equiv & \sum_{i=1}^9 f^i_\mu f_S^i, \\
\label{fmu}
f^i_{\mu} & \equiv & \frac{n^+_{i} + n^-_{i}}
{\sum_{i=1}^{9}{(n^+_{i} + n^-_{i})}}, \\
\sum_{i=1}^{9}{f^i_{\mu}} & = & 1.
\end{eqnarray}

\subsection{Like-sign dimuon charge asymmetry}

We now consider like-sign dimuon events in a given (IP$_1$,IP$_2$) sample.
All of the following equations are given for a particular
(IP$_1$,IP$_2$) sample. However, to simplify the presentation, we drop the index
(IP$_1$,IP$_2$) from all of them.
The main principles of the measurement,
namely applying the background corrections to
the measured raw asymmetry to obtain the underlying CP
asymmetry, are the same as for the
inclusive single muon asymmetry. However,
the dimuon
measurement is more complex because the two muons can arise from
different sources, and be in different $(p_T, |\eta|)$ and IP bins.

Consider first the case when IP$_1$ = IP$_2$.
The number of events with two positive or two negative muons, when
one muon is in the {\pteta} bin $i$ and another is in bin $j$, is $N^{++}_{ij}$
and $N^{--}_{ij}$, respectively.
The like-sign dimuon asymmetry is defined as
\begin{equation}
A_{ij} = \frac{N^{++}_{ij} - N^{--}_{ij}}{N^{++}_{ij} + N^{--}_{ij}}.
\end{equation}
The number of events $N^{\pm \pm}_{ij}$ can be expressed as
\begin{equation}
N^{\pm \pm}_{ij} \equiv N_{ij} (1 \pm A_{\rm CP}^{ij}) (1 \pm a^i_{\rm {bkg}}) (1 \pm a^j_{\rm {bkg}}).
\label{nmm}
\end{equation}
The total number of events in a given (IP$_1$,IP$_2$) sample is $N^{++}_{ij} + N^{--}_{ij} = 2 N_{ij}$ when
higher-order terms in asymmetries are neglected.
By definition $N_{ij} = N_{ji}$.
The quantity $A_{\rm CP}^{ij}$ is the residual charge asymmetry produced by $S$ muons.

The muon background asymmetry in a given $\pteta$ bin $i$ in the dimuon events is
\begin{eqnarray}
\label{abkg2}
a^i_{\rm bkg} & = & \frac{1}{2} A_\mu^i
+ \frac{1}{2} F^i_K a^i_K
+ \frac{1}{2} F^i_\pi a^i_\pi + \frac{1}{2} F^i_p a^i_p. \\
A_\mu^i & \equiv & (2 - F^i_{\rm {bkg}}) \delta_i.
\end{eqnarray}
Here, $\frac{1}{2}F^i_{K}$, $\frac{1}{2}F^i_{\pi}$ and $\frac{1}{2}F^i_{p}$ are the fractions
of muons produced by kaons, pions and protons
reconstructed by the central tracker
in a given $\pteta$ bin $i$ but identified as muons.
Following the definitions in Refs. \cite{D02, D03}, for like-sign dimuon events
the background fractions $F^i_{K}$, $F^i_{\pi}$ and $F^i_{p}$
are normalized per event (not per muon); this is the reason for the
factors $1/2$ in Eq.~(\ref{abkg2}).
The quantity $F_{\rm{bkg}}^i$ is defined as $F_{\rm{bkg}}^i = F^i_{K} + F^i_{\pi} + F^i_{p}$.
The asymmetries $a^i_K$, $a^i_\pi$, $a^i_p$, and $\delta_i$ are the same as in the
inclusive muon sample.

The number of positive and negative muons from the like-sign dimuon events in the $\pteta$ bin $i$ is
\begin{equation}
N^\pm_i = N^{\pm \pm}_{ii} + \sum_{j=1}^9 N^{\pm \pm}_{ij}.
\label{ni}
\end{equation}

The charge asymmetry $A^i$ of muons in the $\pteta$ bin $i$,
to first order in the asymmetries is
\begin{eqnarray}
\label{ai2mu}
A^i & \equiv & \frac{N_i^+ - N_i^-}{N_i^+ + N_i^-}
    = A_{\rm CP}^i + A^i_{\rm bkg}, \\
\label{acpi}
A^i_{\rm CP} & = & \frac{N_{ii} A^{ii}_{\rm CP} + \sum_j N_{ij} A_{\rm CP}^{ij}}
     {N_{ii} + \sum_{j=1}^9 N_{ij}}, \\
\label{ai2mu-1}
A^i_{\rm bkg} & = & \frac{2 N_{ii} a^i_{\rm bkg} + \sum_j N_{ij} (a^i_{\rm bkg} + a^j_{\rm bkg})}
     {N_{ii} + \sum_{j=1}^9 N_{ij}}.
\end{eqnarray}

To interpret the asymmetry $A^{ij}_{\rm CP}$ in terms of CP violation, it is expressed as
\begin{equation}
A_{\rm CP}^{ij} = F_{SS}^{ij} A_S + F_{SL}^{ij} a_S.
\label{acpij}
\end{equation}
The quantity $A_S$ is the charge asymmetry in the events with two like-sign $S$ muons.
Its dependence on the parameters describing CP violation in mixing and
CP violation in interference is discussed in Section~\ref{expectation}.
The quantity $a_S$ is defined in Eq.~(\ref{aCP}).
The quantity $F_{SS}^{ij}$ is the fraction of like-sign dimuon events
with two $S$ muons, and $F_{SL}^{ij}$ is the fraction of like-sign dimuon events
with one $S$ and one $L$ muon
in given {\pteta} bins $i$ and $j$. Equation~(\ref{acpij}) reflects the fact
that the events with two $S$ muons produce the charge asymmetry $A_S$; the events
with one $S$ and one $L$ muon produce the charge asymmetry $a_S$, while the events
with both $L$ muons do not produce a CP-related charge asymmetry.

Multiplying Eq.~(\ref{ai2mu}) by the fraction $F^i_\mu$ of muons in a given $\pteta$ bin $i$,
and summing over $i$, we reproduce the expression of the like-sign dimuon
charge asymmetry in Refs. \cite{D02, D03}:
\begin{eqnarray}
\label{ai2mu-2}
A & \equiv & \sum_{i=1}^9 F^i_\mu A^i = A_{\rm CP} + A_{\rm bkg}, \\
\label{acp}
A_{\rm CP} & \equiv & \sum_{i=1}^9 F^i_\mu A^i_{\rm CP} = F_{SS} A_S + F_{SL} a_S, \\
\label{ai2mu-3}
A_{\rm bkg} & \equiv & \sum_{i=1}^9 F^i_\mu \{ A_\mu^i + F^i_K a^i_K + F^i_\pi a^i_\pi + F^i_p a^i_p \}.
\end{eqnarray}
Here the fraction $F^i_\mu$ is defined as
\begin{equation}
\label{Fmu}
F^i_{\mu} \equiv \frac{N^+_{i} + N^-_{i}}
{\sum_{j=1}^{9}{(N^+_{j} + N^-_{j})}},
\qquad \sum_{i=1}^{9}{F^i_{\mu}} = 1.
\end{equation}

From Eqs.~(\ref{acpi}), (\ref{acpij}), (\ref{acp}), and
(\ref{Fmu}) it follows that
\begin{eqnarray}
F_{SS} & = & \frac{ \sum_{i=1}^9 (N_{ii} F_{SS}^{ii} + \sum_{j=1}^9 N_{ij} F_{SS}^{ij}) }
     {N_{\rm tot}}, \\
F_{SL} & = & \frac{ \sum_{i=1}^9 (N_{ii} F_{SL}^{ii} + \sum_{j=1}^9 N_{ij} F_{SL}^{ij}) }
     {N_{\rm tot}}, \\
N_{\rm tot} & \equiv & \sum_{i=1}^9 N_i
= \sum_{i=1}^9 { \left( N_{ii} + \sum_{j=1}^9 N_{ij} \right) }.
\end{eqnarray}
Here $N_{\rm tot}$ is the total number of dimuon events in a given IP$_1$,IP$_2$ sample.
The quantity $F_{LL}$ gives the fraction of the like-sign dimuon events with
two $L$ muons.
\begin{eqnarray}
\label{fss1}
F_{SS} + F_{SL} + F_{LL} & \equiv & 1, \\
\label{fss2}
2 F_{LL} + F_{SL} & \equiv & \sum_{i=1}^9
\left( \frac{F_K^i}{C_K} + \frac{F_\pi^i}{C_\pi} + F_p^i \right) .
\end{eqnarray}
We solve for the fractions $F_{SS}$, $F_{SL}$ and $F_{LL}$ using Eqs.~(\ref{fss1}),~(\ref{fss2})
and the ratio
\begin{equation}
\label{rll}
R_{LL} = \frac{F_{LL}} {F_{SL}+F_{LL}},
\end{equation}
which is obtained from the simulation. The simulation used in this analysis
is discussed in Section~\ref{simulation}.

For the sample of dimuon events with $\mbox{IP}_1 \neq \mbox{IP}_2$,
it can be shown that the expressions~(\ref{ai2mu}) --~(\ref{ai2mu-3}) remain
the same with background fractions defined as
\begin{eqnarray}
\label{fk2}
F_K^i(\mbox{IP}_1,\mbox{IP}_2) & = & \frac{1}{2} [ F_K^i( \mbox{IP}_1) + F_K^i(\mbox{IP}_2) ], \\
\label{fpi2}
F_\pi^i(\mbox{IP}_1,\mbox{IP}_2) & = & \frac{1}{2} [ F_\pi^i( \mbox{IP}_1) + F_\pi^i(\mbox{IP}_2) ], \\
\label{fp2}
F_p^i(\mbox{IP}_1,\mbox{IP}_2) & = & \frac{1}{2} [ F_p^i( \mbox{IP}_1) + F_p^i(\mbox{IP}_2) ].
\end{eqnarray}
Here, $F_K^i( \mbox{IP}_1)$ is the number of $K \to \mu$ muons
with the impact
parameter in the IP$_1$ range normalized to the total number of events in the (IP$_1$,IP$_2$) sample.
The fractions $F_\pi^i( \mbox{IP}_1)$ and $F_p^i( \mbox{IP}_1)$ are defined
similarly for $\pi \to \mu$ and $p \to \mu$ muons.

Hence, in order to extract the CP-violating asymmetries $a_{\rm CP}$ and $A_{\rm CP}$ from the
binned raw asymmetries $a^i$ and $A^i$, the following quantities are required:
\begin{itemize}
\item
The fractions $f^i_K$, $f^i_{\pi}$, $f^i_p$, $f^i_\mu$,
$F^i_K$, $F^i_{\pi}$, $F^i_p$ and $F^i_\mu$, in bins $i$ of
$(p_T, |\eta|)$, in each IP sample (Tables~\ref{tab1} and \ref{tab2}).
\item
The background asymmetries $a^i_K$, $a^i_{\pi}$, $a^i_p$ and $\delta_i$, in bins $i$
of $(p_T, |\eta|)$. They do not depend on the IP sample, see Section~\ref{sec_abkg} for details.
\end{itemize}
All these quantities are extracted directly from data with minimal contribution
from the MC simulation.

The remainder of this article describes the extraction of these
parameters, and the subsequent interpretation of the asymmetries $A_{\rm CP}$
and $a_{\rm CP}$.

\section{Muon selection}
\label{selection}

The D0 detector is described
in Refs.~\cite{muid,run2det,run2muon,layer0}.
It consists of a magnetic central-tracking system that
comprises a silicon microstrip tracker (SMT) and a central fiber
tracker (CFT), both located within a $2$~T superconducting solenoidal
magnet~\cite{run2det}.
The muon system~\cite{run2muon,muid} is located beyond the liquid argon-uranium
calorimeters that surround the central tracking system, and consists of
a layer A of tracking detectors and scintillation trigger counters
before 1.8~T iron toroids, followed by two similar layers B and C after
the toroids. Tracking for $|\eta|<1$ relies on 10-cm wide drift
tubes, while 1-cm minidrift tubes are used for
$1<|\eta|<2$.

The polarities of the toroidal and solenoidal magnetic fields were reversed
on average every two weeks so that the four solenoid-toroid polarity
combinations were exposed to approximately the same
integrated luminosity. This allows for a cancellation of first-order
effects related to the instrumental asymmetries~\cite{D01}.
To ensure more complete cancellation,
the events are weighted according to the number of events for each
data sample corresponding to a different configuration of the magnet polarities.
These weights are given in Table~\ref{tab00}. The weights for inclusive
muon and the like-sign dimuon samples are different due to different
trigger requirements. The effective reduction of statistics
of the like-sign dimuon sample due to this weighting is less than 2\%.

\begin{table}[b]
\caption{\label{tab00}
Weights assigned to the events recorded with different solenoid and toroid polarities
in the inclusive muon and like-sign dimuon samples.
}
\begin{ruledtabular}
\newcolumntype{A}{D{A}{\pm}{-1}}
\newcolumntype{B}{D{B}{-}{-1}}
\begin{tabular}{cccc}
Solenoid & Toroid   & Weight & Weight \\
polarity & polarity & inclusive muon & like-sign dimuon \\
\hline
$-1$ & $-1$ & 0.954 & 0.967 \\
$-1$ & +1 & 0.953 & 0.983 \\
+1 & $-1$ & 1.000 & 1.000 \\
+1 & +1 & 0.951 & 0.984
\end{tabular}
\end{ruledtabular}
\end{table}


As discussed previously in Section~\ref{Method}, 
the inclusive muon and like-sign dimuon samples are
obtained from data collected with single and dimuon triggers, respectively.
Charged particles with transverse momentum in the range $1.5 < p_T < 25$ GeV and with pseudorapidity
$|\eta| < 2.2$ are considered as muon candidates.
We also require either $p_T > 4.2$ GeV or a longitudinal momentum
component $|p_z| > 5.4$ GeV. Muon candidates are selected
by matching central tracks with a segment reconstructed
in the muon system and by applying tight quality requirements
aimed at reducing false matching and background
from cosmic rays and beam halo \cite{muid}. The transverse IP
of the charged track matched to the muon relative to the reconstructed
$p \bar p$ interaction vertex must be smaller than 0.3 cm, with
the longitudinal distance from the point of closest approach
to this vertex smaller than 0.5 cm.
We use track parameters of the track reconstructed in the CFT and SMT
and do not use the muon momentum and azimuthal angle measurements provided by the muon system.
Strict quality requirements
are also applied to the tracks and to
the reconstructed $p \bar p$ interaction vertex.
The details of these requirements can be found in Ref.~\cite{D02}.
The inclusive muon sample contains all muons passing the selection requirements.
If an event contains more than one muon,
each muon is included in the inclusive muon sample.

\begin{figure}
\begin{center}
\includegraphics[width=0.48\textwidth]{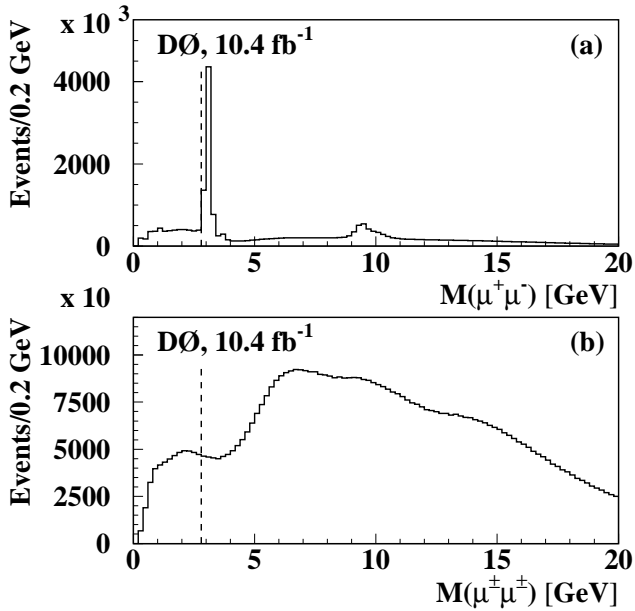}
\caption{Invariant mass $M(\mu \mu)$ of two muons in (a) opposite-sign
dimuon sample, and (b) like-sign dimuon sample. The requirement
$M(\mu^\pm \mu^\pm) > 2.8$ GeV is also shown.}
\label{mmm}
\end{center}
\end{figure}

The like-sign dimuon sample contains all events with at least
two muon candidates with the same charge. These two
muons are required to have the same associated $p \bar{p}$
interaction vertex, and
an invariant mass larger
than 2.8 GeV to minimize the number of events in which
both muons originate from the same $b$ quark.
The invariant mass of two muons in the opposite-sign and like-sign
dimuon sample is shown in Fig.~\ref{mmm}.
If more than two muons pass the single muon selection, the classification
into like-sign or opposite-sign is done using the two muons
with the highest $p_T$. In the like-sign dimuon sample $\approx 0.7$\% of
the events have more than two muons.


\begin{figure}
\begin{center}
\includegraphics[width=0.48\textwidth]{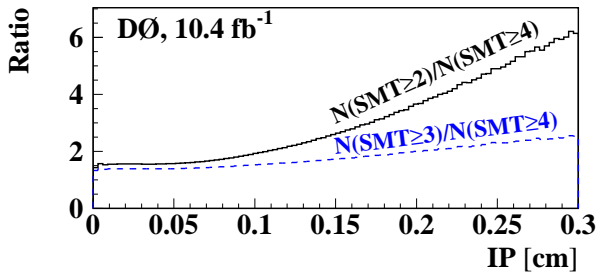}
\caption{
Ratio of number of muon tracks with $\ge 2$ (upper curve)
or $\ge 3$ (lower dashed curve) axial SMT hits to
the number of muon tracks with $\ge 4$ axial SMT hits,
as a function of IP, in the inclusive muon sample.
}
\label{fig-smtcut}
\end{center}
\end{figure}

In addition to these selections, which are identical to the selections
of Refs. \cite{D02,D03}, we apply a stronger
requirement on the number of hits in the SMT
included in the track associated with the muon.
The SMT \cite{layer0}  has axial detector strips parallel to the
beam, and stereo detector strips at an angle to the beam.
We require that the muon track contains at least
three axial SMT hits, instead of the
requirement of two such hits in \cite{D03}.
On average, a track passing through the SMT has hits in four layers of the SMT.
The SMT measurements in axial strips determine
the IP precision, and this stronger requirement substantially reduces
the number of muons with incorrectly measured IP.
The tracks with exactly two axial SMT hits include tracks
with one of the two SMT hits incorrectly associated. 
For such tracks, the IP can be measured to be large.
As a result, muons produced with small IP
migrate to the sample with large IP, as can be seen in Fig.~\ref{fig-smtcut}.
Since the fraction of $L$ muons with
small IP is much larger than that with large IP, this migration results in an
increase of the background $L$ muons in the sample with large IP. Therefore,
the tighter selection on the number of axial SMT hits helps to reduce
the number of background muons with large IP,
which is important for our measurement.

\section{Monte Carlo Simulation}
\label{simulation}

 Most of the quantities required for the measurement of $a_{\rm CP}$ and $A_{\rm CP}$ are extracted directly from data.
The MC simulations are used in a limited way, as discussed in Section~\ref{expectation}.
To produce the simulated events we use the {\sc pythia} v6.409 event
generator \cite{pythia}, interfaced to the {\sc evtgen} decay package \cite{evtgen},
and {\sc CTEQ6L1} parton distribution functions \cite{pdf}. The
generated events are propagated through the D0
detector using a {\sc GEANT}-based program \cite{geant} with full detector simulation.
The response in the detector is digitized, and
the effects of multiple interactions at high luminosity are
modeled by overlaying hits from randomly selected $p \bar p$
collisions on the digitized hits from MC. The complete
events are reconstructed with the same program as used
for data, and, finally, analyzed using the same selection
criteria described above for data.

In this analysis two types of MC sample are used:
\begin{itemize}
\item Inclusive $p \bar p$ collisions with minimum interaction transverse energy
at the generator level $E^{\rm min}_T = 20$ GeV.
\item A simulation of $p \bar p \to b \bar{b} X$ and $p \bar p \to c \bar{c} X $
final states, with $E^{\rm min}_T = 0$ GeV, producing two muons with an additional requirement that
the produced muons have $p_T > 1.5$ GeV and $|\eta| < 2.2$.
\end{itemize}
The second sample is especially useful to extract quantities for
the signal inclusive muon and dimuon events, because it is generated with almost no kinematic bias,
as discussed in Section~\ref{expectation}.

\section{Measurement of background fractions}
\label{sec_bck}

\subsection{The $\bm{K^{*0}}$ method}
\label{secfk}

A kaon, pion, or proton can be misidentified as a muon and thus contribute to the inclusive muon
and the like-sign dimuon samples.
This can happen because of pion and kaon decays in flight, or
punch-through.
We do not distinguish these individual processes, but rather measure the total fraction
of such particles using data.
In the following, the notation ``\ktomu'' stands for ``kaon misidentified as a muon,"
and the notations ``\pitomu'' and ``\ptomu'' have corresponding meanings for pions and protons.

The fraction $f_K$ is measured by reconstructing
the decays $K^{*0} \to K^+ \pi^-$ with $\ktomu$, and
$K^{*+} \to \ks \pi^+$ with one of the pions from $\ks \to \pi^+ \pi^-$ decay misidentified as a muon.
This method is described in detail in \cite{D02,D03}. The main features of this method
are repeated here.

The relation between the fraction $f_{K^{*0}}^i$ of $\ktomu$ originating from
the decay $K^{*0} \to K^+ \pi^-$ and the fraction $f_K^i$ in each $\pteta$ bin $i$ is
\begin{equation}
f_{K^{*0}}^i =  \varepsilon_0^i R^i(K^{*0}) f_K^i.
\label{kstf}
\end{equation}
Here $R^i(K^{*0})$ is the fraction of all kaons that result from $K^{*0} \to K^+ \pi^-$ decays, and
$\varepsilon_0^i$ is the efficiency to reconstruct the charged pion from the $K^{*0} \to K^+ \pi^-$
decay, provided that the \ktomu~ track is reconstructed. The kinematic
parameters of the charged kaon are required to be in the $\pteta$ bin $i$.

We also select $\ks$ mesons and reconstruct $K^{*+} \to K_S \pi^+$ decays. One of the pions
from the decay $\ks \to \pi^+ \pi^-$ is required to be misidentified as a muon.
This requirement ensures that the flavor composition of the samples containing $\ktomu$ and
$\ks \to \pi^+ \pi^- \to \pi^\pm \mu^\mp$ is the same \cite{D03}.
The number of $K^{*+} \to K_S \pi^+$ decays in each $\pteta$ bin $i$ is
\begin{equation}
\label{ksplus}
N^i(K^{*+} \to \ks \pi^+) =  \varepsilon_c^i N^i(\ks) R^i(K^{*+}) ,
\end{equation}
where $R^i(K^{*+})$ is the fraction of $\ks$ mesons that result from $K^{*+} \to K_S \pi^+$ decays, and
$\varepsilon_c^i$ is the efficiency to reconstruct the charged pion in the
$K^{*+} \to K_S \pi^+$ decay, provided that the $K_S$ meson is reconstructed.
The kinematic
parameters of the $\ks$ meson are required to be in the $\pteta$ bin $i$.
We use isospin invariance to set
\begin{equation}
R^i(K^{*0}) = R^i(K^{*+}).
\label{assume1}
\end{equation}
This relation is also confirmed by data from LEP as discussed in Ref.~\cite{D02}.
We apply the same kinematic selection criteria to the charged kaon and $K_S$ candidates,
and use exactly the same criteria to select an additional
pion and reconstruct the $K^{*0} \to K^+ \pi^-$ and $K^{*+} \to K_S \pi^+$ decays.
We therefore assume that
\begin{equation}
\varepsilon_0^i = \varepsilon_c^i.
\label{assume2}
\end{equation}
We assign the systematic uncertainty related to this assumption, see Section~\ref{sec_systematic} 
for details.

From Eqs.~(\ref{kstf})--(\ref{assume2}), we obtain
\begin{equation}
f_K^i = \frac{N^i(\ks)}{N^i(K^{*+} \to \ks \pi^+)} f_{K^{*0}}^i.
\label{fkst}
\end{equation}
This expression is used to measure the kaon fraction $f_K^i$ in the inclusive muon sample without dividing
it into the IP samples. It is based on the equality~(\ref{assume2}) of the efficiencies to reconstruct
the $K^{*0} \to K^+ \pi^-$ and $K^{*+} \to \ks \pi^+$ decays,
provided that the $K^+$ and $\ks$ candidates are reconstructed.
This equality is verified in simulation for a full data sample \cite{D02}.
However, in a given IP sample the efficiencies $\varepsilon_0^i$  and $\varepsilon_c^i$
become unequal because of the differences between the $\ks$ and $K^+$ tracks explained below.

\begin{figure}
\begin{center}
\includegraphics[width=0.48\textwidth]{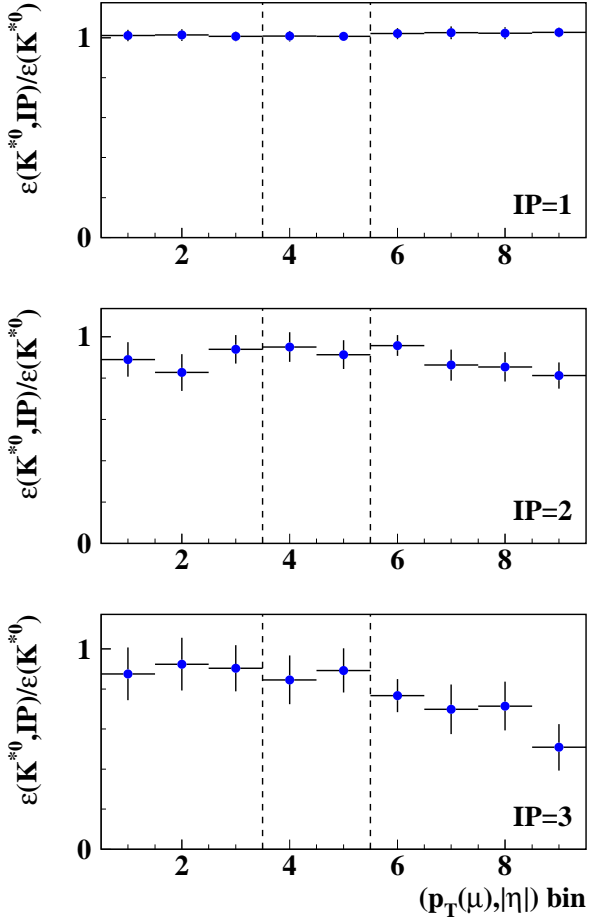}
\caption{Ratio of the $K^{*0} \to K^+ \pi^-$ reconstruction efficiencies
$\varepsilon^i(K^{*0},{\rm IP}) / \varepsilon^i(K^{*0})$
in a given IP sample relative
to that in the total inclusive muon sample.
The vertical dashed lines separate the $\pteta$
bins corresponding to the central, intermediate, and forward regions of the D0 detector, respectively.}
\label{fig-ckst}
\end{center}
\end{figure}

If the $K^\pm \to \mu^\pm \nu$ decay occurs within the tracking volume,
the track parameters of charged $K$ meson can be biased due to the kink
in the $\ktomu$ trajectory.
Such biased $\ktomu$ tracks tend to populate the sample with
large IP. The bias in the $K$ meson track parameters
propagates into a reduced efficiency of $K^{*0} \to K^+ \pi^-$ reconstruction.
This reduction can be seen in Fig.~\ref{fig-ckst} where the ratio of the
$K^{*0} \to K^+ \pi^-$ reconstruction efficiencies
$\varepsilon(K^{*0},{\rm IP}) / \varepsilon(K^{*0})$
in a given IP sample and in the total
inclusive muon sample is shown. These ratios are obtained in simulation.

\begin{table*}
\caption{\label{tab1}
Background and signal fractions in the IP samples of the inclusive
muon sample. The column ``All IP'' corresponds to the full inclusive muon sample
without dividing it into the IP samples. Only statistical uncertainties are given.
}
\begin{ruledtabular}
\newcolumntype{A}{D{A}{\pm}{-1}}
\newcolumntype{B}{D{B}{-}{-1}}
\begin{tabular}{lAAAA}
Quantity &  \multicolumn{1}{c}{All IP} & \multicolumn{1}{c}{IP=1} & \multicolumn{1}{c}{IP=2} & \multicolumn{1}{c}{IP=3} \\
\hline
$f_K \times 10^2$         & 15.73\ A \ 0.21 & 20.30\ A \ 0.34 & 7.71\ A \ 0.24 & 2.69\ A \ 0.14 \\
$f_K/C_K \times 10^2$     & 16.91\ A \ 0.23 & 20.50\ A \ 0.34 & 8.38\ A \ 0.26 & 7.47\ A \ 0.39 \\
$f_\pi \times 10^2$       & 30.43\ A \ 1.60 & 39.13\ A \ 2.09 & 15.39\ A \ 0.83 & 5.32\ A \ 0.30 \\
$f_\pi/C_\pi \times 10^2$ & 32.37\ A \ 1.70 & 40.77\ A \ 2.18 & 18.11\ A \ 0.98 & 7.71\ A \ 0.43 \\
$f_p \times 10^2$         & 0.57\ A \ 0.16  &  0.73\ A \ 0.21 & 0.28\ A \ 0.09  & 0.08\ A \ 0.03 \\
$f_S \times 10^2$         & 49.97\ A \ 1.86 & 38.04\ A \ 2.30 & 73.23\ A \ 1.15 & 84.82\ A \ 0.85 \\
\end{tabular}
\end{ruledtabular}
\end{table*}

The $\ks$ meson is reconstructed from $\pi^+$ and $\pi^-$
tracks with one of the pions required to be misidentified as a muon.
The quality of the $K_S \to \pi^+ \pi^-$ vertex, and the condition that
the $\pi^+ \pi^-$ mass be consistent with the $\ks$ mass, are imposed
to select the $\ks$ candidate. As a result, the sample of $\ks$ candidates
with large IP does not contain an increased contribution from the biased
$\ks$ track measurement. Therefore, the $K^{*+} \to \ks \pi^+$ reconstruction efficiency
in the sample with large $\ks$ track IP can be different from the
$K^{*0} \to K^+ \pi^-$ reconstruction efficiency, and the estimate
of $f_K^i({\rm IP})$ in the large $K^+$ IP sample
using
Eq.~(\ref{fkst}) is biased.

To avoid this bias, the fractions $f_K^i(\rm IP)$
in a given IP sample are measured using the following expression:
\begin{equation}
f_K^i({\rm IP}) = f_K^i \frac{ f_{K^{*0}}^i({\rm IP})}{f_{K^{*0}}^i}
\frac{\varepsilon^i (K^{*0})}{\varepsilon^i(K^{*0},{\rm IP})}.
\label{fkst-ip}
\end{equation}
The fractions $f_{K^{*0}}^i({\rm IP})$ and $f_{K^{*0}}^i$
are measured in the IP sample and in the total inclusive muon sample, respectively.
The fraction $f_K^i$ is obtained using Eq.~(\ref{fkst}). The ratio of efficiencies
$\varepsilon^i(K^{*0},{\rm IP}) / \varepsilon^i(K^{*0})$ is taken from simulation
and is shown in Fig.~\ref{fig-ckst}. The mean value
of $\varepsilon^i(K^{*0},{\rm IP}) / \varepsilon^i(K^{*0})$ is $1.01 \pm 0.01$ for
the IP=1 sample, $0.90 \pm 0.03$ for the IP=2 sample and $0.79 \pm 0.06$ for the IP=3 sample.
The uncertainties are due to limited MC statistics.
%

\begin{figure}
\begin{center}
\includegraphics[width=0.48\textwidth]{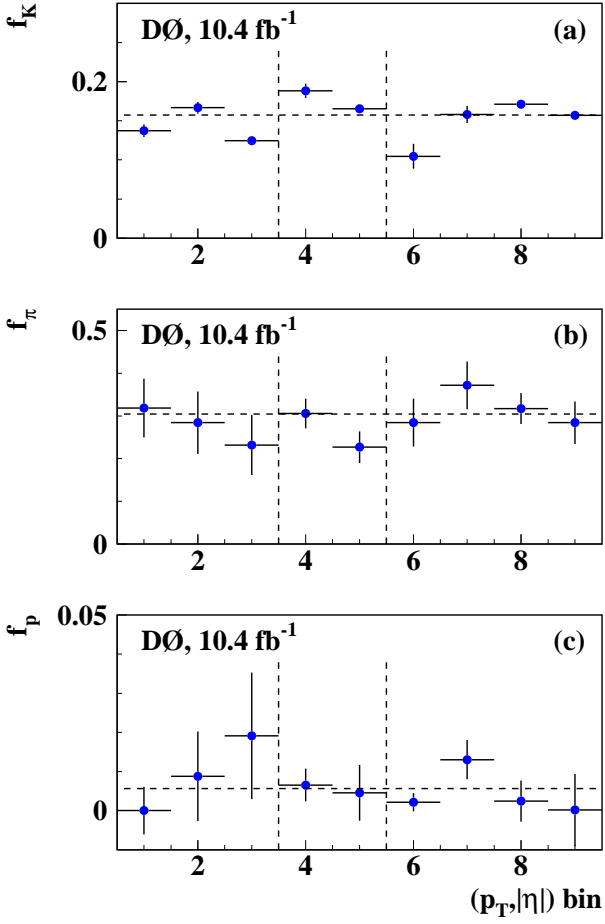}
\caption{Fraction of (a) $K \to \mu$ tracks, (b) $\pi \to \mu$ tracks and
(c) $p \to \mu$ tracks in the inclusive muon sample as a function of the
kaon, pion and proton $\pteta$ bin $i$, respectively. Only statistical uncertainties are shown.
The horizontal dashed lines show the mean values.
The vertical dashed lines separate the $\pteta$
bins corresponding to the central, intermediate, and forward regions of the detector, respectively.
}
\label{fig-bkg}
\end{center}
\end{figure}

The procedure to measure the related background fractions $f^i_\pi$ and $f^i_p$
is the same as in Refs. \cite{D02,D03}.
The values of $f^i_K$, $f^i_\pi$ and $f^i_p$ in the total inclusive muon sample
are shown in Fig.~\ref{fig-bkg}. The background fractions in different
IP samples are given in Table~\ref{tab1}.
For reference, we also give in Table~\ref{tab1} the values $f_K/C_K$ and $f_\pi/C_\pi$ \cite{ck}. These
values for all inclusive muon events can be compared directly
with the corresponding background fractions $(15.96 \pm 0.24)$\%
and $(30.1 \pm 1.6)$\%, respectively in \cite{D03}.

Approximately 17\% (32\%) of muons in the inclusive muon sample are determined
to arise from kaon (pion) misidentification,
with less than 1\% due to proton punch through and fakes.
The remaining $\approx 50$\% of the sample are muons from heavy-flavor decay.

The background fractions vary by a factor of more than five between
the IP=1 and IP=3 samples. Such a large variation is expected.
The parents of $L$ muons are dominantly produced in the primary interaction
and decay outside the tracking volume. The $S$ muons are dominantly produced
in decays of heavy quarks and their tracks have large IP.
Therefore, the fraction of $L$ muons in the sample
with small IP is substantially enhanced. They give the
main contribution to the background asymmetry in this sample.
On the contrary, the fraction of $L$ muons in the large IP sample is suppressed,
and the kaon and detector asymmetries have approximately the same magnitude,
see Table~\ref{tab1a} for details.
The comparison of our prediction and the observed raw asymmetry
in different $(p_T, |\eta|)$ and IP bins
therefore allows us to verify our background measurement method.

The procedure to measure the background fractions in the like-sign dimuon sample
is described in \cite{D03} and is not changed for this analysis.
To obtain the quantity $F_K^i$ of \ktomu~ tracks in the like-sign
dimuon sample we use the relation similar to Eq.~(\ref{fkst}):
\begin{equation}
F_K^i = \frac{N^i(\ks)}{N^i(K^{*+} \to \ks \pi^+)} F_{K^{*0}}^i.
\label{Fkst}
\end{equation}
Here $F_{K^{*0}}^i$ is the fraction of $\kstneu \to K^+ \pi^-$ decays with
\ktomu~ in the $\pteta$ bin $i$ in the like-sign dimuon sample. The numbers $N^i(\ks)$ and $N^i(K^{*+} \to \ks \pi^+)$
are obtained from the inclusive muon sample. The kinematic parameters of the charged
kaon and $\ks$ meson are required to be in the $\pteta$ bin $i$.

In the samples
with the (IP$_1$,IP$_2$) selection the background fractions
$F_K^i(\mbox{IP}_1)$ and $F_K^i(\mbox{IP}_2)$ are determined separately for the IP$_1$ and IP$_2$ kaon
and the total background
fractions are obtained using Eq.~(\ref{fk2}). The fractions $F_K^i(\mbox{IP}_1)$ and $F_K^i(\mbox{IP}_2)$
are obtained using the expression
\begin{equation}
F_K^i({\rm IP}_{1,2}) = F_K^i \frac{ F_{K^{*0}}^i({\rm IP}_{1,2})}{F_{K^{*0}}^i}
\frac{\varepsilon^i (K^{*0})}{\varepsilon^i(K^{*0},{\rm IP}_{1,2})}.
\label{Fkst-ip}
\end{equation}
The fractions $F_{K^{*0}}^i({\rm IP}_1)$ and  $F_{K^{*0}}^i({\rm IP}_2)$
are measured using the IP$_1$ and IP$_2$ kaons, respectively. The
fraction $F_{K^{*0}}^i$ is measured in the total like-sign dimuon sample.
The fraction $F_K^i$ is obtained using Eq.~(\ref{Fkst}).

\begin{figure}
\begin{center}
\includegraphics[width=0.48\textwidth]{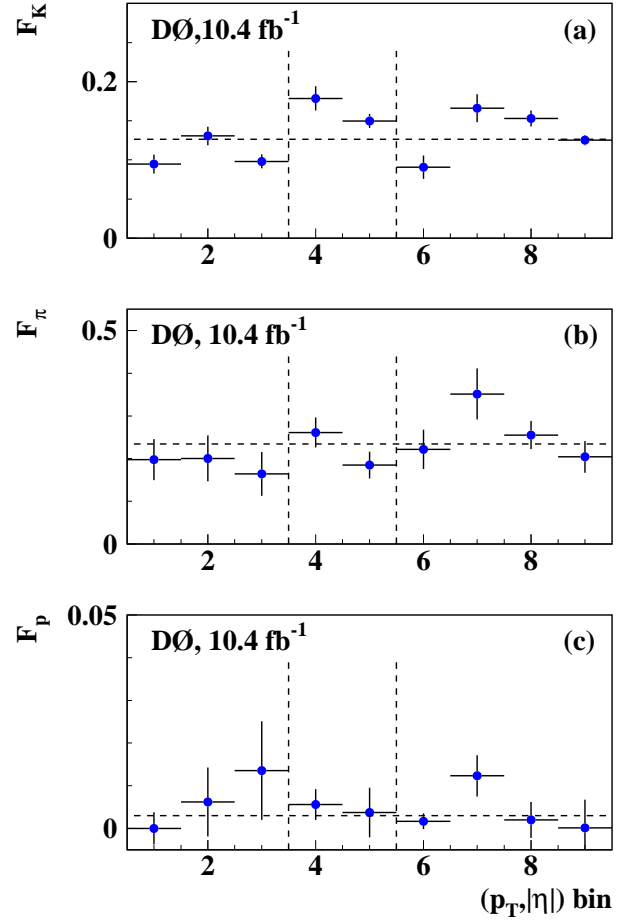}
\caption{Fraction of (a) $K \to \mu$ tracks, (b) $\pi \to \mu$ tracks and
(c) $p \to \mu$ tracks per event in the like-sign dimuon sample as a function of the
kaon, pion and proton $\pteta$ bin $i$, respectively. Only statistical uncertainties are shown.
The horizontal dashed lines
show the mean values.
The vertical dashed lines separate the $\pteta$
bins corresponding to the central, intermediate, and forward regions of the D0 detector, respectively.
}
\label{fig-bkg2}
\end{center}
\end{figure}

The values of $F^i_K$, $F^i_\pi$ and $F^i_p$ in the total like-sign  dimuon sample
are shown in Fig.~\ref{fig-bkg2}. The background fractions in different
(IP$_1$,IP$_2$) samples are given in Table~\ref{tab2}. For reference, we also give
the values $F_K/C_K$ and $F_\pi / C_\pi$ \cite{ck}. These values
for all like-sign dimuon events can be compared directly with
the corresponding background fractions $(13.78 \pm 0.38)$\% and $(24.81 \pm 1.34)$\%,
respectively, in Ref. \cite{D03}.
Table~\ref{tab2} also contains the values
of $F_{SS}$ and $F_{SL}$ for each (IP$_1$,IP$_2$) sample and for the total sample of
like-sign dimuon events.

For the like-sign dimuon sample
approximately 6.5\% (12.5\%) of muons arise from kaon (pion) misidentification,
with less than 0.25\% from proton punch through or fakes.
These values are derived from Table~\ref{tab2} taking into account that
the background fractions given in this table are defined per dimuon event.
We find that 69\% of the events have both muons from heavy-flavor decays,
and a further 23\% have one muon from heavy-flavor decay.
Similar to the inclusive muon events, the background
fractions are considerably reduced and the signal contribution is increased
in the samples with large muon IP.

\begin{table*}
\caption{\label{tab2}
Background and signal fractions in the (IP$_1$,IP$_2$) samples of the like-sign
dimuon sample. The column ``All IP'' corresponds to the full like-sign dimuon sample
without dividing it into the (IP$_1$,IP$_2$) samples.
Only statistical uncertainties are given.
}
\begin{ruledtabular}
\newcolumntype{A}{D{A}{\pm}{-1}}
\newcolumntype{B}{D{B}{-}{-1}}
\begin{tabular}{lAAAAAAA}
Quantity &  \multicolumn{1}{c}{All IP} & \multicolumn{1}{c}{IP$_1$,IP$_2$=11} & \multicolumn{1}{c}{IP$_1$,IP$_2$=12} &
\multicolumn{1}{c}{IP$_1$,IP$_2$=13} & \multicolumn{1}{c}{IP$_1$,IP$_2$=22} & \multicolumn{1}{c}{IP$_1$,IP$_2$=23} &
\multicolumn{1}{c}{IP$_1$,IP$_2$=33} \\
\hline
$F_K \times 10^2$ & 12.63\ A \ 0.35 & 26.77\ A \ 1.32 & 15.04\ A \ 1.51 & 9.73\ A \ 1.20
                                    & 10.34\ A \ 3.17 & 4.13\ A \ 1.82 & 2.39\ A \ 2.08 \\
$F_K/C_K \times 10^2$ & 13.44\ A \ 0.38 & 27.04\ A \ 1.33 & 15.78\ A \ 1.58 &14.11\ A \ 1.74
                                    & 11.24\ A \ 3.45 &10.21\ A \ 4.50 & 6.65\ A \ 5.78 \\
$F_\pi \times 10^2$ & 23.42\ A \ 1.36 & 48.71\ A \ 3.46 & 27.28\ A \ 3.10 & 18.26\ A \ 2.30
                                      & 19.77\ A \ 5.98 & 8.48\ A \ 3.28 & 3.94\ A \ 3.71 \\
$F_\pi/C_\pi \times 10^2$ & 24.91\ A \ 1.45 & 50.74\ A \ 3.60 & 30.00\ A \ 3.41 & 21.23\ A \ 2.67
                                      & 23.26\ A \ 7.04 &12.16\ A \ 4.70 & 5.71\ A \ 5.38 \\
$F_p \times 10^2$ & 0.41\ A \ 0.13 &  0.84\ A \ 0.26 & 0.52\ A \ 0.13 & 0.27\ A \ 0.10
                                   & 0.18\ A \ 0.14 & 0.15\ A \ 0.08 & 0.04\ A \ 0.06 \\
$F_{SS} \times 10^2$ & 69.14\ A \ 1.49 & 45.83\ A \ 3.25 & 63.83\ A \ 2.74 & 68.75\ A \ 3.03
                                    & 67.24\ A \ 9.37 & 78.34\ A \ 5.45 & 88.01\ A \ 5.81 \\
$F_{SL} \times 10^2$ & 22.69\ A \ 1.10 & 29.79\ A \ 1.79 & 26.05\ A \ 0.84 & 26.98\ A \ 2.20
                                    & 30.83\ A \ 8.82 & 20.90\ A \ 4.97 & 11.66\ A \ 5.65 \\
\end{tabular}
\end{ruledtabular}
\end{table*}

\subsection{The local variables method}
\label{sec_local}

The $K^{*0}$ method presented in Section~\ref{secfk} depends on the validity of Eqs.~(\ref{assume1}) and~(\ref{assume2}) and on the ratio $\varepsilon^i(K^{*0},{\rm IP}) / \varepsilon^i(K^{*0})$,
which cannot be verified directly in our data. To assign the systematic uncertainties due
 to these inputs we develop
a complimentary method of local variables presented below. The systematic
uncertainty on the background fractions is assigned following the comparison of these two fully
independent methods. It is discussed in Section~\ref{sec_systematic}.

The D0 muon detection system \cite{run2muon} is
capable of measuring the local momentum of the
identified muon.
A distinctive feature of the muons included in the background fractions
$f_K$, $f_\pi$, $F_K$ and $F_\pi$ is that their track parameters
measured by the tracking system (referred to as ``central" track parameters)
correspond to the original kaon or pion,
while the track parameters measured by the muon system (referred to as ``local" track parameters)
correspond to the muon produced in kaon or pion decay.
Thus, these two measurements are intrinsically different.
We exploit this feature in our event selection by selecting muons with
$\chi^2 < 12$ for 4 d.o.f. \cite{D03},
where $\chi^2$ is calculated from the difference between the track parameters
measured in the central tracker and in the local muon system.
In addition to this selection, in the present analysis
we develop a method of measuring the background
fractions using the difference in the central and local measurements of the muon track parameters.

\begin{figure}
\begin{center}
\includegraphics[width=0.48\textwidth]{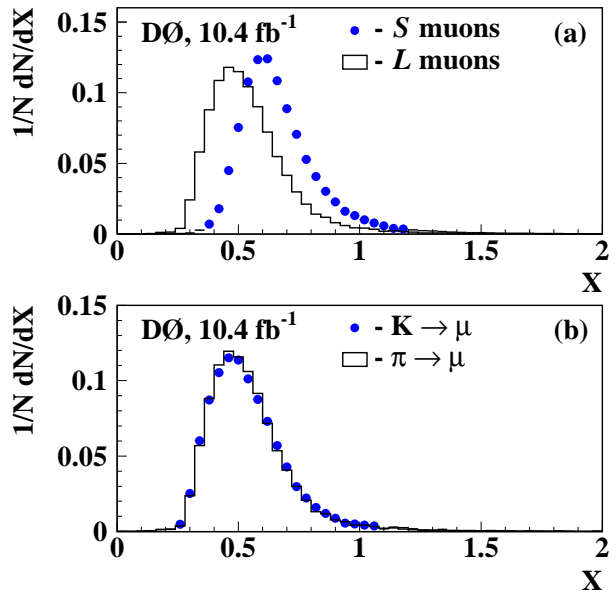}
\caption{(a)
Normalised distributions of $X$ for $S$ and $L$ muons in $\pteta$ bin 2 of the inclusive muon sample.
(b) Normalised distributions of $X$ for $K \to \mu$ and $\pi \to \mu$ muons.
}
\label{locx}
\end{center}
\end{figure}

We define a variable $X$ as
\begin{equation}
X = \frac{p({\rm local})}{p({\rm central})}.
\end{equation}
Here $p({\rm local})$ and $p({\rm central})$ are the momenta measurements of the local
and central tracks, respectively.
Figure~\ref{locx}(a) shows the normalised distributions of this variable for $S$ muons and $L$ muons
in the $\pteta$ bin 2 of the inclusive muon sample.
The distribution for $S$ muons is obtained using identified muons
from the decay $D^0 \to K^- \mu^+ \nu$. The distribution for $L$ muons
is obtained as a linear combination of the distributions of $K \to \mu$ tracks
and $\pi \to \mu$ tracks with the coefficients corresponding to their
fractions in the inclusive muon sample. These two distributions are shown separately
in Fig.~\ref{locx}(b). The distribution for $K \to \mu$ tracks is
obtained using kaons produced in the $\phi \to K^+ K^-$ decay and misidentified as muons.
The distribution for $\pi \to \mu$ tracks is obtained using pions produced
in the $\ks \to \pi^+ \pi^-$ decay and misidentified as muons. 
Since we select muons with at least 3 hits in SMT, the $\ks$ decay is forced to be
within the beam pipe.
All these distributions
are obtained using exclusively the events in a given $\pteta$ bin. 

Figure~\ref{locx}(a) shows that the distribution for $L$ muons is shifted towards
lower $X$ values reflecting the fact that a part of the total momentum of the kaon or pion
is taken away by the neutrino. The difference
between the distributions for $K \to \mu$ muons and $\pi \to \mu$ muons is relatively small.
This observation corresponds to the expectation that the fraction of momentum in the laboratory frame
taken away by the neutrino is similar in $K \to \mu$ and $\pi \to \mu$ decays.
The position of the maximum of the distribution of the $X$ variable for $S$ muons
is lower than 1 because of the muon energy loss in the detector material.
The typical energy loss of muons in the material of D0 detector is 3--4 GeV
depending on muon $\eta$ \cite{muid}.

\begin{figure}
\begin{center}
\includegraphics[width=0.48\textwidth]{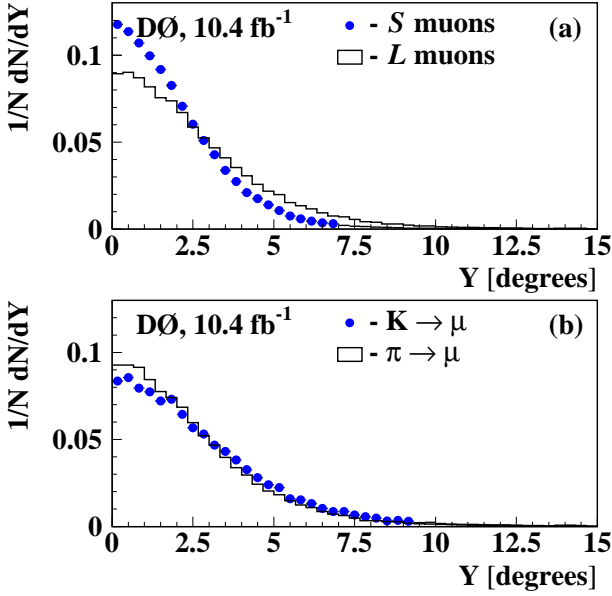}
\caption{(a)
Normalised distributions of $Y$ for $S$ and $L$ muons in $\pteta$ bin 2 of the inclusive muon sample.
(b) Normalised distributions of $Y$ for $K \to \mu$ and $\pi \to \mu$ muons.
}
\label{locy}
\end{center}
\end{figure}

Another variable used in this study is the difference between
the polar angles of the local and central tracks
\begin{equation}
Y = | \theta({\rm local}) - \theta({\rm central})|.
\end{equation}
Figure~\ref{locy}(a) shows the normalised distributions of this variable for $S$ muons and $L$ muons
in the $\pteta$ bin 2 of the inclusive muon sample.
Figure~\ref{locy}(b) presents the separate distributions of $K \to \mu$ and $\pi \to \mu$ tracks.
The distribution for $L$ muons is wider than that for $S$ muons.
A part of the four-momentum of $L$ muons is taken away by an invisible
neutrino. This missing momentum results in a kink in the $K \to \mu$ or $\pi \to \mu$ track,
which produces a wider $Y$ distribution.

\begin{figure}
\begin{center}
\includegraphics[width=0.48\textwidth]{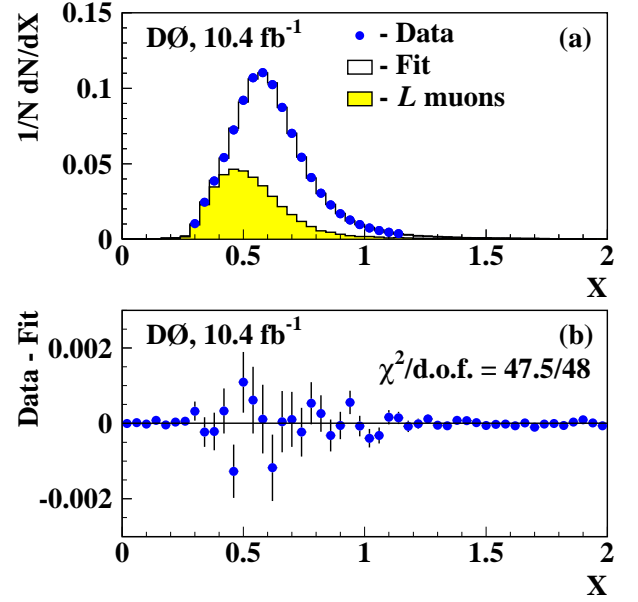}
\caption{(color online). (a)
Normalised distributions of $X$ in $\pteta$ bin 2 of the inclusive muon sample.
Both the data and fitted distributions are shown. The filled histogram shows
the contribution of $L$ muons.
(b) Difference between data and fitted normalised distributions
of $X$. The $\chi^2$/d.o.f. of the fit of these differences to their average is also shown.
}
\label{locx-fit}
\end{center}
\end{figure}

\begin{figure}
\begin{center}
\includegraphics[width=0.48\textwidth]{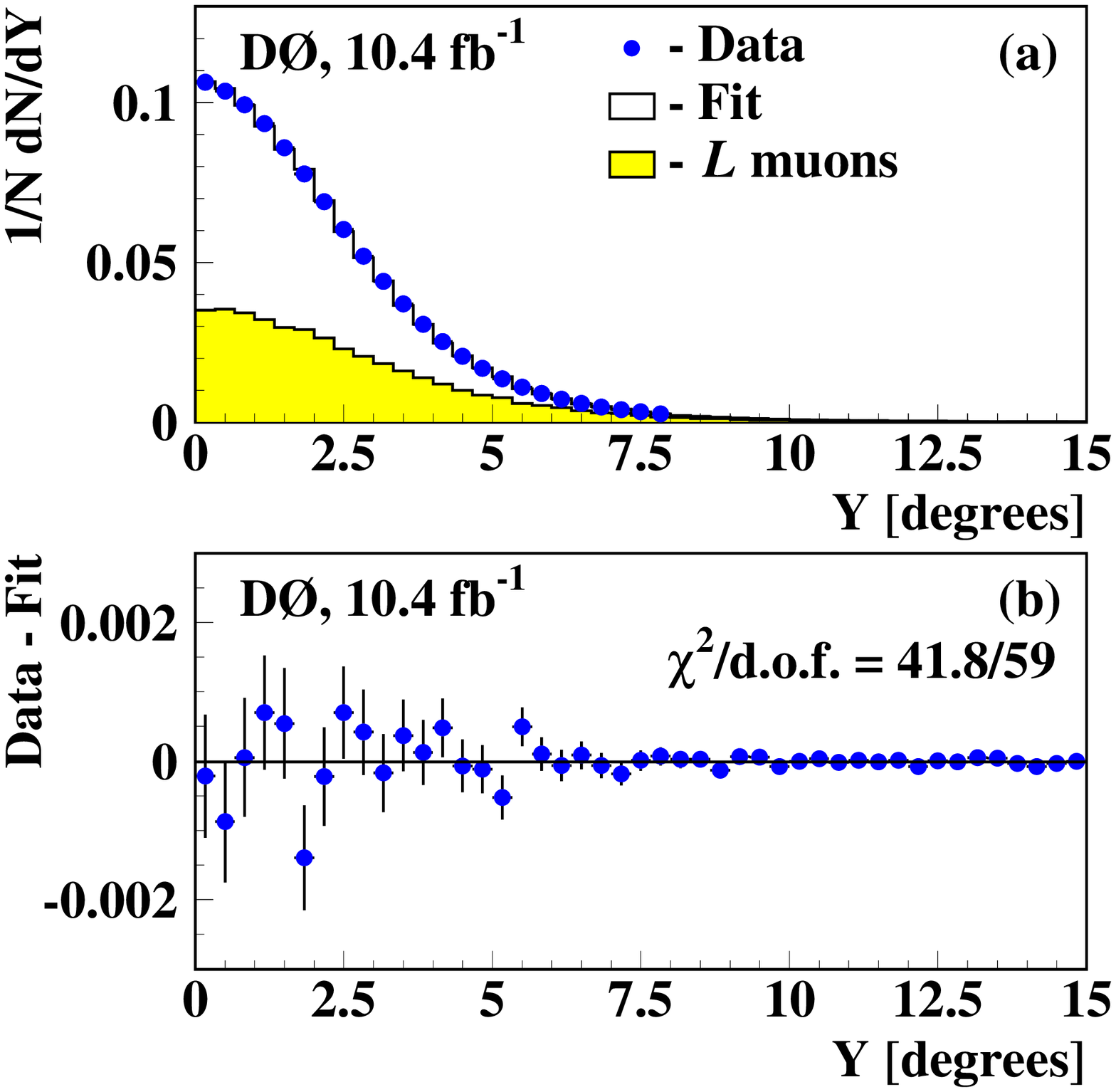}
\caption{(color online). (a)
Normalised distributions of $Y$ in $\pteta$ bin 2 of the inclusive muon sample.
Both the data and fitted distributions are shown. The filled histogram shows
the contribution of $L$ muons.
(b) Difference between data and fitted normalised distributions
of $Y$. The $\chi^2$/d.o.f. of the fit of these differences to their average is also shown.
}
\label{locy-fit}
\end{center}
\end{figure}

We fit the distribution of $X$ and $Y$ variables in each $\pteta$ bin $i$
of each IP sample of the inclusive muon sample using the templates
for $S$ muons and $L$ muons and determine the background fraction $f_{\rm bkg}^i$ for this
IP sample. Since the distributions for $K \to \mu$ tracks and $\pi \to \mu$ tracks are similar, this
method is not sensitive to the separate fractions $f_K$ and $f_\pi$. Therefore, the ratio of these
two fractions is fixed to the value measured in data using the $K^{*0}$ method.
The templates for each $\pteta$ bin $i$
are built using exclusively the events in a given $\pteta$ bin.
The background fraction in a given IP sample is computed using the relation
\begin{equation}
f_{\rm bkg} = \sum_{i=1}^9 f_\mu^i f_{\rm bkg}^i,
\end{equation}
where the sum is taken over all $\pteta$ bins.
Figures~\ref{locx-fit} and \ref{locy-fit}  show an example of this fit for the $X$ and $Y$ variables, respectively, in the $\pteta$ bin 2.
Figures~\ref{locx-fit}(a) and \ref{locy-fit}(a) show the normalised distributions of $X$ and $Y$ in the
inclusive muon sample, and the expected distributions obtained from the fit.
These distributions are indistinguishable on this scale, since the statistics in the inclusive muon sample
is very large. Figures~\ref{locx-fit}(b) and \ref{locy-fit}(b) show the difference between
the observed and expected normalised distributions. The quality of the description
of the observed distributions is very good. The fit of these differences to their average
gives $\chi^2$/d.o.f. = 48/48 for $X$ and 42/59 for $Y$.

\begin{table*}
\caption{\label{tab9}
Comparison of background fractions measured using local variables with the background
fractions obtained using $K^{*0}$ production. The relative difference $\delta_f$
between two independent measurements is also shown. This quantity is defined in Eq.~(\ref{df}).
Only statistical uncertainties are shown.
}
\begin{ruledtabular}
\newcolumntype{A}{D{A}{\pm}{-1}}
\newcolumntype{B}{D{B}{-}{-1}}
\begin{tabular}{lAAAA}
Quantity &  \multicolumn{1}{c}{All IP} & \multicolumn{1}{c}{IP=1} & \multicolumn{1}{c}{IP=2} & \multicolumn{1}{c}{IP=3} \\
\hline
$f_{\rm bkg}$(local)$\times 10^2$ from $X$ & 42.70\ A \ 0.09 & 55.28\ A \ 0.09 & 22.89\ A \ 0.10 & 8.49\ A \ 0.12 \\
$f_{\rm bkg}$(local)$\times 10^2$ from $Y$ & 40.97\ A \ 0.30 & 53.41\ A \ 0.28 & 20.60\ A \ 0.37 & 8.76\ A \ 0.40 \\
\hline
Average $f_{\rm bkg}$(local)$ \times 10^2$    & 42.56\ A \ 0.87 & 55.10\ A \ 0.94 & 22.73\ A \ 1.15 & 8.51\ A \ 0.14 \\
\hline
$f_{\rm bkg}(K^{*0}) \times 10^2$     & 46.73\ A \ 1.76 & 60.19\ A \ 2.21 & 23.38\ A \ 1.01 & 8.09\ A \ 0.47 \\
$\delta_f \times 10^2$ &
                                        -8.92\ A \ 3.90 & -8.46\ A \ 3.71 & -2.78\ A \ 6.47 & 5.19\ A \ 6.35 \\
\end{tabular}
\end{ruledtabular}
\end{table*}

The resulting background fractions in different IP samples are given in Table~\ref{tab9}.
Only the statistical uncertainties are given.
The statistical uncertainty of the measurements with $X$ and $Y$ variables
is less than the difference between them.
Therefore, we take the weighted average of these two measurements as the
central value of the background fraction $f_{\rm bkg}$(local)
and assign half of the difference between them as its uncertainty.

The obtained values $f_{\rm bkg}$(local) can be compared with the background fractions $f_{\rm bkg}(K^{*0})$
measured using the $K^{*0}$ method described in the previous section. These fractions, as well as the
relative difference
\begin{equation}
\label{df}
\delta_f \equiv \frac{f_{\rm bkg}\mbox{(local)} - f_{\rm bkg}(K^{*0})}
{f_{\rm bkg}(K^{*0})}
\end{equation}
are also given in Table~\ref{tab9}.

All templates for the measurement of background fractions with local variables
are obtained using the inclusive muon sample.
It makes this measurement self-consistent. The available statistics of the dimuon events
is insufficient to obtain the corresponding templates for the measurement in the dimuon sample.
Therefore, the background
fractions are measured only in the inclusive muon sample, and the method of local variables is
used as a cross check of the corresponding quantities obtained with the $K^{*0}$ method.

The background measurements with these two methods are fully independent. They are based
on different assumptions and are subject to different systematic uncertainties, which are not
included in the uncertainty of $\delta_f$ shown in Table~\ref{tab9}.
The background fraction changes by more than six times between the
samples with small and large IP. Nevertheless, the two methods
give consistent results for all IP samples. The remaining difference between them, which exceeds
two standard deviations only for the sample with small IP, is assigned as a
systematic uncertainty, and is discussed in Section~\ref{sec_systematic}.
Thus, the background measurement
with local variables provides an independent and important
confirmation of the validity of the analysis procedure used
to determine the background fractions.


\subsection{Systematic uncertainties on backgrounds}
\label{sec_systematic}

The systematic uncertainties for the background fractions are discussed in Refs.~\cite{D02, D03}.
Here we describe the changes applied in the present analysis.
In our previous measurement the systematic uncertainty of the fraction $f_K$ was set to 9\% \cite{D02,D03}.
In the present analysis we perform an alternative measurement of background fractions using the local
variables. The results of two independent measurements, given in Table \ref{tab9}, are
statistically different only for the IP=1 sample. We attribute this difference to the systematic
uncertainties of the two measurements.
Since the background fractions $f_\pi$ and $f_p$ are derived
using the measured fraction $f_K$ \cite{D02},
we set the relative systematic uncertainty of $f_K$,
 $f_\pi$, and $f_p$ in each IP sample
to $\delta_f / \sqrt{2}$, or to $\sigma(\delta_f)/ \sqrt{2}$, whichever value is larger.
Here, $\sigma(\delta_f)$ is the uncertainty of $\delta_f$.
We assume the full correlation of this uncertainty between $f_K$, $f_\pi$, and $f_p$.
Numerically, the value of the systematic uncertainty of $f_K$ is about 6.3\% for the IP=1
sample and 4.5\% for the IP=3 sample, which is smaller than,
but consistent with, our previous assignment \cite{D03} of the
systematic uncertainty on the $f_K$ fraction.

The procedure to determine the relative systematic uncertainty on the ratio $F_K/f_K$ is discussed in Ref.~\cite{D03}.
Following this procedure we set this uncertainty to 2.9\%,
compared to 3.0\% uncertainty applied in \cite{D03} where the change is due to the addition of the final 1.4 fb$^{-1}$ of data.

Other systematic uncertainties remain as in \cite{D02, D03}.
Namely, the systematic uncertainties on the ratios of multiplicities
$n_\pi / n_K$ and $n_p / n_K$, required to compute $f_\pi$ and $f_p$,
are set to 4\%.
The systematic uncertainties on	the ratios of multiplicities
$N_\pi / N_K$ and $N_p / N_K$, required	to compute $F_\pi$ and $F_p$,
are also set to 4\%.

\section{Measurement of background asymmetries}
\label{sec_abkg}

The background asymmetries arise from the difference of interaction
cross-section of positive and
negative particles with the detector material. The
asymmetries for kaons, pions and protons are denoted as $a_K$, $a_\pi$ and $a_p$, respectively.
The origin
of different asymmetries and their measurement techniques
are discussed in detail in Ref. \cite{D02,D03}.
The asymmetry $a_K$ is measured by reconstructing exclusive decays
$K^{*0} \rightarrow K^+ \pi^-$ and $\phi \rightarrow K^+ K^-$ with
$\ktomu$. The asymmetry $a_\pi$ is measured
using the reconstructed $K_S \rightarrow \pi^+ \pi^-$ decay with
$\pitomu$. The asymmetry $a_p$
is measured by reconstructing the $\Lambda \rightarrow p \pi^-$ decay with the proton
misidentified as a muon. All these asymmetries are measured directly in data and therefore
they include the possible asymmetry induced by the trigger.

Another source of background asymmetry is the difference between positive and negative
muon detection, identification and track reconstruction. This asymmetry $\delta$
is measured by reconstructing decays $J/\psi \rightarrow \mu^+ \mu^-$
using track information only and then counting the tracks that have been
identified as muons.
Due to the measurement method, the asymmetry $\delta$ does not include the possible
track reconstruction asymmetry. A separate study presented in Ref.~\cite{D02}
shows that track reconstruction asymmetry is consistent with zero within the experimental uncertainties.
This is a direct consequence of the regular reversal of magnet polarities discussed
in Section~\ref{selection}.

\begin{figure}
\begin{center}
\includegraphics[width=0.48\textwidth]{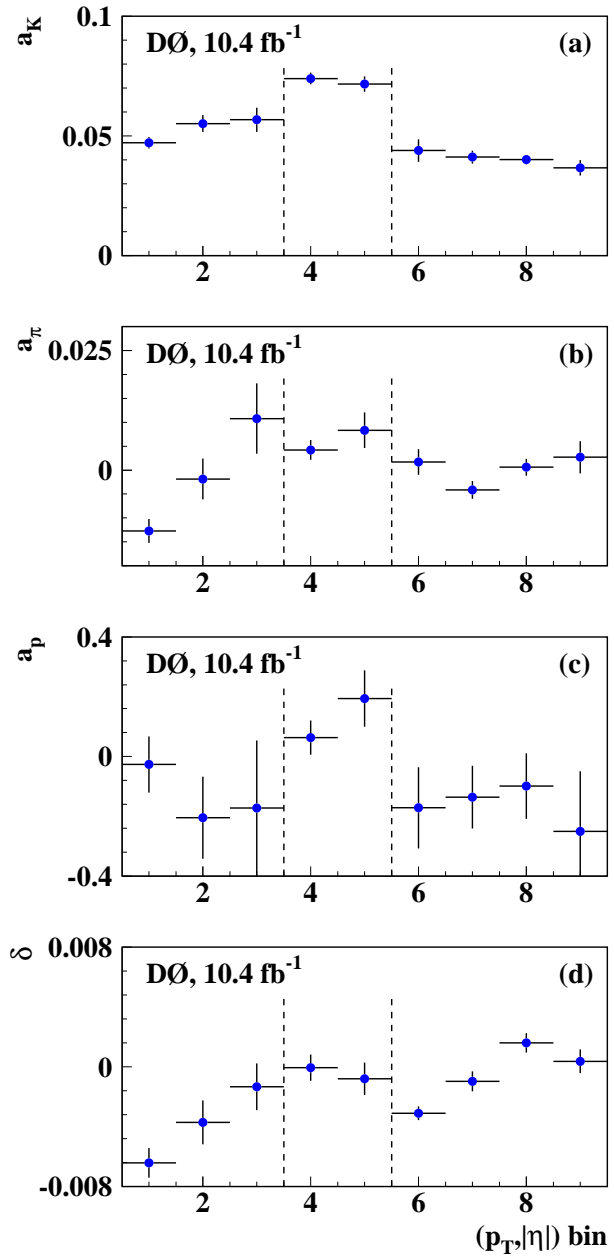}
\caption{Asymmetries (a) $a_K$, (b) $a_\pi$,
(c) $a_p$, and (d) $\delta$ as a function of the
kaon, pion, proton and muon $\pteta$ bin $i$, respectively.
Only statistical uncertainties are shown.
The vertical dashed lines separate the $\pteta$
bins corresponding to the central, intermediate, and forward regions of the D0 detector, respectively.
}
\label{fig-abkg}
\end{center}
\end{figure}

In this analysis all background
asymmetries are measured in $\pteta$ bins.
It was verified in Ref. \cite{D03} that the background asymmetries do not depend
on the particle IP within the statistical uncertainties of their measurement.
Therefore, the same values of background asymmetries are used for different
IP samples. The background asymmetries obtained are shown in Fig.~\ref{fig-abkg}.
The values of the background asymmetries averaged over all $\pteta$ bins are:
\begin{eqnarray}
a_K   & = & +0.0510 \pm 0.0010,\\
a_\pi & = & -0.0006 \pm 0.0008, \\
a_p   & = & -0.0143 \pm 0.0342, \\
\delta & = & -0.0013 \pm 0.0002.
\end{eqnarray}

\section{Measurement of asymmetries $\bm{a_{\rm CP}}$ and $\bm{A_{\rm CP}}$}
\label{sec_results}

Using the full $10.4$ fb$^{-1}$ of integrated luminosity collected by the D0 experiment
in Run II we select
$2.17 \times 10^9$ inclusive muon events and
$6.24 \times 10^6$ like-sign dimuon events. For comparison, the number
of opposite-sign dimuon events with the same selections is $2.18 \times 10^7$.
The fraction of like-sign dimuon events in the present analysis common with the events used in Ref.~\cite{D03}
is 74\%. This value reflects the changes of the sample size
due to the luminosity increased from 9 fb$^{-1}$
to 10 fb$^{-1}$, and due to the additional requirement
of the number of SMT hits associated with a muon,
see section~\ref{selection} for details.

The raw asymmetries $a^i$ and $A^i$ in a given $\pteta$ bin $i$ are determined
using Eqs.~(\ref{ai}) and~(\ref{ai2mu}), respectively. The raw asymmetries
$a$
and $A$
are obtained using Eqs.~(\ref{atot}) and~(\ref{ai2mu-2}), respectively.
The background asymmetries $a_{\rm bkg}$ and $A_{\rm bkg}$ are obtained
using the methods presented in Sections~\ref{sec_bck} and~\ref{sec_abkg}.
They are subtracted from the raw asymmetries
$a$ and $A$ to obtain the residual asymmetries $a_{\rm CP}$
and $A_{\rm CP}$.

The raw asymmetry $a$, the contribution of
different background sources, and the residual asymmetry $a_{\rm CP}$
for the total inclusive muon sample and
for different IP samples are given in Table~\ref{tab1a}.
This table gives the values with statistical uncertainties only.
The asymmetry $a_{\rm CP}$ with both statistical and systematic uncertainties
is given in Table~\ref{tab3}.

\begin{table*}
\caption{\label{tab1a}
Contributions to background asymmetry $a_{\rm bkg}$, the raw asymmetry $a$, and
the residual charge asymmetry $a_{\rm CP}$ in the IP samples of the inclusive
muon sample. The column ``All IP'' corresponds to the full inclusive muon sample
without dividing it into the IP samples. Only statistical uncertainties are given.
}
\begin{ruledtabular}
\newcolumntype{A}{D{A}{\pm}{-1}}
\newcolumntype{B}{D{B}{-}{-1}}
\begin{tabular}{lAAAA}
Quantity &  \multicolumn{1}{c}{All IP} & \multicolumn{1}{c}{IP=1} & \multicolumn{1}{c}{IP=2} & \multicolumn{1}{c}{IP=3} \\
\hline
$f_K a_K \times 10^3$ &
     7.99\ A \  0.21  &  10.37\ A \  0.29 &  3.85\ A \  0.17  &  1.34\ A \  0.10  \\
$f_\pi a_\pi \times 10^3$ &
     -0.19\ A \  0.31 &  -0.22\ A \  0.40 &  -0.19\ A \  0.16 &  -0.07\ A \  0.06 \\
$f_p a_p \times 10^3$ &
     -0.08\ A \  0.09 &  -0.10\ A \  0.12 &  -0.04\ A \  0.05 &  -0.01\ A \  0.01 \\
$a_\mu \times 10^3$ &
     -0.70\ A \  0.12 &  -0.50\ A \  0.09 &  -1.02\ A \  0.17 &  -1.28\ A \  0.21 \\
$a \times 10^3$       &
      6.70\ A \  0.02 &   9.30\ A \  0.03 &  2.77\ A \  0.06  &  -0.49\ A \  0.05 \\
$a_{\rm bkg} \times 10^3$         &
      7.02\ A \  0.42 &   9.54\ A \  0.53 &   2.59\ A \  0.27 &  -0.01\ A \  0.23 \\
$a_{\rm CP} \times 10^3$     &
      -0.32\ A \  0.42 &  -0.24\ A \  0.53&   0.18\ A \  0.28 &  -0.48\ A \  0.24 \\
\end{tabular}
\end{ruledtabular}
\end{table*}

\begin{figure}
\begin{center}
\includegraphics[width=0.48\textwidth]{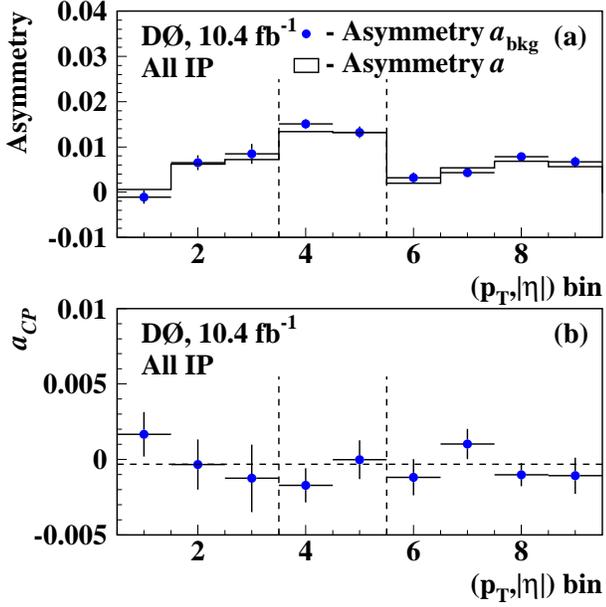}
\caption{(a)
The asymmetry $a_{\rm bkg}^i$ (points with error
bars representing the statistical uncertainties), shown in each $\pteta$ bin $i$,
is compared to the measured asymmetry $a^i$
for the total inclusive muon sample (shown as a histogram, since the statistical uncertainties
are negligible).
The asymmetry from CP violation is negligible
compared to the background uncertainty in the inclusive muon sample.
The vertical dashed lines separate the $\pteta$
bins corresponding to the central, intermediate, and forward regions of the D0 detector, respectively.
(b)~The asymmetry $a_{\rm CP}^i$. The horizontal dashed line shows the value of $a_{\rm CP}$
defined as the weighted sum in Eq.~(\ref{atot1}).
}
\label{fig-acomp-all}
\end{center}
\end{figure}

\begin{figure}
\begin{center}
\includegraphics[width=0.48\textwidth]{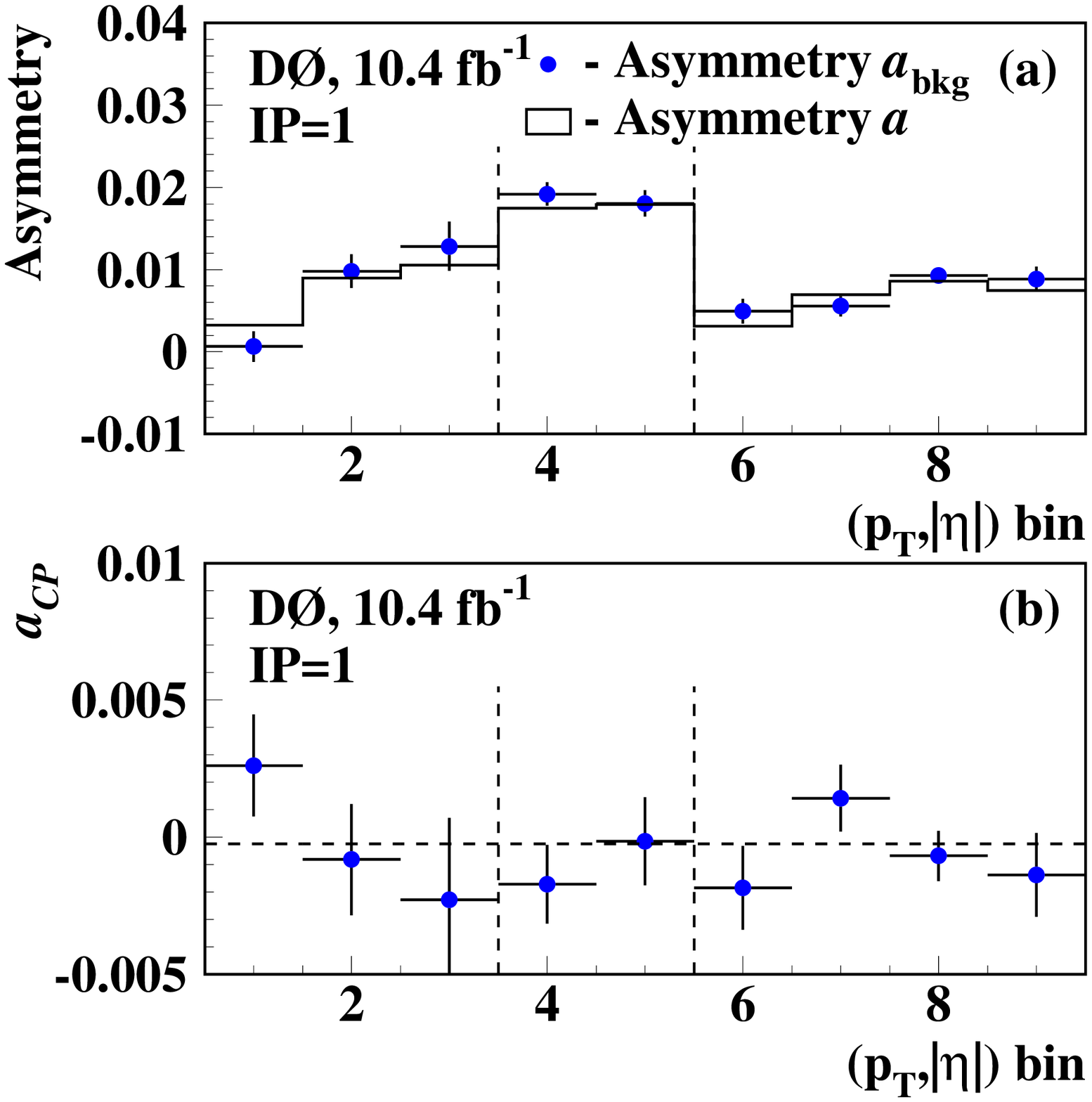}
\caption{Same as Fig.~\ref{fig-acomp-all} for IP=1 sample.
}
\label{fig-acomp-ip1}
\end{center}
\end{figure}

\begin{figure}
\begin{center}
\includegraphics[width=0.48\textwidth]{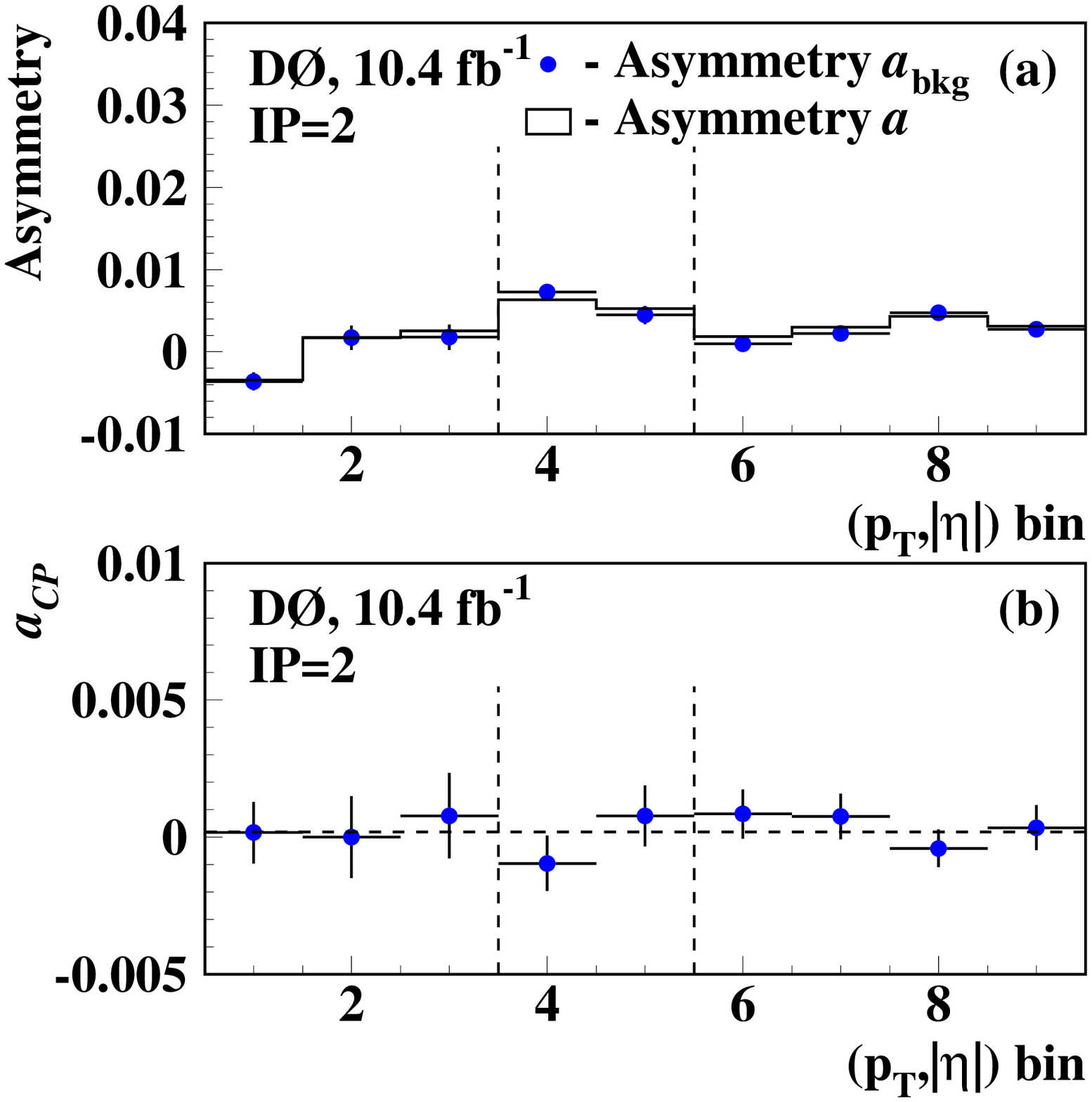}
\caption{Same as Fig.~\ref{fig-acomp-all} for IP=2 sample.
}
\label{fig-acomp-ip2}
\end{center}
\end{figure}

\begin{figure}
\begin{center}
\includegraphics[width=0.48\textwidth]{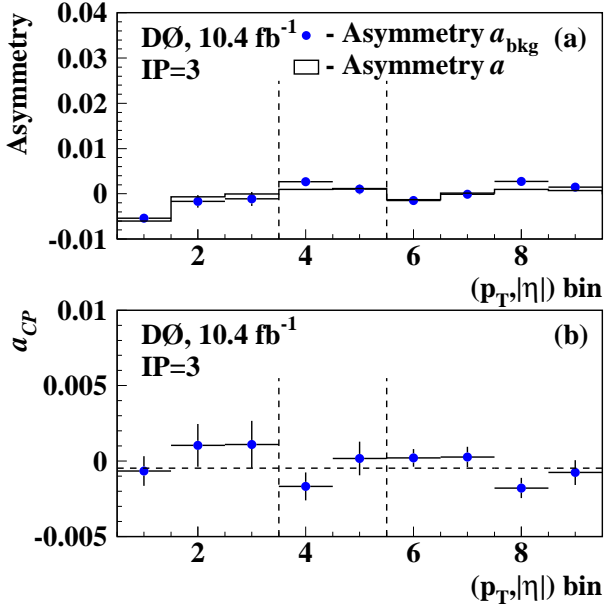}
\caption{Same as Fig.~\ref{fig-acomp-all} for IP=3 sample.
}
\label{fig-acomp-ip3}
\end{center}
\end{figure}

The charge asymmetry of $S$ muons in the inclusive muon sample is expected to be small,
see Section~\ref{expectation} for details.
Thus, the observed inclusive single muon asymmetry
is expected to be consistent with the estimated background within its uncertainties.
Therefore, the comparison of the observed and expected inclusive single muon asymmetries
provides a stringent closure test and validates the method of background calculation.
In the present analysis such a comparison is performed both for the total inclusive muon
sample and for the IP samples. The results are shown in
Figs.~\ref{fig-acomp-all}--\ref{fig-acomp-ip3}.
The $\chi^2(a_{\rm CP})$ of the fits of the
differences $a^i - a^i_{\rm bkg}$ to their averages are given in Table~\ref{tab3}.
For each fit the number of degrees of freedom is equal to eight.
Only the statistical uncertainties of $a^i$ and $a^i_{\rm bkg}$ are used to compute
$\chi^2(a_{\rm CP})$.

The comparison shows an excellent agreement between the observed
and expected asymmetries in different kinematic $\pteta$ bins and in different IP samples.
The difference for the total sample
is consistent with zero within 0.042\% accuracy,
while the raw asymmetry varies as much as 1.5\% between $\pteta$ bins.
This result agrees with the expectation that the charge
asymmetry of $S$ muons in the inclusive muon sample should be negligible compared
to the uncertainty of the background asymmetry, see Tables~\ref{tab1a} and~\ref{tab5}.

The comparison of observed and expected asymmetries
in the three non-overlapping IP samples
does not reveal any bias with the change of the muon IP.
The values of $\chi^2(a_{\rm CP})$ in Table~\ref{tab3} are obtained
with statistical uncertainties only. The compatibility of these values
with the statistical $\chi^2$ distribution indicates
that the systematic uncertainties do not depend on the kinematic properties of the event.
For the IP=3 sample the contribution of the background asymmetry
is strongly suppressed. Therefore, the observed asymmetry is sensitive to a possible
charge asymmetry of $S$ muons, which could be reflected in the
deviation of $a_{\rm CP}$ from zero for this sample. Still, this deviation,
taking into account the systematic uncertainty, is less than two standard deviations.
The obtained values of $a_{\rm CP}$ in the total inclusive muon sample and in the three
non-overlapping IP samples, including the systematic uncertainties, are given in Table~\ref{tab3}.
\begin{table}
\caption{\label{tab3}
Residual asymmetry $a_{\rm CP}$ in the full
inclusive muon sample (row ``All IP''), and in different IP samples.
The first uncertainty is statistical and the second
uncertainty is systematic. The last column gives the $\chi^2$
of the fit of the asymmetries $a^i_{\rm CP}$ in
nine $\pteta$ bins $i$ to their average.
}
\begin{ruledtabular}
\newcolumntype{A}{D{A}{\pm}{-1}}
\newcolumntype{B}{D{B}{-}{-1}}
\begin{tabular}{ccc}
Sample & $a_{\rm CP}$ & $\chi^2(a_{\rm CP})$/d.o.f. \\
\hline
All IP & $(-0.032 \pm 0.042 \pm 0.061)\%$ & 6.93/8 \\
\hline
IP=1   & $(-0.024 \pm 0.053 \pm 0.075)\%$ & 7.54/8 \\
IP=2   & $(+0.018 \pm 0.028 \pm 0.024)\%$ & 3.48/8 \\
IP=3   & $(-0.048 \pm 0.024 \pm 0.011)\%$ & 10.8/8 \\
\end{tabular}
\end{ruledtabular}
\end{table}

The closure test performed in the total inclusive muon sample and in three IP samples
validates the adopted method of the background measurement and demonstrates its
robustness in different kinematic $\pteta$ and IP regions. For the IP=1 sample the kaon asymmetry is the
dominant background source, while for the IP=3 sample
the kaon and detector asymmetries have approximately the same magnitude,
see Table~\ref{tab1a}. In both cases the expected asymmetry follows the
variation of the observed asymmetry in different kinematic bins, so that the prediction and
the observation agree within statistical uncertainties. Thus, the closure
test provides the confidence
in the measurement of the like-sign dimuon charge asymmetry, where the same method
of background measurement is applied.

The dimuon raw asymmetry $A$, the contribution of
different background sources, and the residual asymmetry $A_{\rm CP}$
for the total like-sign dimuon sample and
for different (IP$_1$,IP$_2$) samples are given in Table~\ref{tab2a}.

\begin{table*}
\caption{\label{tab2a}
Contributions to background asymmetry $A_{\rm bkg}$, the raw asymmetry $A$, and
the residual charge asymmetry $A_{\rm CP}$ in the (IP$_1$,IP$_2$) samples of the like-sign
dimuon sample. The column ``All IP'' corresponds to the full like-sign dimuon sample
without dividing it into the (IP$_1$,IP$_2$) samples.
Only statistical uncertainties are given.
}
\begin{ruledtabular}
\newcolumntype{A}{D{A}{\pm}{-1}}
\newcolumntype{B}{D{B}{-}{-1}}
\begin{tabular}{lAAAAAAA}
Quantity &  \multicolumn{1}{c}{All IP} & \multicolumn{1}{c}{IP$_1$,IP$_2$=11} & \multicolumn{1}{c}{IP$_1$,IP$_2$=12} &
\multicolumn{1}{c}{IP$_1$,IP$_2$=13} & \multicolumn{1}{c}{IP$_1$,IP$_2$=22} & \multicolumn{1}{c}{IP$_1$,IP$_2$=23} &
\multicolumn{1}{c}{IP$_1$,IP$_2$=33} \\
\hline
$F_K a_K \times 10^3$ &
    6.25\ A \ 0.29  & 13.45\ A \ 0.78 & 7.76\ A \ 0.78  & 4.69\ A \ 0.62
    & 5.25\ A \ 1.66 & 2.05\ A \ 0.95 & 1.18\ A \ 1.08 \\
$F_\pi a_\pi \times 10^3$ &
     0.04\ A \ 0.25 &  0.36\ A \ 0.53 &  0.09\ A \ 0.32 &  0.00\ A \ 0.23
    & 0.25\ A \ 0.43 &-0.12\ A \ 0.21 & 0.07\ A \ 0.20 \\
$F_p a_p \times 10^3$ &
    -0.06\ A \ 0.07 & -0.12\ A \ 0.13 & -0.07\ A \ 0.09 & -0.04\ A \ 0.04
    &-0.03\ A \ 0.03 &-0.02\ A \ 0.03 &-0.01\ A \ 0.01 \\
$A_\mu \times 10^3$ &
    -2.88\ A \ 0.30 & -2.38\ A \ 0.22 & -2.80\ A \ 0.28 & -2.96\ A \ 0.32
    & -3.05\ A \ 0.31  & -3.11\ A \ 0.35  & -3.43\ A \ 0.36  \\
$A \times 10^3$ &
     1.01\ A \ 0.40 &  6.90\ A \ 0.79 &  3.90\ A \ 0.94 & -1.96\ A \ 0.77
    & -0.21\ A \ 2.12  & -2.68\ A \ 1.15  & -5.29\ A \ 1.18  \\
$A_{\rm bkg} \times 10^3$ &
     3.36\ A \ 0.50 & 11.31\ A \ 1.01 &  4.97\ A \ 1.07 &  1.70\ A \ 0.75
    &  2.43\ A \ 1.87  & -1.20\ A \ 1.52  & -2.17\ A \ 1.24  \\
$A_{\rm CP} \times 10^3$ &
    -2.35\ A \ 0.64 & -4.41\ A \ 1.28 & -1.08\ A \ 1.43 & -3.65\ A \ 1.07
    & -2.64\ A \ 2.83  & -1.48\ A \ 1.91  & -3.12\ A \ 1.71  \\
\end{tabular}
\end{ruledtabular}
\end{table*}

\begin{figure}
\begin{center}
\includegraphics[width=0.48\textwidth]{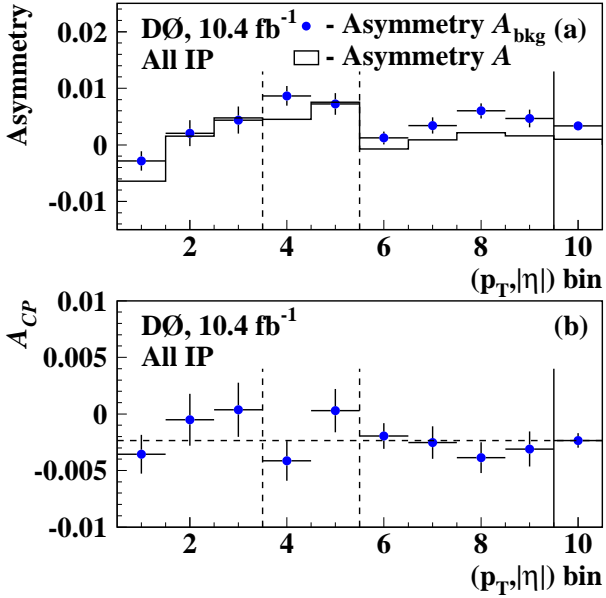}
\caption{(a) The asymmetry $A_{\rm bkg}^i$ (points with error
bars), shown in each $\pteta$ bin~$i$,
is compared to the measured asymmetry $A^i$
for the like-sign dimuon sample (shown as a histogram).
The error bars represent the statistical uncertainty of the difference
$A^i - A_{\rm bkg}^i$.
The vertical dashed lines separate the $\pteta$
bins corresponding to the central, intermediate, and forward regions of the D0 detector, respectively.
The last bin separated by the vertical line shows the values of $A_{\rm bkg}$ defined
as the weighted sum in Eq.~(\ref{ai2mu-3})
and $A$ defined as the weighted sum in Eq.~(\ref{ai2mu-2}) and their statistical uncertainties.
(b)~The asymmetry $A_{\rm CP}^i$.
The last bin separated by the vertical line shows the value of $A_{\rm CP}$ defined
as the weighted sum in Eq.~(\ref{acp}) and its statistical uncertainty.
The horizontal dashed line corresponds to this value of $A_{\rm CP}$.
}
\label{acomp2-all}
\end{center}
\end{figure}

\begin{figure}
\begin{center}
\includegraphics[width=0.48\textwidth]{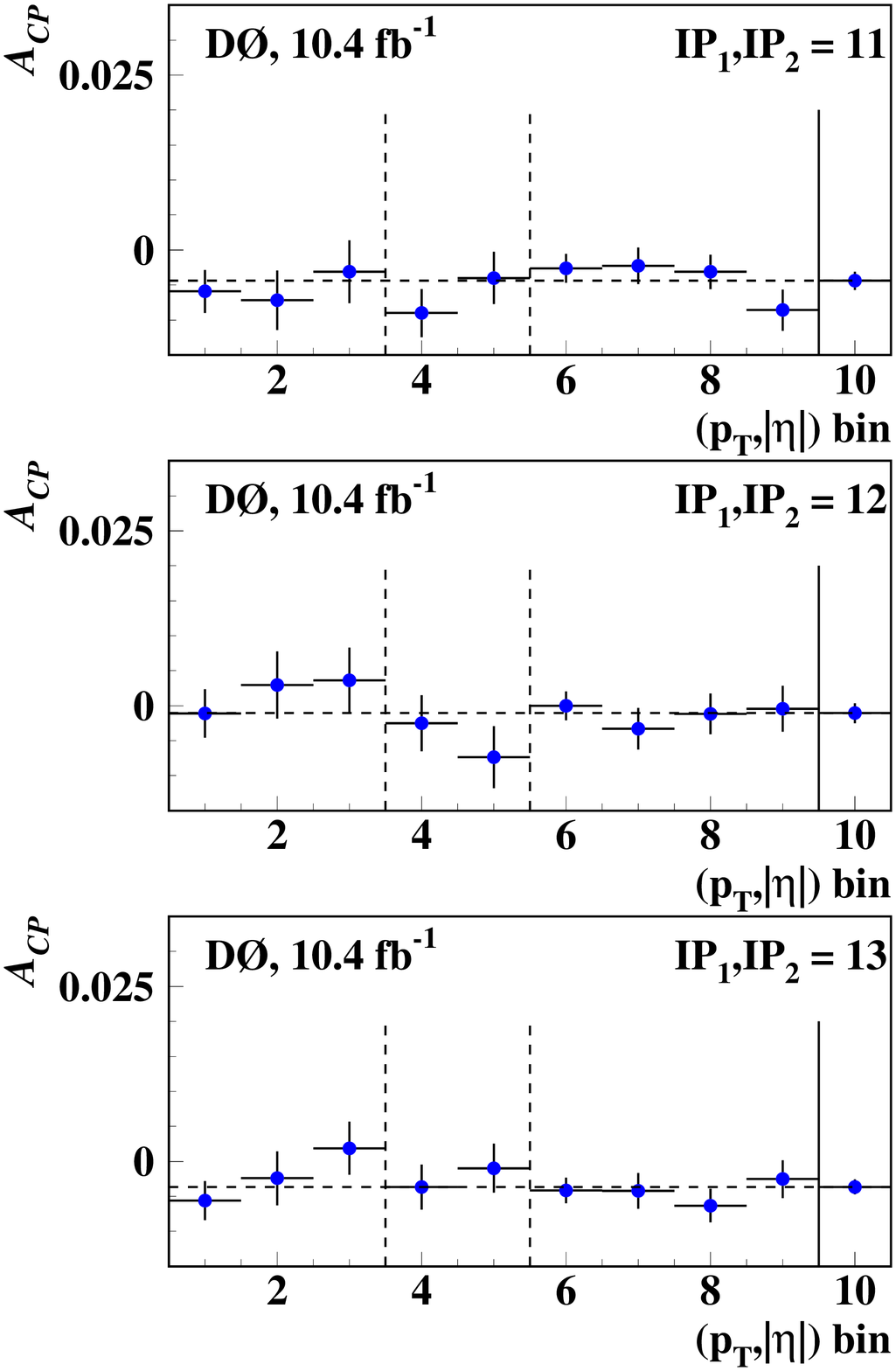}
\caption{The asymmetry $A_{\rm CP}^i$ as a function of $\pteta$ bin~$i$
in different (IP$_1$,IP$_2$) samples.
The error bars represent its statistical uncertainty.
The vertical dashed lines separate the $\pteta$
bins corresponding to the central, intermediate, and forward regions of the D0 detector, respectively.
The last bin separated by the vertical line shows the value of $A_{\rm CP}$ defined
as the weighted sum in Eq.~(\ref{acp}) and its statistical uncertainty.
The horizontal dashed line corresponds to this value of $A_{\rm CP}$.
}
\label{acomp3-1}
\end{center}
\end{figure}

\begin{figure}
\begin{center}
\includegraphics[width=0.48\textwidth]{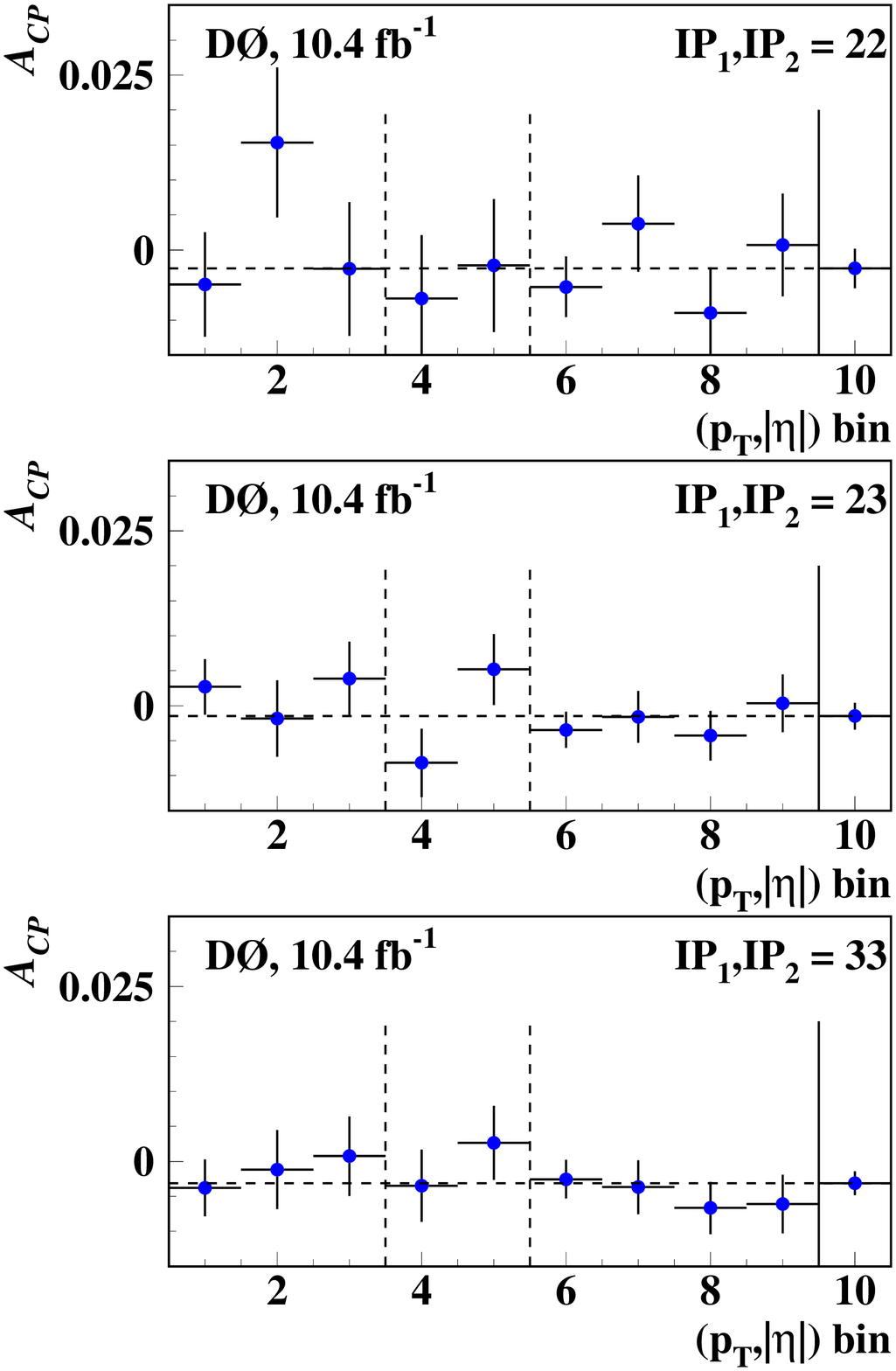}
\caption{The asymmetry $A_{\rm CP}^i$ as a function of $\pteta$ bin~$i$
in different (IP$_1$,IP$_2$)
samples. The error bars represent its statistical uncertainty.
The vertical dashed lines separate the $\pteta$
bins corresponding to the central, intermediate, and forward regions of the D0 detector, respectively.
The last bin separated by the vertical line shows the value of $A_{\rm CP}$ defined
as the weighted sum in Eq.~(\ref{acp}) and its statistical uncertainty.
The horizontal dashed line corresponds to this value of $A_{\rm CP}$.
}
\label{acomp3-2}
\end{center}
\end{figure}

The comparison of the observed and expected background asymmetries
in different kinematic bins is shown in Fig.~\ref{acomp2-all}.
The asymmetry $A^i$ in each $\pteta$ bin is defined in Eq.~(\ref{ai2mu}).
The expected background asymmetry $A^i_{\rm bkg}$ is computed using Eq.~(\ref{ai2mu-1}).
There are two entries per like-sign dimuon event corresponding to the $\pteta$ values of
each muon. Figures~\ref{acomp3-1} and~\ref{acomp3-2} show the values of $A_{\rm CP}^i$
in each $\pteta$ bin for different IP$_1$,IP$_2$ samples.
The last bin separated by the vertical line shows the value of $A_{\rm CP}$ defined
as the weighted sum in Eq.~(\ref{acp}) and its statistical uncertainty.

The quality of agreement between the observed and expected background asymmetries
in different kinematic bins $\pteta$ is given by $\chi^2(A_{\rm CP})$, which is
obtained from the fit of the
differences $A^i - A^i_{\rm bkg}$ to their average.
The values of $\chi^2(A_{\rm CP})$ are given in Table~\ref{tab4}.
The correlation of the $A_i$ and $A^i_{\rm bkg}$ between different $\pteta$ bins is taken
into account in these $\chi^2(A_{\rm CP})$ values.
For each sample the number of degrees of freedom is equal to eight.
Only the statistical uncertainties of $A^i$ and $A^i_{\rm bkg}$ are used to compute
$\chi^2(A_{\rm CP})$.

The comparison shown in Fig.~\ref{acomp2-all}
demonstrates that the expected background asymmetry $A^i_{\rm bkg}$
follows the changes of the observed asymmetry $A^i$ in different kinematic bins $\pteta$
within their statistical uncertainties. However,
the overall deviation of $A - A_{\rm bkg}$ from zero
exceeds three times the statistical uncertainty for the total like-sign dimuon sample, and is
present with less significance in each (IP$_1$,IP$_2$) sample.

The obtained values of $A_{\rm CP}$ are given in Table~\ref{tab4}.
The correlation matrix of the measured asymmetries
$a_{\rm CP}$ and $A_{\rm CP}$ is given in Table~\ref{tab-cor}.
The large correlation between some measurements is because of the common
statistical and systematic uncertainties of the background, see Appendix~\ref{sec_fit} for details.
The asymmetries $a_{\rm CP}$ and $A_{\rm CP}$ measured with full inclusive muon and like-sign dimuon
samples without dividing them into IP samples are given in rows ``All IP'' of Tables~\ref{tab3}
and~\ref{tab4}. The correlation between these measurements is
\begin{equation}
\label{coraA}
\rho = 0.782.
\end{equation}
Figure~\ref{summary1} presents the asymmetries $a_{\rm CP}$ and $A_{\rm CP}$.

\begin{table}
\caption{\label{tab4}
Residual asymmetry $A_{\rm CP}$ in the full
like-sign dimuon sample (row ``All IP''), and
in different (IP$_1$,IP$_2$) samples.
The first uncertainty is statistical and the second
uncertainty is systematic. The last column gives the $\chi^2$
of the fit of the asymmetries $A^i_{\rm CP}$ in
nine $\pteta$ bins $i$ to their average.
}
\begin{ruledtabular}
\newcolumntype{A}{D{A}{\pm}{-1}}
\newcolumntype{B}{D{B}{-}{-1}}
\begin{tabular}{ccc}
Sample & $A_{\rm CP}$ & $\chi^2(A_{\rm CP})$/d.o.f.\\
\hline
All IP  & $(-0.235 \pm 0.064 \pm 0.055)\%$ & 7.57/8 \\
\hline
IP$_1$,IP$_2 = 11$    & $(-0.441 \pm 0.128 \pm 0.113)\%$ & 6.68/8 \\
IP$_1$,IP$_2 = 12$    & $(-0.108 \pm 0.143 \pm 0.061)\%$ & 5.04/8 \\
IP$_1$,IP$_2 = 13$    & $(-0.365 \pm 0.107 \pm 0.036)\%$ & 5.00/8 \\
IP$_1$,IP$_2 = 22$    & $(-0.264 \pm 0.283 \pm 0.039)\%$ & 5.80/8 \\
IP$_1$,IP$_2 = 23$    & $(-0.148 \pm 0.191 \pm 0.033)\%$ & 7.50/8 \\
IP$_1$,IP$_2 = 33$    & $(-0.312 \pm 0.171 \pm 0.012)\%$ & 3.49/8 \\
\end{tabular}
\end{ruledtabular}
\end{table}

\begin{table*}
\caption{\label{tab-cor}
Correlation matrix of the measured values of $a_{\rm CP}$ and $A_{\rm CP}$ in different IP samples.
}
\begin{ruledtabular}
\newcolumntype{A}{D{A}{\pm}{-1}}
\newcolumntype{B}{D{B}{-}{-1}}
\begin{tabular}{lccccccccc}
\multirow{3}{*}{Asymmetry}   & \multicolumn{3}{|c}{$a_{\rm CP}$} & \multicolumn{6}{|c}{$A_{\rm CP}$} \\
\cline{2-10}
                            & \multicolumn{3}{|c}{ }         & \multicolumn{1}{|c}{ } \\
                            & \multicolumn{1}{|c}{IP=1}      & IP=2     & IP=3      & \multicolumn{1}{|c}{IP$_1$,IP$_2$=11}     & =12      & =13      & =22      & =23      & =33      \\
\cline{1-10}
$a_{\rm CP}$ IP=1               & \multicolumn{1}{|c}{1.000}  & 0.785 &  0.459 & \multicolumn{1}{|c}{0.753} & 0.494 & 0.432 & 0.178 & 0.194 & 0.070 \\
$a_{\rm CP}$ IP=2               & \multicolumn{1}{|c}{ 0.785} & 1.000 &  0.686 & \multicolumn{1}{|c}{0.616} &  0.501 &  0.447 &  0.212 &  0.304 &  0.139 \\
$a_{\rm CP}$ IP=3               & \multicolumn{1}{|c}{ 0.459} & 0.686 &  1.000 & \multicolumn{1}{|c}{0.388} &  0.332 &  0.429 &  0.158 &  0.280 &  0.210 \\
\hline
$A_{\rm CP}$ IP$_1$,IP$_2$=11   & \multicolumn{1}{|c}{ 0.753} & 0.616 &  0.388 & \multicolumn{1}{|c}{1.000} &  0.396 &  0.354 &  0.145 &  0.176 &  0.065 \\
$A_{\rm CP}$ IP$_1$,IP$_2$=12   & \multicolumn{1}{|c}{ 0.494} & 0.501 &  0.331 & \multicolumn{1}{|c}{0.396} &  1.000 &  0.294 &  0.121 &  0.213 &  0.063 \\
$A_{\rm CP}$ IP$_1$,IP$_2$=13   & \multicolumn{1}{|c}{ 0.432} & 0.447 &  0.429 & \multicolumn{1}{|c}{0.354} &  0.294 &  1.000 &  0.112 &  0.211 &  0.088 \\
$A_{\rm CP}$ IP$_1$,IP$_2$=22   & \multicolumn{1}{|c}{ 0.178} & 0.212 &  0.158 & \multicolumn{1}{|c}{0.145} &  0.121 &  0.112 &  1.000 &  0.082 &  0.033 \\
$A_{\rm CP}$ IP$_1$,IP$_2$=23   & \multicolumn{1}{|c}{ 0.194} & 0.304 &  0.280 & \multicolumn{1}{|c}{0.176} &  0.213 &  0.211 &  0.082 &  1.000 &  0.059 \\
$A_{\rm CP}$ IP$_1$,IP$_2$=33   & \multicolumn{1}{|c}{ 0.070} & 0.139 &  0.210 & \multicolumn{1}{|c}{0.065} &  0.063 &  0.088 &  0.033 &  0.059 &  1.000 \\
\end{tabular}
\end{ruledtabular}
\end{table*}

\begin{figure}
\begin{center}
\includegraphics[width=0.48\textwidth]{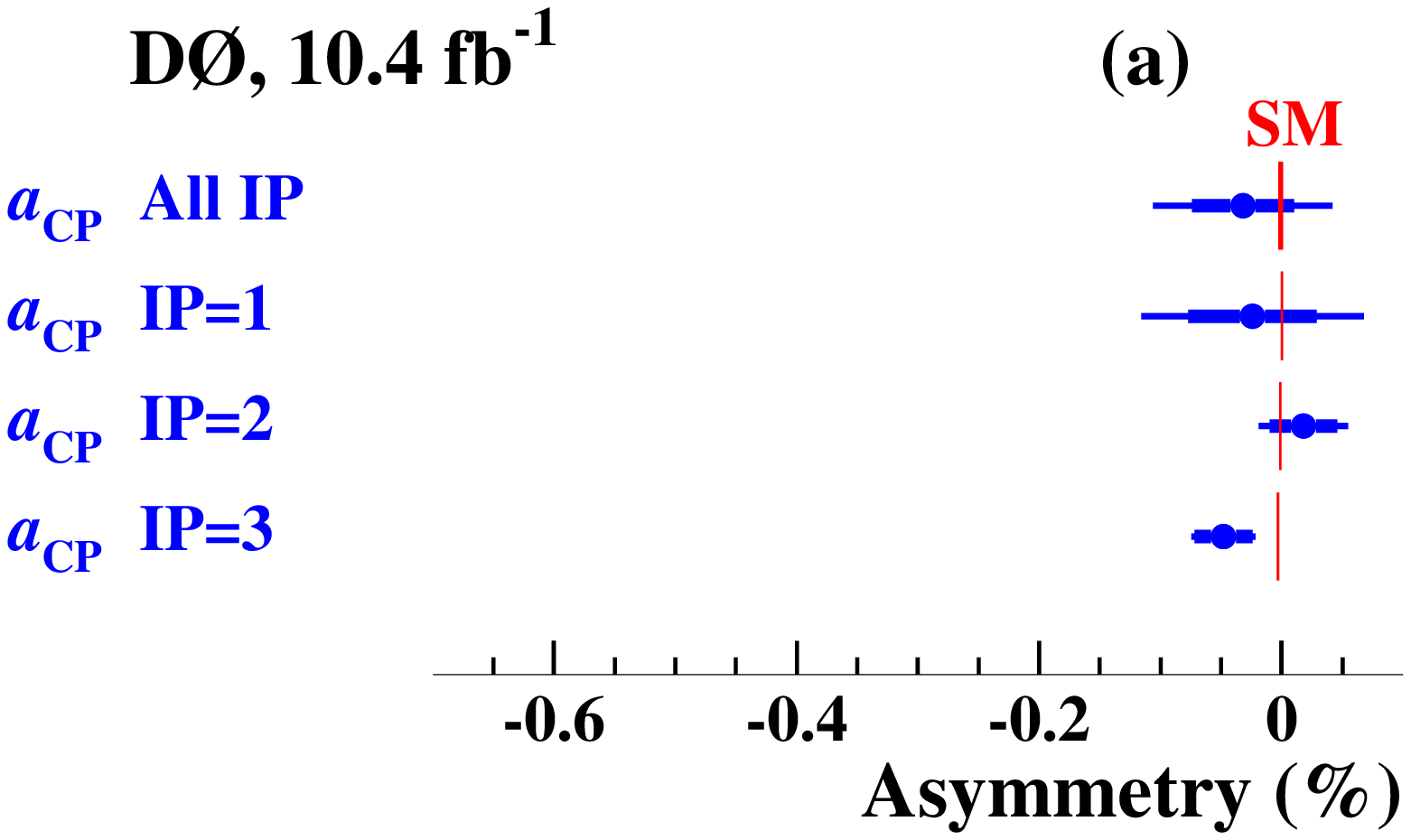}
\includegraphics[width=0.48\textwidth]{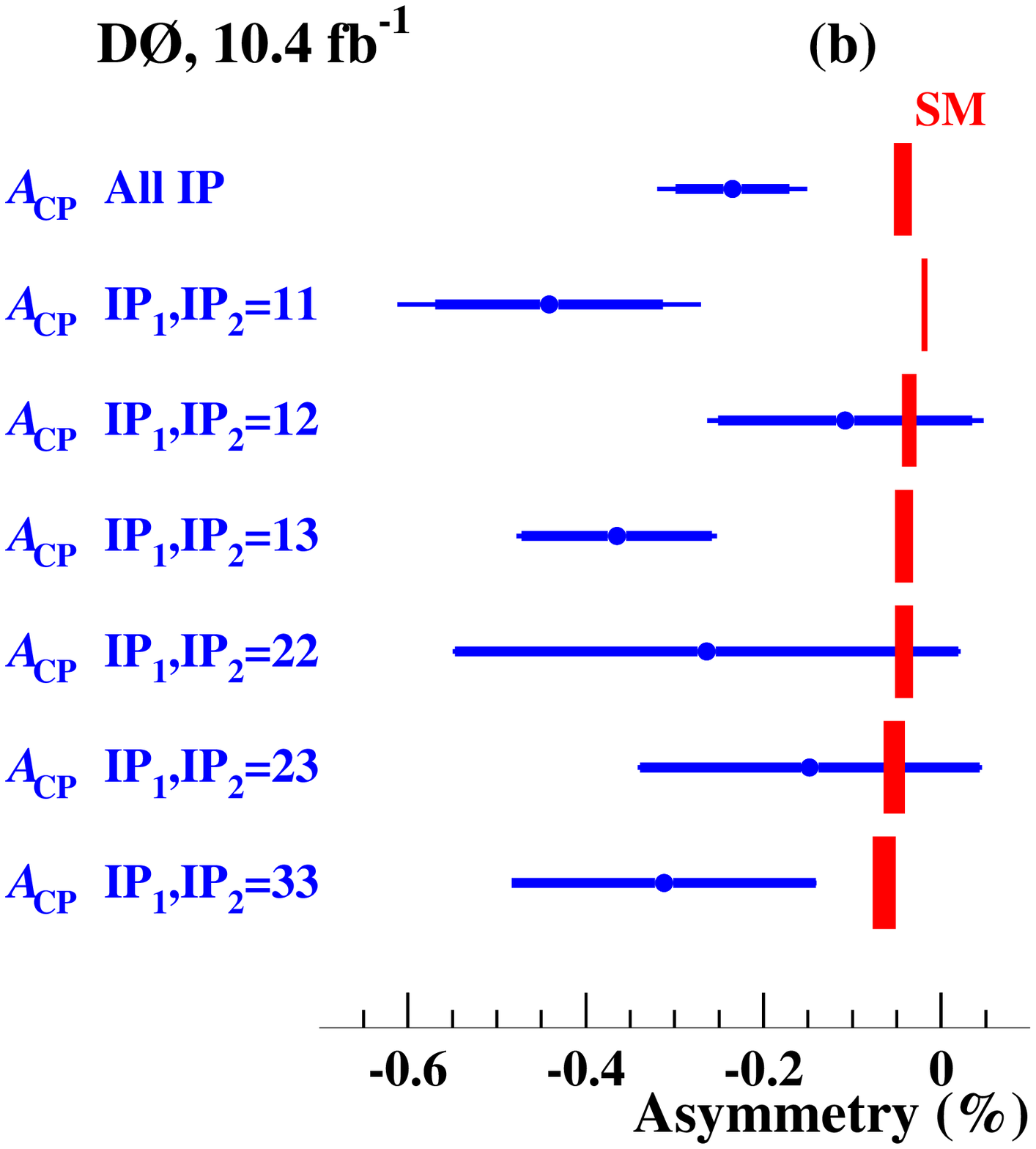}
\caption{(color online).
(a) Asymmetry $a_{\rm CP}$ measured in different IP samples.
(b) Asymmetry $A_{\rm CP}$ measured in different IP$_1$,IP$_2$ samples.
The thick error bar for each measurement presents the statistical uncertainty,
while the thin error bar shows the total uncertainty.
The filled boxes
show the SM prediction. The half width of each box corresponds to the theoretical uncertainty.
}
\label{summary1}
\end{center}
\end{figure}


The obtained values $a_{\rm CP}$ and $A_{\rm CP}$ for all events
in Tables~\ref{tab3} and \ref{tab4} can be compared with our previous results~\cite{D03}
obtained with 9.1 fb$^{-1}$ of integrated luminosity presented in Tables~\ref{tab8a} and \ref{tab8}.
The previous and new results are consistent and the difference between them
is attributed to a more detailed measurement of the background asymmetry
using $\pteta$ bins in the present analysis.
During data taking in Run II we published several measurements of $a_{\rm CP}$ and $A_{\rm CP}$
\cite{asyst}.
The value of $a_{\rm CP}$ is changed between Ref.~\cite{D02} and~\cite{D03} because of
the change in the method of background measurement. This change does not exceed the assigned
systematic uncertainty.
Otherwise, the results demonstrate a good stability
despite the increase by an order of magnitude in the integrated luminosity, and
the many improvements of the analysis methods over the years.
\begin{table}
\caption{\label{tab8a}
Residual asymmetry $a_{\rm CP} = a - a_{\rm bkg}$ measured with different
integrated luminosities $\int L dt$. }
\begin{ruledtabular}
\newcolumntype{A}{D{A}{\pm}{-1}}
\newcolumntype{B}{D{B}{-}{-1}}
\begin{tabular}{ccc}
$\int L dt$ & $a_{\rm CP}$ & Ref. \\
\hline
6.1 fb$^{-1}$  & $(+0.038 \pm 0.047 \pm 0.089)\%$ & \cite{D02}, Table XII \\
9.0 fb$^{-1}$  & $(-0.034 \pm 0.042 \pm 0.073)\%$ & \cite{D03}, Table XII \\
10.4 fb$^{-1}$ & $(-0.032 \pm 0.042 \pm 0.061)\%$ &  this paper \\
\end{tabular}
\end{ruledtabular}
\end{table}

\begin{table}
\caption{\label{tab8}
Residual asymmetry $A_{\rm CP} = A - A_{\rm bkg}$ measured with different
integrated luminosities $\int L dt$. }
\begin{ruledtabular}
\newcolumntype{A}{D{A}{\pm}{-1}}
\newcolumntype{B}{D{B}{-}{-1}}
\begin{tabular}{ccc}
$\int L dt$ & $A_{\rm CP}$ & Ref. \\
\hline
1.0 fb$^{-1}$  & $(-0.28 \pm 0.13 \pm 0.09)\%$ & \cite{D01}, Eq. (11) \\
6.1 fb$^{-1}$  & $(-0.252 \pm 0.088 \pm 0.092)\%$ & \cite{D02}, Table XII \\
9.0 fb$^{-1}$  & $(-0.276 \pm 0.067 \pm 0.063)\%$ & \cite{D03}, Table XII \\
10.4 fb$^{-1}$ & $(-0.235 \pm 0.064 \pm 0.055)\%$ &  this paper \\
\end{tabular}
\end{ruledtabular}
\end{table}

The values given in Tables~\ref{tab3}, \ref{tab4} and \ref{tab-cor} constitute the main
model-independent results of this analysis.

\section{Sources of charge asymmetry}
\label{expectation}

This analysis is performed at a $p \bar p$ collider.
Due to CP-invariant initial state, we assume no production asymmetry of muons.
In the following, we consider the contributions to the charge asymmetries $a_{\rm CP}$
and $A_{\rm CP}$ coming from CP violation in both mixing of neutral $B$ mesons
and in interference of $B$ decays with and without mixing.
Because our measurements are inclusive,
other as yet unknown sources of CP violation could contribute
to the asymmetries $a_{\rm CP}$ and $A_{\rm CP}$ as well.
These sources are not discussed in this paper.

Assuming that the only source of the inclusive single muon charge asymmetry is
CP violation in $\mixBd$ and $\mixBs$ mixing, the asymmetry $a_{S}$ defined
in Eq.~(\ref{atot1}) can
be expressed as
\begin{equation}
\label{asinc}
a_{S} = c_{b} \aslb.
\end{equation}
The coefficient $c_b$, obtained from simulation, represents the fraction of muons produced in the
semileptonic decay of $B$ mesons that have oscillated among all $S$ muons.
This fraction is typically 3\% -- 11\% depending on IP as shown in Table~\ref{tab5}.
The semileptonic charge asymmetry $\aslb$ has contributions from the semileptonic
charge asymmetries $\asld$ and $\asls$ of $\Bd$ and $\Bs$ mesons \cite{Grossman}, respectively:
\begin{equation}
\label{aslb}
\aslb = C_d \asld + C_s \asls.
\end{equation}

The charge asymmetry \aslq~ $(q=d,s)$ of ``wrong-charge"
semileptonic $B^0_q$-meson decay induced by oscillations is defined as
\begin{equation}
\aslq = \frac{\Gamma(\bar{B}^0_q(t)\rightarrow \mu^+ X) -
              \Gamma(    {B}^0_q(t)\rightarrow \mu^- X)}
             {\Gamma(\bar{B}^0_q(t)\rightarrow \mu^+ X) +
              \Gamma(    {B}^0_q(t)\rightarrow \mu^- X)}.
\end{equation}
This quantity is independent of the proper decay time $t$ \cite{proper}.

The semileptonic charge asymmetry $\aslq$ $(q=d,s)$ depends on the complex non-diagonal parameters
of the mass mixing matrix
$ \bm{M}_q + i \bm{\Gamma}_q$
of the neutral $(B^{0,L}_q, B^{0,H}_q)$ meson system \cite{Branco} as
\begin{equation}
\label{aslq}
\aslq = \frac{\Delta \Gamma_q}{\Delta m_q} \tan(\phi_q^{12}),
\end{equation}
where
\begin{eqnarray}
\Delta m_q & \equiv & m_q^H - m_q^L = 2 |m_q^{12}|, \\
\Delta \Gamma_q & \equiv & \Gamma_q^L - \Gamma_q^H = 2 |\Gamma_q^{12}| \cos(\phi_q^{12}), \\
\phi_q^{12} & \equiv & \arg \left( -\frac{m_q^{12}}{\Gamma_q^{12}} \right).
\end{eqnarray}
Here $m^{L,H}_q$ and $\Gamma^{L,H}_q$ are the mass and width of the light ($L$) and
heavy ($H$) member of the $\Bq$ system, respectively. $\phi_q^{12}$ is the CP-violating
phase of the $(B^{0,L}_q, B^{0,H}_q)$ mass matrix. With this sign convention,
both $\Delta m_q$ and $\Delta \Gamma_q$ are positive in the SM.

The asymmetries $\asld$ and $\asls$ within the SM
are predicted \cite{Nierste} to be significantly smaller than the background asymmetries
and current experimental precision:
\begin{equation}
\label{aslsm}
\asld = (-4.1 \pm 0.6) \times 10^{-4}, ~~~~~ \asls = (1.9 \pm 0.3) \times 10^{-5}.
\end{equation}
Measurements of $\asld$ and $\asls$
\cite{hfag,d0-asld,d0-asls,lhcb-asls} agree well with the SM expectation.

The coefficients $C_d$ and $C_s$ depend on the mean mixing probabilities $\chi_d$
and $\chi_s$
and on the production fractions $f_d$ and $f_s$ of $\Bd$ and $\Bs$ mesons, respectively.
The mixing probability of a neutral $\Bq$ meson
is proportional to $1 - \cos(\Delta m_q t)$,
where $t$ is the proper decay time \cite{proper} of the $\Bq$ meson.
The mean proper decay time of $\Bq$ mesons is increased in the samples
with large IP. Because the value of $\Delta m_d$ is comparable
to the width $\Gamma_d$,
selecting muons with large IP results in an increase of the
mean mixing probability $\chi_d$.
The values of $\chi_d$ in different IP samples
are obtained using simulation and are given in Tables~\ref{tab5} and~\ref{tab6}.
On the contrary, the mass difference $\Delta m_s$ of the $\Bs$ meson
is very large compared to its width $\Gamma_s$,
and the different IP samples have approximately the
same value of $\chi_s \approx 0.5$.
The coefficients
$C_d$ and $C_s$ in a given sample are computed using the following expressions:
\begin{eqnarray}
C_d & = & f_d \chi_d / (f_d \chi_d + f_s \chi_s), \\
C_s & = & 1 - C_d.
\end{eqnarray}
Thus, the contribution of the asymmetries $\asld$ and $\asls$ to the
asymmetry $\aslb$ is different for different IP samples,
with $C_d$ increasing in the range 31\% -- 73\% when moving from
smaller to larger IP (see Tables~\ref{tab5} and \ref{tab6}).
We use the values of $f_d$ and $f_s$ measured at LEP and at Tevatron as averaged by
the Heavy Flavor Averaging Group (HFAG) \cite{hfag}:
\begin{eqnarray}
f_d & = & 0.401 \pm 0.007, \\
f_s & = & 0.107 \pm 0.005.
\end{eqnarray}


\begin{table*}
\caption{\label{tab5}
Quantities extracted from the simulation and used to interpret
the residual asymmetry $a_{\rm CP}$ in terms of CP violation in mixing.
}
\begin{ruledtabular}
\newcolumntype{A}{D{A}{\pm}{-1}}
\newcolumntype{B}{D{B}{-}{-1}}
\begin{tabular}{lAAAA}
Quantity &  \multicolumn{1}{c}{All IP} & \multicolumn{1}{c}{IP=1} & \multicolumn{1}{c}{IP=2} & \multicolumn{1}{c}{IP=3} \\
\hline
$\chi_d \times 10^2$          & 18.62\ A \ 0.23 & 6.00\ A \ 0.18 & 13.58\ A \ 0.41 & 35.14\ A \ 1.05 \\
$C_d \times 10^2$             & 58.3\ A \ 1.5 & 31.0\ A \ 1.1 & 50.4\ A \ 1.6 & 72.5\ A \ 2.2 \\
$c_b \times 10^2$             & 6.3\ A \ 0.7 & 3.4\ A \ 1.1 & 5.3\ A \ 0.8 & 10.9\ A \ 1.1 \\
$C_K$             & 0.93\ A \ 0.01 & 0.99\ A \ 0.01 & 0.92\ A \ 0.02 & 0.36\ A \ 0.06  \\
$C_\pi$           & 0.94\ A \ 0.01 & 0.96\ A \ 0.01 & 0.85\ A \ 0.02 & 0.69\ A \ 0.05 \\
\hline
$a_S\mbox{(SM)} \times 10^5$  &-1.5\ A \ 0.3 &-0.4\ A \ 0.1 &-1.0\ A \ 0.2 & -3.2\ A \ 0.7 \\
$a_{\rm CP}\mbox{(SM)} \times 10^5$     &
      -0.7\ A \  0.2 &  -0.2\ A \  0.1&  -0.8\ A \  0.1 &  -2.7\ A \  0.6 \\
\end{tabular}
\end{ruledtabular}
\end{table*}

\begin{table*}
\caption{\label{tab6}
Quantities extracted from the simulation
and used to interpret
the residual asymmetry $A_{\rm CP}$ in terms of CP violation in mixing and
CP violation in interference of decays with and without mixing.
}
\begin{ruledtabular}
\newcolumntype{A}{D{A}{\pm}{-1}}
\newcolumntype{B}{D{B}{-}{-1}}
\begin{tabular}{lAAAAAAA}
Quantity &  \multicolumn{1}{c}{All IP} & \multicolumn{1}{c}{IP$_1$,IP$_2$=11} & \multicolumn{1}{c}{IP$_1$,IP$_2$=12} &
\multicolumn{1}{c}{IP$_1$,IP$_2$=13} & \multicolumn{1}{c}{IP$_1$,IP$_2$=22} & \multicolumn{1}{c}{IP$_1$,IP$_2$=23} &
\multicolumn{1}{c}{IP$_1$,IP$_2$=33} \\
\hline
$\chi_d \times 10^2$ & 18.62\ A \ 0.23 & 6.00\ A \ 0.18 & 9.79 \ A \ 0.31 & 20.57\ A \ 0.62
                                    & 13.58\ A \ 0.41 & 24.36\ A \ 0.77 & 35.14\ A \ 1.05 \\
$C_d \times 10^2$ & 58.3\ A \ 1.5 & 31.0\ A \ 1.1 & 42.3\ A \ 1.3 & 60.7\ A \ 1.9
                                      & 50.4\ A \ 1.6 & 64.6\ A \ 2.0 & 72.5\ A \ 2.2 \\
$C_b \times 10^2$ & 52.4\ A \ 4.0  & 45.2 \ A \ 3.2  & 46.7\ A \ 3.2  & 54.2\ A \ 4.2
                                   & 45.6\ A \ 3.2  & 53.6\ A \ 4.1  & 57.6\ A \ 4.2  \\
$I \times 10^2$ & 48.3 \ A \ 0.4  & 27.9 \ A \ 0.6  & 38.1 \ A \ 0.9  & 49.2 \ A \ 1.0
                                    & 48.3 \ A \ 1.1  & 59.4 \ A \ 1.3  & 70.5 \ A \ 1.2  \\
$R_P     \times 10^2$ &
    19.3\ A \ 0.6   & 24.1 \ A \ 0.7  & 24.1\ A \ 0.7   & 18.7\ A \ 0.6
    & 20.9\ A \ 0.6  & 17.2\ A \ 0.6  & 14.9\ A \ 0.5  \\
$R_{LL} \times 10^2$ & 26.5 \ A \ 2.1  & 45.0 \ A \ 4.2  & 40.9 \ A \ 4.1  & 21.6 \ A \ 3.3
                                    &  5.9 \ A \ 5.9  &  8.0 \ A \ 8.0  & 2.8  \ A \ 2.8  \\
\hline
$\amix \mbox{(SM)}      \times 10^4$ &
    -1.2 \ A \ 0.2  & -0.5 \ A \ 0.1  & -0.8 \ A \ 0.2  & -1.3 \ A \ 0.3
    &-0.9 \ A \ 0.2  &-1.4 \ A \ 0.3  &-1.7 \ A \ 0.4  \\
$\aint \mbox{(SM)}      \times 10^4$ &
    -5.0 \ A \ 1.2  & -3.6 \ A \ 0.8  & -4.9 \ A \ 1.2  & -4.9 \ A \ 1.2
    &-5.4 \ A \ 1.3  &-5.4 \ A \ 1.3  &-5.6 \ A \ 1.4  \\
$A_{\rm CP}\mbox{(SM)} \times 10^4$ &
    -4.3\ A \ 1.0 & -1.9\ A \ 0.3 & -3.6\ A \ 0.8 & -4.2\ A \ 1.0
    & -4.2\ A \ 1.0  & -5.3\ A \ 1.2  & -6.4\ A \ 1.3  \\
\end{tabular}
\end{ruledtabular}
\end{table*}

The two largest SM contributions to the like-sign dimuon charge asymmetry
are CP violation in $\mixBd$ and $\mixBs$ mixing, \amix \cite{Branco},
and CP violation in interference
of $\Bd$ and $\Bs$ decay amplitudes
with and without mixing, $\aint$ \cite{cpv-source}. Thus, the asymmetry $A_S$ defined in
Eq.~(\ref{acp}) is expressed as
\begin{eqnarray}
\label{As_SM}
A_S & = & \amix + \aint, \\
\label{As_SM_2}
\amix & = & C_b \aslb.
\end{eqnarray}
The first contribution, $\amix$, due to CP violation in mixing, is proportional to $\aslb$,
with the coefficient $C_b$ typically 45\% - 58\% (see Table~\ref{tab6}).
The second contribution, $\aint$, is generated by the CP violation in the
decay $\Bd (\barBd) \to c \bar c d \bar d$. This final state is accessible for
both $\Bd$ and $\barBd$, and the interference of decay amplitudes to these final states
with and without $\mixBd$ mixing results in CP violation.
This contribution was not included before in the SM estimate of the dimuon
charge asymmetry.
It can be shown \cite{cpv-source}
that this CP violation in interference produces a like-sign dimuon charge asymmetry, while it does not
contribute to the inclusive single muon charge asymmetry. An example of the final state
produced in $\Bd (\barBd) \to c \bar c d \bar d$ decay is $\Bd (\barBd) \to D^{(*)+}D^{(*)-}$.
Similar contribution of CP violation in $\Bs \to c \bar c s \bar s$ decay is found to be
negligible \cite{cpv-source} and is not considered in our analysis.

The value of $\aint$ is obtained using the following expression \cite{cpv-source}:
\begin{equation}
\label{aint}
\aint = - \frac{0.5 f_d R_d R_P}{\mbox{\cal{B}}(b \to c \bar c X) }
\frac{\Delta \Gamma_d}{\Gamma_d} \sin(2 \beta) I,
\end{equation}
where
\begin{eqnarray}
R_d & \equiv & \frac{\mbox{\cal{B}}(c \bar c d \bar d \to \mu X)}{\mbox{\cal{B}} (c \bar c X \to \mu X)}, \\
R_P & \equiv & \frac{P(b \to c \bar c X \to \mu X) (P(b) - P(\bar b))}{P(b)P(\bar b)}, \\
\label{integral}
I & \equiv & \Gamma_d \int \exp(-\Gamma_d t) \sin (\Delta m_d t) dt.
\end{eqnarray}
Here the ratio $R_d$ reflects the fact that the final state of the decay $\Bd \to c \bar c d \bar d$
contains more $D^\pm$ mesons than the generic $b \to c \bar c X$ final state, and that
the branching fraction of $D^\pm  \to \mu^\pm X$ decays is much larger than that of
all other charmed hadrons. Using the known branching fractions of $b$- and $c$-hadron
decays taken from Ref. \cite{PDG}, we estimate
\begin{equation}
R_d = 1.5 \pm 0.2.
\end{equation}

In the expression for $R_P$ the quantity $P(b \to c \bar c X \to \mu X)$ is the probability to reconstruct
a muon coming from the decay $b \to c \bar c X \to \mu X$. It depends on
the muon reconstruction efficiency, including all fiducial requirements,
and on the branching fractions of the decays
$b \to c \bar c X$ and $c \to \mu X$. The quantity $P(b)$ is the probability to reconstruct
a ``right-sign" muon from the $b \to \mu^-$ decay. It includes both the muon reconstruction
efficiency and the branching fractions of all possible decay modes of $b$ quarks producing a ``right-sign" muon.
Similarly, the quantity $P(\bar b)$ is the probability to reconstruct
a ``wrong-sign" muon from the $\bar b \to \mu^-$ decay.
All these probabilities depend on the IP requirement.
They are determined using simulation. The values of $R_P$ for
different IP samples are given in Table~\ref{tab6}.

The branching fraction $\mbox{\cal{B}}(b \to c \bar c X)$ of $b$-hadron decays producing a $c \bar c$ pair
is obtained using the experimental value of \cal{B}($b$-hadron mixture $\to c / \bar c X)$ measured
at LEP~\cite{PDG}:
\begin{equation}
\label{bcc}
\mbox{\cal{B}}(B \mbox{ mixture} \to c / \bar c X) = (116.2 \pm 3.2)\%,
\end{equation}
where ``$c/\bar{c}$" counts multiple charm quarks per decay.
Assuming a negligible fraction of charmless $b$-hadron decays, we derive
from Eq.~(\ref{bcc}) the following value for the branching fraction of decay of $b$ quark
into two charm quarks:
\begin{equation}
\mbox{\cal{B}}(b \to c \bar c X)  =  (16.2 \pm 3.2)\%.
\end{equation}
The angle $\beta$ is one of the angles of the unitarity triangle defined as
\begin{equation}
\beta = \arg \left( -\frac{V_{cd} V_{cb}^*}{V_{td} V_{tb}^*} \right),
\end{equation}
where the quantities $V_{q q'}$ are the parameters of the CKM matrix.
The world average value
of $\sin (2 \beta)$ \cite{PDG} is
\begin{equation}
\sin (2 \beta) = 0.679 \pm 0.020.
\end{equation}
The SM prediction \cite{Nierste}
\begin{equation}
\label{dgsm}
\frac{\Delta \Gamma_d} {\Gamma_d}\mbox{(SM)} = (0.42 \pm 0.08) \times 10^{-2},
\end{equation}
is used in our estimate of the SM expectation of the $\aint$ asymmetry.
The precision of the measured world average of $\Delta \Gamma_d / \Gamma_d$ \cite{PDG}
is about 20 times larger:
\begin{equation}
\label{dgamma_ex}
\frac{\Delta \Gamma_d} {\Gamma_d} = (1.5 \pm 1.8) \times 10^{-2}.
\end{equation}

Finally, the integration in Eq.~(\ref{integral}) is taken over all $\Bd$ decays in a given IP sample.
For the total dimuon sample it can be obtained analytically with the result
\begin{eqnarray}
I & = & \frac{x_d}{1+x_d^2}, \\
x_d & \equiv & \frac{\Delta m_d} {\Gamma_d}.
\end{eqnarray}
For the IP samples the value of $I$ is obtained in simulation with simulation and the results
are given in Table~\ref{tab6}.

The CP violation in interference of $\Bs$ decay amplitudes with and without mixing is expected
to be significantly smaller than the contribution from $\Bd$ mesons \cite{cpv-source}
due to the relatively small values of $x_s/(1+x_s^2)$ and $\sin(2 \beta_s)$.
The contribution due to $\Bs$ mesons is neglected in this analysis.

Hence, to determine the expected SM values of
asymmetries $a_S$ and $A_S$ we need the following quantities, all
extracted from simulation, and all listed in Tables~\ref{tab5} and~\ref{tab6}:
\begin{itemize}
\item
The fractions $c_b$ and $C_b$, in different IP samples.
\item
The coefficient $C_d$, itself derived from the average mixing
probability $\chi_d$, in different IP samples.
\item
The quantities $R_P$ and $I$, required to evaluate the contribution
$\aint$, in different IP samples.
\end{itemize}

The coefficients $C_b$ and $R_P$ are determined using the simulation
of $b \bar b$ and $c \bar c$ events
producing two muons. This simulation allows an estimate of these coefficients
taking into account the possible correlation in the detection of two muons.
This simulation was not available for our previous measurement \cite{D03}.
For comparison, the value of $C_b$ used in Ref. \cite{D03}
for the full sample of dimuon events was $C_b = 0.474 \pm 0.032$.
The uncertainty on all quantities listed in Tables~\ref{tab5} and~\ref{tab6}
include the uncertainty on the input quantities taken from Ref.~\cite{PDG}
and the limited simulation statistics. In addition, the uncertainty on the
coefficients $c_b$, $C_b$, and $R_P$ includes
the uncertainty on the momentum of the generated $b$ hadrons.

In addition, in order to convert the asymmetries $a_S$ and $A_S$ into the asymmetries
$a_{\rm CP}$ and $A_{\rm CP}$ using Eqs.~(\ref{atot1}) and~(\ref{acp}),
the fractions $f_S$, $F_{SS}$ and $F_{SL}$ are required.
These quantities are obtained using the values $f_K$, $f_\pi$, $f_p$, $F_K$,
$F_\pi$, and $F_P$. All of them are measured in data and given in Tables~\ref{tab1} and \ref{tab2}.
We also need the following
quantities extracted from simulation and listed in Tables~\ref{tab5} and \ref{tab6}:
\begin{itemize}
\item
The quantities $C_K$, and $C_{\pi}$ in different IP samples. They are defined in Eq.~(\ref{ck}).
\item The quantity $R_{LL}$ in different IP samples. It is defined in Eq.~(\ref{rll}).
\end{itemize}

The coefficients $C_K$ and $C_\pi$ are defined as
the fraction of $K \to \mu$
and $\pi \to \mu$ tracks with the reconstructed track parameters
corresponding to the track parameters of the kaon or pion, respectively.
Since the kaons and pions are mainly produced in the primary interactions,
such muons have small IP.
If, on the contrary, the reconstructed muon track parameters
correspond to the track parameters of the
muon from $K^\pm \to \mu^\pm \nu$ and $\pi^\pm \to \mu^\pm \nu$ decay,
the IP of such muons is large
because the kaons and pions decay at a distance from the primary interaction and the muon track
has a kink with respect to the hadron's trajectory. Therefore, the fraction
of such muons increases with increasing IP,
and the coefficients $C_K$ and $C_\pi$ become small for the samples with large IP.

Tables~\ref{tab5} and \ref{tab6} also include the SM expectation for $a_S$,
$\amix$, $\aint$, $a_{\rm CP}$, and $A_{\rm CP}$. The expected value of $a_S$ is smaller
than that of $\amix$. The contribution $\aint$ due to CP violation in interference
of decay amplitudes with and without mixing exceeds that from $\amix$.

\section{Interpretation of results}
\label{sec_interpretation}

We measure the asymmetry $a^i_{\rm CP} = a^i - a^i_{\rm bkg}$
in 27 bins with different $\pteta$ and IP,
and the asymmetry $A^i_{\rm CP} = A^i - A^i_{\rm bkg}$
in 54 bins with different $\pteta$, IP$_1$, and IP$_2$,
and compare the result with the SM prediction.

The largest SM contributions to the inclusive single muon and like-sign dimuon charge
asymmetries come from CP violation in $\mixBd$ and $\mixBs$ mixing,
and CP violation in interference
of $\Bd$ and $\Bs$ decay amplitudes
with and without mixing. The expected numerical values of these contributions
to the asymmetries $a_S$ and $A_S$ are given in Tables~\ref{tab5} and \ref{tab6}.
The asymmetries $a_S$ and $A_S$ are related to the residual asymmetries $a_{\rm CP}$
and $A_{\rm CP}$ as
\begin{eqnarray}
\label{aCP_as}
a_{\rm CP}& = & f_S a_S, \\
\label{ACP_AS}
A_{\rm CP}& = & F_{SS} A_S + F_{SL} a_S,
\end{eqnarray}
see Eqs.~(\ref{atot1}) and~(\ref{acp}).
The fractions $f_S$, $F_{SS}$, $F_{SL}$ are given in Tables~\ref{tab1} and \ref{tab2}.

Using all these values we determine the consistency of our measurements with the
SM expectation. The SM expectation for $a_{\rm CP}$ and $A_{\rm CP}$ are given in Tables
\ref{tab5} and \ref{tab6}, respectively.
The expectation
for $a_{\rm CP}$(SM) is significantly smaller in magnitude
than the experimental uncertainty for all IP samples.
The measured $A_{\rm CP}$ are systematically larger in amplitude than
their corresponding $A_{\rm CP}$(SM) expectations.

Using the measurements
with full samples of inclusive muon and like-sign dimuon events
given in rows ``All IP'' in Tables~\ref{tab3} and~\ref{tab4} and taking into
account the correlation between them given in Eq.~(\ref{coraA}), we obtain
the $\chi^2$ of the difference between these measurements
and their SM expectations
\begin{eqnarray}
\label{res_chi2}
\chi^2\mbox{/d.o.f.} & = & 9.9/2, \\
\label{res_pv}
p\mbox{(SM)} & = & 7.1 \times 10^{-3}.
\end{eqnarray}
This result, that uses no IP information,
corresponds to $2.7$ standard deviations from the SM expectation.

The values of $\chi^2$ in Eq.~(\ref{res_chi2}), and throughout this section, include
both statistical and systematic uncertainties. These $\chi^2$ values are minimized
by a fit that takes into account all correlations between the uncertainties,
see Appendix A.
The $p$ value quoted in Eq.~(\ref{res_pv}), and throughout this section, is the
probability that the $\chi^2$ for a given number of degrees of freedom (d.o.f.)
exceeds the observed $\chi^2$.
These $p$ values are translated to the equivalent
number of standard deviations for a single variable.

%

Using the same measurements $a_{\rm CP}$ and $A_{\rm CP}$ obtained
with full inclusive muon and like-sign dimuon samples we obtain the value of
the charge asymmetry $\aslb$ defined in Eq.~(\ref{aslb}).
Assuming that the contribution of CP violation in interference corresponds
to the SM expectation given in Table~\ref{tab6}, we get
\begin{equation}
\aslb = (-0.496 \pm 0.153 \pm 0.072) \times 10^{-2}.
\end{equation}
This value differs from the SM expectation
$\aslb = (-0.023 \pm 0.004) \times 10^{-2}$ obtained from Eq.~(\ref{aslb})
by $2.8$ standard deviations.

The change
in the central value and the uncertainty compared to our previous result \cite{D03}
is due to several factors.
The contribution of the CP violation in interference was not considered in Ref.~\cite{D03}.
The simulation of $b \bar b$ and $c \bar c$ events producing two muons, which was not
available for our previous measurement, allows a better estimate of the coefficient $C_b$.
Finally, a more accurate procedure for measuring background asymmetries
using $\pteta$ bins results in the change of $A_{\rm CP}$ with respect to the
previous result \cite{D03}, which is also reflected in the change of the $\aslb$ asymmetry,
see Table~\ref{tab8}.

The comparison of our result with the SM prediction benefits from the use of
each IP region separately, due to the large variations in the background
fraction in each IP sample.
The three measurements of $a_{\rm CP}$ in different IP samples
and six measurements of $A_{\rm CP}$ in different (IP$_1$,IP$_2$) samples
can be compared with the SM expectation. Both statistical and systematic uncertainties
are used in this comparison. The correlation between different measurements
 given in Table~\ref{tab-cor} are taken into account.
The $\chi^2$(IP) of the difference between the measured
residual asymmetries and the SM expectation is
\begin{eqnarray}
\chi^2\mbox{(IP)/d.o.f.} & = & 31.0/9, \\
p\mbox{(SM)} & = & 3 \times 10^{-4}.
\end{eqnarray}
This result corresponds to $3.6$ standard deviations from the SM expectation.
The $p$ value of the hypothesis that the $a_{\rm CP}$ and $A_{\rm CP}$
asymmetries in all IP samples are equal to zero is
\begin{equation}
p\mbox{(CPV}=0) = 3 \times 10^{-5},
\end{equation}
which corresponds to $4.1$ standard deviations.


If we assume that the observed asymmetries $a_{\rm CP}$ and
$A_{\rm CP}$ are due to the CP violation in mixing,
the results in different IP samples can be used to measure
the semileptonic charge asymmetries $\asld$ and $\asls$. Their contribution
to the asymmetries $a_{\rm CP}$ and $A_{\rm CP}$, determined by the coefficients $C_d$
and $C_s$, varies considerably in different IP samples. Performing this measurement
we assume that the contribution of the CP violation in interference of
decay amplitudes with and without mixing, given by Eq.~(\ref{aint}), corresponds to the SM
expectation presented in Table~\ref{tab6}. In particular, the value of $\dgg$ is set
to its SM expectation given in Eq.~(\ref{dgsm}). We obtain
\begin{eqnarray}
\asld & = & (-0.62 \pm 0.42) \times 10^{-2}, \\
\asls & = & (-0.86 \pm 0.74) \times 10^{-2}. \\
\chi^2\mbox{/d.o.f.} & = & 10.1/7.
\end{eqnarray}
The correlation between the fitted parameters $\asld$ and $\asls$ is
\begin{equation}
\rho_{d,s} =  -0.79.
\end{equation}
The difference between these $\asld$ and $\asls$ values and the combined
SM expectation~(\ref{aslsm}) corresponds to $3.4$ standard deviations.

\begin{figure}
\begin{center}
\includegraphics[width=0.48\textwidth]{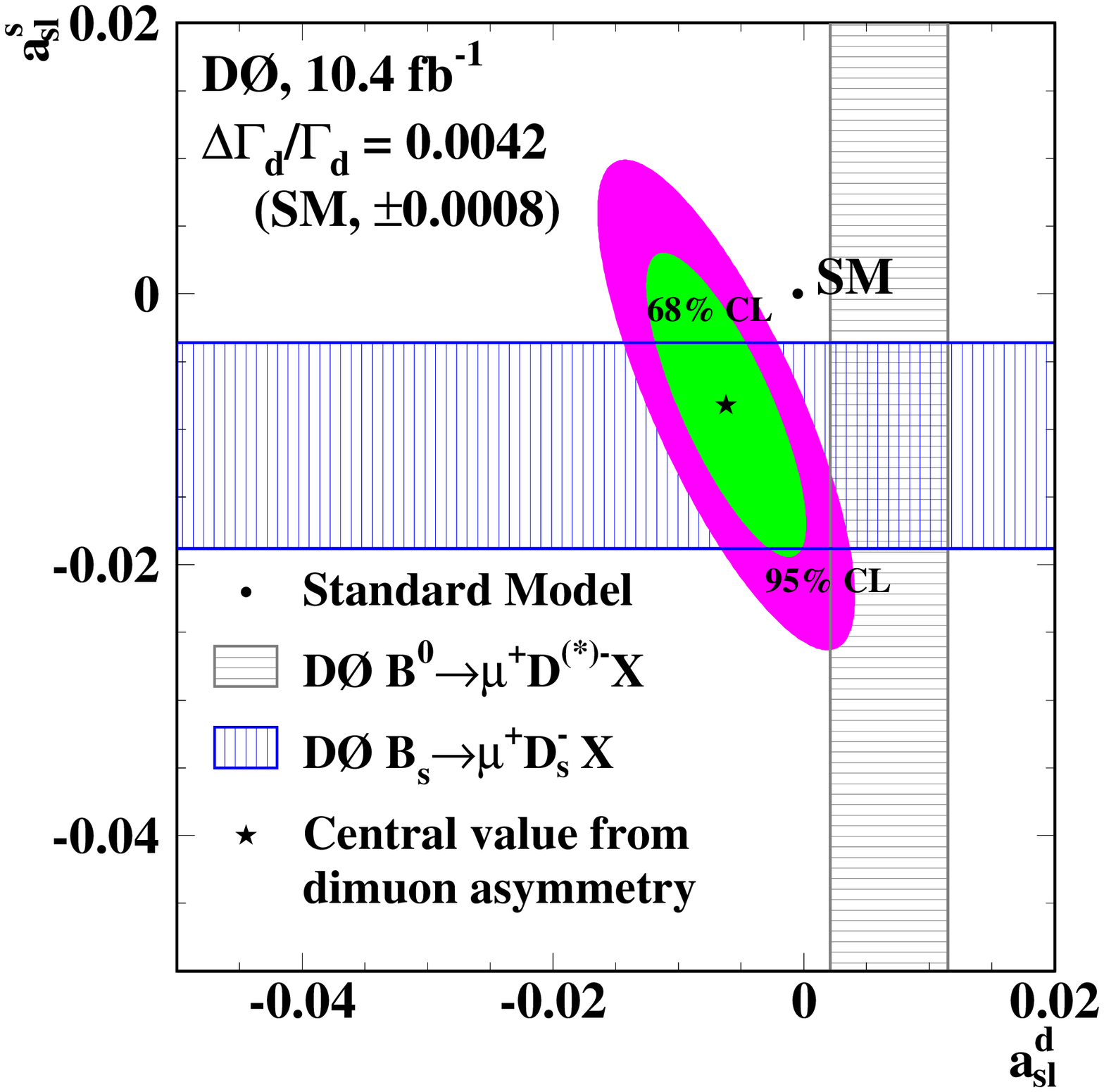}
\caption{(color online).
The 68\% and 95\% confidence level contours in the $\asld - \asls$ plane
obtained from the fit of the inclusive single muon and like-sign dimuon asymmetries
with fixed value of $\Delta \Gamma_d / \Gamma_d = 0.0042$
corresponding to the expected SM value~(\ref{dgsm})
which has an uncertainty $\pm 0.0008$.
The independent measurements of  $\asld$ \cite{d0-asld}
and $\asls$ \cite{d0-asls}
by the D0 collaboration
are also shown.
The error bands represent $\pm 1$ standard deviation uncertainties of these measurements.
}
\label{ads1-0.0042}
\end{center}
\end{figure}

\begin{figure}[t]
\begin{center}
\includegraphics[width=0.48\textwidth]{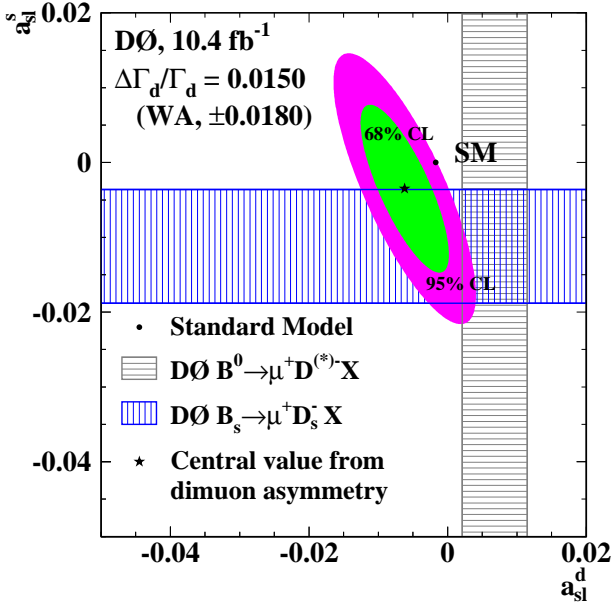}
\caption{(color online).
The 68\% and 95\% confidence level contours in the $\asld - \asls$ plane
obtained from the fit of the inclusive single muon and like-sign dimuon asymmetries
with fixed value of $\Delta \Gamma_d / \Gamma_d = 0.0150$
corresponding to the experimental world average value~(\ref{dgamma_ex})
which has an uncertainty $\pm 0.0180$.
The independent measurements of  $\asld$ \cite{d0-asld}
and $\asls$ \cite{d0-asls}
by the D0 collaboration
are also shown.
The error bands represent $\pm 1$ standard deviation uncertainties of these measurements.
}
\label{ads1-0.0150}
\end{center}
\end{figure}

The like-sign dimuon charge asymmetry depends on the value of $\Delta \Gamma_d/ \Gamma_d$, see
Eqs.~(\ref{aslb},\ref{aslq},\ref{As_SM}--\ref{aint}).
By fixing the values of
$\phi_d^{12}$ and $\asls$
to their SM expectations
$\phi_d^{12} = -0.075 \pm 0.024$ and $\asls = (+1.9 \pm 0.3) \times 10^{-5}$ \cite{Nierste},
we can extract the value of $\Delta \Gamma_d/ \Gamma_d$ from our measurements
of $a_{\rm CP}$ and $A_{\rm CP}$ in different IP samples.
We obtain
\begin{eqnarray}
\Delta \Gamma_d / \Gamma_d & = & (+2.63 \pm 0.66) \times 10^{-2}, \\
\chi^2\mbox{/d.o.f.} & = & 13.8/8.
\end{eqnarray}
This result differs from the SM expectation~(\ref{dgsm}) by $3.3$ standard deviations.
The values of $\phi_d^{12}$ and $\dgg$ determine the value of $\asld$,
see Eq.~(\ref{aslq}).

Finally, we can interpret our results as the measurement of
$\asld$, $\asls$ and $\Delta \Gamma_d / \Gamma_d$, allowing all these quantities to vary
in the fit. We obtain
\begin{eqnarray}
\label{eq_sum_beg}
\asld & = & (-0.62 \pm 0.43) \times 10^{-2}, \\
\asls & = & (-0.82 \pm 0.99) \times 10^{-2}, \\
\label{eq_sum_dg}
\frac{\Delta \Gamma_d} {\Gamma_d} & = & (+0.50 \pm 1.38) \times 10^{-2}, \\
\chi^2\mbox{/d.o.f.} & = & 10.1/6.
\end{eqnarray}
The correlations between the fitted parameters are
\begin{equation}
\label{eq_sum_end}
\rho_{d,s}  =  -0.61,~~ \rho_{d, \Delta \Gamma}  =  -0.03,~~ \rho_{s, \Delta \Gamma}  =  +0.66.
\end{equation}
This result differs from the combined SM expectation for $\asld$, $\asls$, and
$\Delta \Gamma_d / \Gamma_d$ by $3.0$ standard deviations.

Figure~\ref{ads1-0.0042} shows the 68\% and 95\% confidence level contours in the $\asld - \asls$ plane
obtained from the re-fit of the inclusive single muon and like-sign dimuon asymmetries
with a fixed value of $\Delta \Gamma_d / \Gamma_d = 0.0042$
corresponding to the expected SM value~(\ref{dgsm}).
The same plot also shows two bands
of the independent measurements of $\asld$ and $\asls$ by the D0 collaboration \cite{d0-asld,d0-asls}.
Figure~\ref{ads1-0.0150} presents the result
of the fit of the inclusive single muon and like-sign dimuon asymmetries
with fixed value of $\Delta \Gamma_d / \Gamma_d = 0.0150$
corresponding to the experimental world average value~(\ref{dgamma_ex}).
These two plots show that if the currently imprecise experimental value of $\dgg$
is used instead of the SM prediction,
the values of $\asld$ and $\asls$  become consistent
with the SM expectation within two standard deviations.
This observation demonstrates the importance for independent
measurements of $\dgg$ which have not been a high priority of experimentalists before \cite{Gershon}.


The combination of the measurements of the semileptonic
charge asymmetries $\asld$ \cite{d0-asld} and $\asls$ \cite{d0-asls}
by the D0 collaboration
with the present analysis of the inclusive single muon and like-sign dimuon charge asymmetries gives
\begin{eqnarray}
\label{eq_d0_beg}
\asld & = & (-0.09 \pm 0.29) \times 10^{-2}, \\
\asls & = & (-1.33 \pm 0.58) \times 10^{-2}, \\
\label{eq_d0_dg}
\frac{\Delta \Gamma_d} {\Gamma_d} & = & (+0.79 \pm 1.15) \times 10^{-2}, \\
\chi^2\mbox{/d.o.f.} & = & 4.4/2.
\end{eqnarray}
The correlations between the fitted parameters are
\begin{equation}
\label{eq_d0_end}
\rho_{d,s}  =  -0.34,~~ \rho_{d, \Delta \Gamma}  =  +0.24,~~ \rho_{s, \Delta \Gamma}  =  +0.55.
\end{equation}
In this combination we treat all D0 measurements as statistically independent.
This result differs from the combined SM expectation for $\asld$, $\asls$, and
$\Delta \Gamma_d / \Gamma_d$ by $3.1$ standard deviations.
Currently, these are the most precise measurements of $\asld$, $\asls$ and
$\Delta \Gamma_d / \Gamma_d$ by a single experiment.

\begin{figure}
\begin{center}
\includegraphics[width=0.48\textwidth]{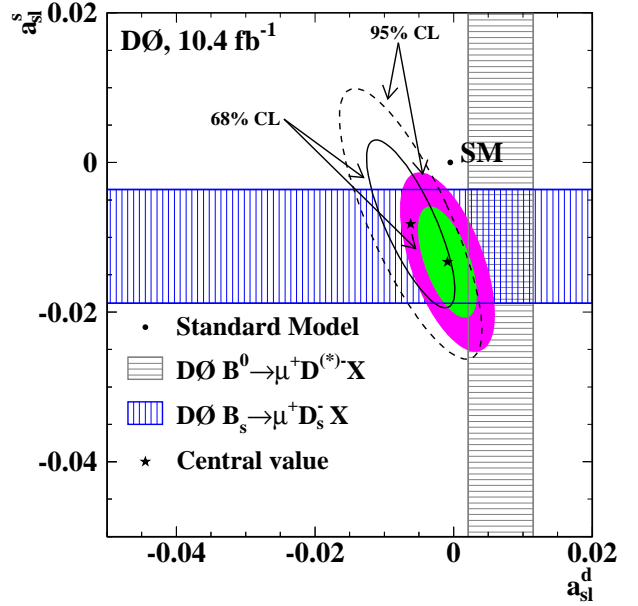}
\caption{(color online).
The 68\% (full line) and 95\% (dashed line) confidence level contours in the $\asld - \asls$ plane
representing the profile of the results given by Eq.~(\ref{eq_sum_beg})--(\ref{eq_sum_end})
at the best fit value of $\Delta \Gamma_d / \Gamma_d = 0.0050$ corresponding to Eq.~(\ref{eq_sum_dg}).
The contours with filled area show
the 68\% and 95\% confidence level contours in the $\asld - \asls$ plane
representing the profile of the results
obtained by the combination of all D0 measurements and
given by Eq.~(\ref{eq_d0_beg})--(\ref{eq_d0_end})
at the best fit value of $\Delta \Gamma_d / \Gamma_d = 0.0079$
corresponding to Eq.~(\ref{eq_d0_dg}).
The independent measurements of  $\asld$ \cite{d0-asld}
and $\asls$ \cite{d0-asls}
by the D0 collaboration
are also shown.
The error bands represent $\pm 1$ standard deviation uncertainties of these measurements.
}
\label{ads1-comb}
\end{center}
\end{figure}

Figure~\ref{ads1-comb} shows the 68\% and 95\% confidence level contours in the $\asld - \asls$ plane
representing the profile of the results given by Eq.~(\ref{eq_sum_beg})--(\ref{eq_sum_end})
at the best fit value of $\Delta \Gamma_d / \Gamma_d = 0.0050$ corresponding to Eq.~(\ref{eq_sum_dg}).
The same figure shows the 68\% and 95\% confidence level contours in the $\asld - \asls$ plane
representing the profile of the results
obtained by the combination of all D0 measurements and
given by Eq.~(\ref{eq_d0_beg})--(\ref{eq_d0_end})
at the best fit value of $\Delta \Gamma_d / \Gamma_d = 0.0079$
corresponding to Eq.~(\ref{eq_d0_dg}).


\section{Conclusions}
\label{sec_conclusions}

We have presented the final measurements of the inclusive single muon and like-sign dimuon
charge asymmetries using the full data set of 10.4 fb$^{-1}$ collected by the D0 experiment in Run II
of the Tevatron collider at Fermilab.
The measurements of the inclusive muon sample are performed in
27 non-overlapping bins of $\pteta$ and IP.
The measurements of the like-sign dimuon sample are performed in
54 non-overlapping bins of $\pteta$, IP$_1$ and IP$_2$.
The background contribution
is measured using two independent methods that give consistent results.
The achieved agreement between the observed asymmetry $a$ and the expected
background asymmetry $a_{\rm bkg}$ in the inclusive muon sample
is at the level of $3 \times 10^{-4}$, see Table~\ref{tab1a}.

The model-independent charge asymmetries $a_{\rm CP}$ and $A_{\rm CP}$, obtained by
subtracting the expected background
contribution from the raw charge asymmetries,
are given in Tables~\ref{tab3}, \ref{tab4} and \ref{tab-cor}, respectively,
and are shown in Fig.~\ref{summary1}.
These measurements provide evidence at the 4.1 standard deviations level for the deviation of the dimuon
charge asymmetry from zero.
The $\chi^2$ of the difference between these measurements and the SM expectation
of CP violation in $\mixBd$ and $\mixBs$ mixing, and in interference of $\Bd$ and $\Bs$ decay
amplitudes with and without mixing,
is 31.0 for 9 d.o.f., which corresponds to $3.6$ standard deviations.

If we interpret all observed asymmetries in terms of anomalous
CP violation in neutral $B$ meson
mixing and interference, we obtain the semileptonic charge asymmetries $\asld$
and $\asls$ of $\Bd$ and $\Bs$ mesons respectively, and the width difference of the $\Bd$ system, $\Delta \Gamma_d$:
\begin{eqnarray}
\asld & = & (-0.62 \pm 0.43) \times 10^{-2}, \\
\asls & = & (-0.82 \pm 0.99) \times 10^{-2}, \\
\frac{\Delta \Gamma_d} {\Gamma_d} & = & (+0.50 \pm 1.38) \times 10^{-2}, \\
\chi^2\mbox{/d.o.f.} & = & 10.1/6.
\end{eqnarray}
The correlations between the fitted parameters are
\begin{equation}
\rho_{d,s}  =  -0.61,~~ \rho_{d, \Delta \Gamma}  =  -0.03,~~ \rho_{s, \Delta \Gamma}  =  +0.66.
\end{equation}
This result differs from the SM expectation by $3.0$ standard deviations.


Because our measurements are inclusive,
other as yet unknown sources of CP violation could contribute
to the asymmetries $a_{\rm CP}$
and $A_{\rm CP}$ as well. Therefore, the model-independent asymmetries $a_{\rm CP}$ and
$A_{\rm CP}$ measured in different IP samples
constitute the main result of our analysis. They are presented in a form
which can be used as an input for alternative interpretations.




%
We thank the staffs at Fermilab and collaborating institutions,
and acknowledge support from the
DOE and NSF (USA);
CEA and CNRS/IN2P3 (France);
MON, NRC KI and RFBR (Russia);
CNPq, FAPERJ, FAPESP and FUNDUNESP (Brazil);
DAE and DST (India);
Colciencias (Colombia);
CONACyT (Mexico);
NRF (Korea);
FOM (The Netherlands);
STFC and the Royal Society (United Kingdom);
MSMT and GACR (Czech Republic);
BMBF and DFG (Germany);
SFI (Ireland);
The Swedish Research Council (Sweden);
and
CAS and CNSF (China).
%

\appendix
\section{Fitting procedure}
\label{sec_fit}
The asymmetries $a_{\rm CP}$ and $A_{\rm CP}$ measured in different IP samples are given in
Tables~\ref{tab3} and \ref{tab4}.
Following Eqs.~(\ref{asinc}),~(\ref{aslb}),~(\ref{As_SM}),~(\ref{As_SM_2}),
(\ref{aint}),~(\ref{aCP_as}) and~(\ref{ACP_AS}), they can be expressed in a given
IP sample as
\begin{eqnarray}
\label{app1}
a_{\rm CP}& = & f_S c_b \aslb, \\
A_{\rm CP}& = & (F_{SS} C_b  + F_{SL} c_b) \aslb + F_{SS} \aint, \\
\aslb & = & C_d \asld + C_s \asls, \\
\label{app-last}
\aint & = & \aint\mbox{(SM)}\frac{\delta_\Gamma}{\delta_\Gamma\mbox{(SM)}}, \\
\delta_\Gamma & \equiv & \frac{\Delta \Gamma_d}{\Gamma_d}.
\end{eqnarray}
The values of $c_b$, $C_b$, $C_d$ and $\aint\mbox{(SM)}$ are given in Tables~\ref{tab5} and \ref{tab6}.
The values of $f_s$, $F_{SS}$, and $F_{SL}$ are given in Tables~\ref{tab1} and \ref{tab2}.
The value of $\delta_\Gamma\mbox{(SM)}$, which does not depend on the IP requirement,
is given in Eq.~(\ref{dgsm}). The value of $C_s$ is defined as $C_s = 1 - C_d$.

Equations~(\ref{app1})-(\ref{app-last}) for a given IP sample $i$ can be rewritten as
\begin{equation}
\label{yi}
y^i = K^i_d \asld + K^i_s \asls + K^i_\delta \delta_\Gamma.
\end{equation}
Index $i$ varies from 1 to 9. The definitions of quantities $y^i$, $K^i_d$, $K^i_s$ and
$K^i_\delta$ are given in Table~\ref{tapp1}.
Definitions of the quantities $a'$, $A'$ and $C_\delta$ used in Table~\ref{tapp1} are given below:
\begin{eqnarray}
a' & \equiv & \frac{a_{\rm CP}}{f_S c_b}, \\
A' & \equiv & \frac{A_{\rm CP}}{F_{SS} C_b + F_{SL} c_b}, \\
C_\delta & \equiv & \frac{F_{SS}}{F_{SS} C_b + F_{SL} c_b} \frac{\aint\mbox{(SM)}}{\delta_\Gamma\mbox{(SM)}}.
\end{eqnarray}
All quantities in these expressions, except $\delta_\Gamma\mbox{(SM)}$,
depend on the IP requirement. The quantities $y^i$ are measured experimentally. The
coefficients $K^i_d$, $K^i_s$ and $K^i_\delta$ are determined using the input from
simulation and from data.
The components necessary for their computation
are given in Tables~\ref{tab1}, \ref{tab2}, \ref{tab5}, and \ref{tab6}.
The values of $c_b$ for different (IP$_1$,IP$_2$) samples are determined as
\begin{equation}
c_b(\mbox{IP}_1,\mbox{IP}_2) = \frac{1}{2}(c_b(\mbox{IP}_1) + c_b(\mbox{IP}_2)).
\end{equation}

\begin{table}
\caption{\label{tapp1}
Definition of $y^i$, $K^i_d$, $K^i_s$ and $K^i_\delta$.}
\begin{ruledtabular}
\newcolumntype{A}{D{A}{\pm}{-1}}
\newcolumntype{B}{D{B}{-}{-1}}
\begin{tabular}{cccc}
$i$ & $y^i$ & $K^i_d = 1 - K^i_s$ &  $K^i_\delta$ \\
\hline
1   & $a'$(IP=1)             & $C_d$(IP=1)             & 0                            \\
2   & $a'$(IP=2)             & $C_d$(IP=2)             & 0                            \\
3   & $a'$(IP=3)             & $C_d$(IP=3)             & 0                            \\
4   & $A'$(IP$_1$,IP$_2$=11) & $C_d$(IP$_1$,IP$_2$=11) & $C_\delta$(IP$_1$,IP$_2$=11) \\
5   & $A'$(IP$_1$,IP$_2$=12) & $C_d$(IP$_1$,IP$_2$=12) & $C_\delta$(IP$_1$,IP$_2$=12) \\
6   & $A'$(IP$_1$,IP$_2$=13) & $C_d$(IP$_1$,IP$_2$=13) & $C_\delta$(IP$_1$,IP$_2$=13) \\
7   & $A'$(IP$_1$,IP$_2$=22) & $C_d$(IP$_1$,IP$_2$=22) & $C_\delta$(IP$_1$,IP$_2$=22) \\
8   & $A'$(IP$_1$,IP$_2$=23) & $C_d$(IP$_1$,IP$_2$=23) & $C_\delta$(IP$_1$,IP$_2$=23) \\
9   & $A'$(IP$_1$,IP$_2$=33) & $C_d$(IP$_1$,IP$_2$=33) & $C_\delta$(IP$_1$,IP$_2$=33) \\
\end{tabular}
\end{ruledtabular}
\end{table}

\begin{table}
\caption{\label{tapp3}
Sources of uncertainty on $y^i$.
The first nine rows contain statistical uncertainties,
while the next five rows reflect contributions from systematic uncertainties.
}
\begin{ruledtabular}
\newcolumntype{A}{D{A}{\pm}{-1}}
\newcolumntype{B}{D{B}{-}{-1}}
\begin{tabular}{cccc}
Index $k$ & Source  & $\rho^k_{12}$ & $\rho^k_{14}$ \\
\hline
1 & $A$ or $a$ (stat)          &  0  & 0 \\
2 & $n(K^{*0})$ or $N(K^{*0})$(stat)       &  0   & 0 \\
3 & $n(K^{*+})$                &  1   & 1 \\
4 & $P(\pitomu)/P(\ktomu)$     &  1   & 1 \\
5 & $P(\ptomu)/P(\ktomu)$      &  1   & 1 \\
6 & $a_K$                      &  1   & 1 \\
7 & $a_\pi$                    &  1   & 1 \\
8 & $a_p$                      &  1   & 1 \\
9 & $\delta$                   &  1   & 1 \\
\hline
10 & $f_K$ (syst)              &  1   & 1 \\
11 & $F_K/f_K$ (syst)          &  0   & 0 \\
12 & $\pi$, $K$, $p$
      multiplicity             &  1   & 1 \\
13 & $c_b$ or $C_b$            &  0   & 0 \\
14 & $\varepsilon(K^{*0})$     &  0   & 1
\end{tabular}
\end{ruledtabular}
\end{table}

\begin{table*}
\caption{\label{tapp2}
Values of $y^i$ ($i=1,...,9$) and the contributions to their uncertainties $\sigma_k^i$
from different sources $k$ ($k = 1,...,14$). The definition of different measurements
is given in Table~\ref{tapp1}. The definition of all sources is given in Table~\ref{tapp3}.
}
\begin{ruledtabular}
\newcolumntype{A}{D{A}{\pm}{-1}}
\newcolumntype{B}{D{B}{-}{-1}}
\begin{tabular}{lccccccccc}
\multirow{2}{*}{Quantity}   & \multicolumn{9}{|c}{index $i$} \\
\cline{2-10}
                            & \multicolumn{1}{|c}{1}      & 2     & 3      & 4      & 5      & 6      & 7      & 8      & 9      \\
\cline{1-10}
$y^i \times 10^2$           & \multicolumn{1}{|c}{$-1.869$} & 0.473 & $-0.519$ & $-2.029$ & $-0.347$ & $-0.936$ & $-0.817$ & $-0.335$ & $-0.600$ \\
$\sigma_1^i \times 10^2$    & \multicolumn{1}{|c}{ 0.204} & 0.146 &  0.058 &  0.365 &  0.303 &  0.196 &  0.657 &  0.259 &  0.228 \\
$\sigma_2^i \times 10^2$    & \multicolumn{1}{|c}{ 0.425} & 0.152 &  0.059 &  0.385 &  0.283 &  0.170 &  0.634 &  0.300 &  0.254 \\
$\sigma_3^i \times 10^2$    & \multicolumn{1}{|c}{ 1.767} & 0.237 &  0.036 &  0.248 &  0.161 &  0.092 &  0.098 &  0.172 &  0.006 \\
$\sigma_4^i \times 10^2$    & \multicolumn{1}{|c}{ 1.569} & 0.196 &  0.034 &  0.139 &  0.035 &  0.023 &  0.029 &  0.010 &  0.003 \\
$\sigma_5^i \times 10^2$    & \multicolumn{1}{|c}{ 0.367} & 0.042 &  0.007 &  0.031 &  0.010 &  0.006 &  0.007 &  0.002 &  0.001 \\
$\sigma_6^i \times 10^2$    & \multicolumn{1}{|c}{ 1.534} & 0.198 &  0.029 &  0.152 &  0.060 &  0.035 &  0.063 &  0.018 &  0.006 \\
$\sigma_7^i \times 10^2$    & \multicolumn{1}{|c}{ 2.765} & 0.349 &  0.051 &  0.227 &  0.089 &  0.050 &  0.087 &  0.025 &  0.009 \\
$\sigma_8^i \times 10^2$    & \multicolumn{1}{|c}{ 0.919} & 0.128 &  0.014 &  0.058 &  0.028 &  0.011 &  0.011 &  0.007 &  0.002 \\
$\sigma_9^i \times 10^2$    & \multicolumn{1}{|c}{ 0.709} & 0.458 &  0.229 &  0.100 &  0.091 &  0.081 &  0.096 &  0.079 &  0.070 \\
$\sigma_{10}^i \times 10^2$ & \multicolumn{1}{|c}{ 5.948} & 0.499 &  0.072 &  0.617 &  0.171 &  0.090 &  0.106 &  0.025 &  0.014 \\
$\sigma_{11}^i \times 10^2$ & \multicolumn{1}{|c}{ 0.000} & 0.000 &  0.000 &  0.259 &  0.082 &  0.045 &  0.061 &  0.015 &  0.008 \\
$\sigma_{12}^i \times 10^2$ & \multicolumn{1}{|c}{ 0.152} & 0.017 &  0.016 &  0.071 &  0.019 &  0.019 &  0.021 &  0.016 &  0.016 \\
$\sigma_{13}^i \times 10^2$ & \multicolumn{1}{|c}{ 0.604} & 0.071 &  0.052 &  0.155 &  0.020 &  0.051 &  0.050 &  0.014 &  0.032 \\
$\sigma_{14}^i \times 10^2$ & \multicolumn{1}{|c}{ 0.973} & 0.358 &  0.103 &  0.098 &  0.068 &  0.039 &  0.064 &  0.072 &  0.022
\end{tabular}
\end{ruledtabular}
\end{table*}

The experimental measurements $a_{\rm CP}$ and $A_{\rm CP}$ therefore
depend linearly on three physics quantities
$\asld$, $\asls$ and $\delta_\Gamma$. There are three measurements
of the inclusive single muon asymmetry, and six measurements of the like-sign dimuon asymmetry. In total there
are nine independent measurements. Since the coefficients in Eq.~(\ref{yi}) are different
for different IP samples, the physics quantities $\asld$, $\asls$ and $\delta_\Gamma$
can be obtained by minimization of the $\chi^2$.

In this $\chi^2$ minimization the correlation between measured values $a_{\rm CP}$, $A_{\rm CP}$,
$F_{SS}$ and $F_{SL}$ are taken into account.
The expression for $\chi^2$, which takes into account this correlation,
can be written as
\begin{eqnarray}
\chi^2 & = \sum_{i,j = 1}^9 & (y^i - K_d^i \asld - K_s^i \asls - K_\delta^i \delta_\Gamma) V^{-1}_{ij} \nonumber \\
       &   &                  (y^j - K_d^j \asld - K_s^j \asls - K_\delta^j \delta_\Gamma).
\end{eqnarray}
The indexes $i$ and $j$ correspond to the IP samples.
The covariance matrix $V_{ij}$ is defined as
\begin{equation}
V_{ij} = \sum_{k=1}^{14} \sigma_k^i \sigma_k^j \rho^k_{ij}.
\end{equation}
$\sigma^i_k$ is the contribution to the uncertainty on $y^i$ from a given source $k$.
The list of the sources of uncertainty on $y^i$ is given in Table~\ref{tapp3}.
The parameters $\rho^k_{ij}$ are the correlation between
the measurements $i$ and $j$ for the source of uncertainty $k$.
The assignment of the correlation of different sources of uncertainties is set based on the
analysis procedure. For example, the same muon detection asymmetry $\delta_i$ is used
to measure both $a_{\rm CP}$ and $A_{\rm CP}$ for each IP.
Therefore the correlation due to this source is
set to 1.
The values of $y^i$ and $\sigma_k^i$ are given
in Table~\ref{tapp2}.

Table~\ref{tapp3} gives the values of the correlation coefficients $\rho^k_{12}$ and $\rho^k_{14}$.
For all other correlation coefficients the following relations apply:
\begin{eqnarray}
\rho^k_{12} & = & \rho^k_{13} = \rho^k_{17} = \rho^k_{18} = \rho^k_{19} = \nonumber \\
            & = & \rho^k_{23} = \rho^k_{24} = \rho^k_{26} = \rho^k_{29} = \nonumber \\
            & = & \rho^k_{34} = \rho^k_{35} = \rho^k_{37} = \rho^k_{47} = \nonumber \\
            & = & \rho^k_{48} = \rho^k_{49} = \rho^k_{59} = \rho^k_{67} = \nonumber \\
            & = & \rho^k_{79}.
\end{eqnarray}
\begin{eqnarray}
\rho^k_{14} & = & \rho^k_{15} = \rho^k_{16} = \rho^k_{25} = \rho^k_{27} = \nonumber \\
            & = & \rho^k_{28} = \rho^k_{36} = \rho^k_{38} = \rho^k_{39} = \nonumber \\
            & = & \rho^k_{45} = \rho^k_{46} = \rho^k_{56} = \rho^k_{57} = \nonumber \\
            & = & \rho^k_{58} = \rho^k_{68} = \rho^k_{69} = \rho^k_{78} = \nonumber \\
            & = & \rho^k_{89}.
\end{eqnarray}


This input is used to obtain the results given in Section~\ref{sec_interpretation}
and the correlation matrix given in Table~\ref{tab-cor}.


\begin{thebibliography}{99}
\bibitem{cpconj}
Throughout the article, we imply the CP-conjugate reactions.
\bibitem{eta}
Pseudorapidity is defined as $\eta \equiv -\ln{ \left[ \tan(\theta/2) \right] }$, where
$\theta$ is the polar angle of the track relative to the
proton beam direction.
\bibitem{ip}
The impact parameter IP is defined as the distance of closest approach of the track to
the $p \bar{p}$ collision point projected onto the
plane transverse to the $p \bar{p}$ beams.
\bibitem{ktomu}
The notation $K \to \mu$ refers to charged kaons reconstructed by the central tracker and
identified as muons. This group of particles
includes kaon decays in flight, punch-through, or muon misidentification.
The corresponding notations for charged pions and protons are $\pi \to \mu$ and $p \to \mu$.
\bibitem{ip1}
Samples of IP are denoted with integer numbers
as defined in Tables~\ref{IP0} and \ref{IP1IP2}.
\bibitem{mucand}
Generally, throughout this article the word ``muon'' is an abbreviation for ``muon candidate
passing the quality requirements''.
\bibitem{ck}
In our previous publications \cite{D02,D03} all background fractions have been
corrected by the coefficients $C_K$ and $C_\pi$.
Therefore, the correspondence
between the present (``new") and previous (``old") notations of the background
fractions $f_K$, $f_\pi$, $F_K$ and $F_\pi$ is, e.g., $f_K({\rm new}) = C_K f_K({\rm old})$.
Similarly, we do not correct the background asymmetries $a_K$ and $a_\pi$ by the coefficients
$C_K$ and $C_\pi$. The correspondence between the present and previous notations
of $a_K$ and $a_\pi$ is, e.g., $a_K({\rm new}) = a_K({\rm old})/C_K$.
The value of the background contributions to the charge asymmetry, e.g., $f_K a_K$,
are not affected by this change of notation.
\bibitem{asyst}
The systematic uncertainties of $a_{\rm CP}$ and $A_{\rm CP}$
for the measurements with $\int L dt=6.1$ fb$^{-1}$ and $\int L dt=9.0$ fb$^{-1}$
were not provided in Refs. \cite{D02} and \cite{D03}. They are provided here for the first time.
\bibitem{proper}
The proper decay time is the decay time of a particle in its rest frame.
\end{thebibliography}
\end{document}